\documentclass[fleqn,usenatbib]{mnras}

\usepackage{newtxtext,newtxmath}

\usepackage[T1]{fontenc}

\DeclareRobustCommand{\VAN}[3]{#2}
\let\VANthebibliography\thebibliography
\def\thebibliography{\DeclareRobustCommand{\VAN}[3]{##3}\VANthebibliography}

\usepackage{graphicx}	
\usepackage{amsmath}	

\usepackage{amssymb}	
\usepackage{makecell}
\usepackage{float}
\usepackage{multirow}
\usepackage{hyperref}
\usepackage{url}
\usepackage[normalem]{ulem}
\usepackage{xparse}

\usepackage{xspace}

\newcommand{\mJy}{\text{\,mJy}\xspace}

\newcommand{\GHz}{\text{\,GHz}\xspace}
\newcommand{\MHz}{\text{\,MHz}\xspace}

\newcommand{\rad}{\text{\,rad}\xspace}
\newcommand{\radPerSqm}{\text{\,rad\,m}\mbox{$^{-2}$\xspace}}

\newcommand{\kpc}{\text{\,kpc}\xspace}
\newcommand{\cms}{\text{\,cm}\xspace}
\newcommand{\kms}{\text{\,km\,s}\mbox{$^{-1}$}\xspace}

\newcommand{\src}{Swift J1727\xspace}
\newcommand{\primary}{PKS B1934$-$638\xspace}
\newcommand{\secondary}{J1733$-$1304\xspace}
\newcommand{\polcal}{3C286\xspace}

\NewDocumentCommand{\relto}{o m m g}{%
  \ensuremath{%
    \IfNoValueTF{#1}{}{#1\,}%
    {#2}\,{#3}%
    \IfNoValueTF{#4}{}{\,\text{#4}}%
  }%
}

\newcommand{\SB}[1]{\,{#1}\,}

\title[Swift J1727.8$-$1613: Spectropolarimetry]{Spectropolarimetric detection of baryonic mass loading in a transient relativistic jet: application to the black hole X-ray binary Swift J1727.8$-$1613}

\newcommand{\AuthorList}{%
A.~K.~Hughes,$^{1}$\thanks{E-mail: hughesakh@gmail.com}
R.~P.~Fender,$^{1,2}$
G.~R.~Sivakoff,$^{3}$
F.~J.~Cowie,$^{1}$
I.~Heywood,$^{4,1,5,6,7}$
J.~H.~Matthews,$^{1}$\newauthor
K.~Savard,$^{1}$ 
F.~Carotenuto,$^{8}$
T.~D.~Russell,$^{9}$
C.~M.~Wood,$^{10}$
M.~C.~Baglio,$^{11}$
S.~Corbel,$^{12}$
S.~E.~Motta$^{11}$ 
}

\newcommand{\EndAffil}{%
$^{1}$Department of Physics, University of Oxford, Denys Wilkinson Building, Keble Road, Oxford OX1 3RH, UK\\
$^{2}$Department of Astronomy, University of Cape Town, Private Bag X3, Rondebosch 7701, South Africa\\
$^{3}$Department of Physics, University of Alberta, CCIS 4-181, Edmonton, AB T6G 2E1\\
$^{4}$SKA Observatory, Jodrell Bank, Lower Withington, Macclesfield, SK11 9FT, UK\\
$^{5}$Centre for Radio Astronomy Techniques and Technologies, Department of Physics and Electronics, Rhodes University, Makhanda 6140, South Africa\\
$^{6}$South African Radio Astronomy Observatory, 2 Fir Street, Observatory, 7925, South Africa\\
$^{7}$Jodrell Bank Centre for Astrophysics, Dept. of Physics \& Astronomy, University of Manchester, Manchester M13 9PL, UK\\
$^{8}$INAF-Osservatorio Astronomico di Roma, Via Frascati 33, I-00078, Monte Porzio Catone (RM), Italy\\
$^{9}$INAF, Istituto di Astrofisica Spaziale e Fisica Cosmica, Via U. La Malfa 153, I-90146 Palermo, Italy\\
$^{10}$International Centre for Radio Astronomy Research, Curtin University, GPO Box U1987, Perth, WA 6845, Australia\\
$^{11}$INAF–Osservatorio Astronomico di Brera, Via Bianchi 46, I-23807 Merate (LC), Italy\\
$^{12}$Université Paris Cité and Université Paris Saclay, CEA, CNRS, AIM, F-91190 Gif-sur-Yvette, France
}

\author[A. K. Hughes et al.]{%
\AuthorList
\\%
Affiliations are listed at the end of the paper
}

\date{Accepted XXX. Received YYY; in original form ZZZ}

\pubyear{\the\year{}}

\begin{document}
\label{firstpage}
\pagerange{\pageref{firstpage}--\pageref{lastpage}}
\maketitle

\begin{abstract}
Radio emission during X-ray binary outbursts is dominated by synchrotron radiation from relativistic jets, but is usually studied through total-intensity diagnostics such as flux density, spectra, variability, and proper motion. Radio spectropolarimetry provides a complementary probe of the magneto-ionic plasma through Faraday rotation and depolarisation. When the Faraday rotating material is local to the source, these effects can constrain the jet plasma composition and mass content, but this approach is rarely applied to transient jetted sources. We present MeerKAT L-band spectropolarimetry of the black hole X-ray binary \src\ during its 2023 outburst, focusing on the brightest radio flaring interval, when relativistic jets were being launched intermittently. Using multiple spectropolarimetric techniques, we identify transient Faraday-complex structure coincident with the major radio flares. The close temporal association with the flaring activity, together with the stability of the foreground Faraday screen, favours an origin local to the jet rather than in the ISM or in a separate local screen external to the emitting plasma. Since internal Faraday rotation is suppressed in a pure electron--positron plasma, the data favour a dominant electron--proton component. Interpreting the characteristic Faraday thickness as internal rotation, and anchoring the magnetic-field and size scales with synchrotron self-absorption arguments, we infer a characteristic Faraday-rotating mass of order $M_{\rm rot}\sim10^{21}{\rm\,g}$, corresponding to only a small fraction, $f_{\rm rot}\sim10^{-3}$, of the accreted mass available during the flare. These results show that time-domain spectropolarimetry can turn transient Faraday complexity into a diagnostic of jet composition, mass loading, and plasma evolution in X-ray binary outbursts, and potentially other transient jetted sources.
\end{abstract}




\section{Introduction}
X-ray binaries (XRBs) are accreting stellar systems in which a compact object---either a neutron star (NS) or a black hole (BH)---draws material from a non-degenerate companion star. They are commonly classified by donor-star mass; this work focuses on a low-mass BH XRB, in which a stellar-mass black hole \citep[typically ${\sim}7\,{\rm M_\odot}$;][]{BlackCAT} accretes from a donor of mass ${\lesssim}1\,{\rm M_\odot}$. Observationally, these systems are characterised by sporadic, luminous, multi-wavelength outbursts, during which the accretion flow evolves through a sequence of distinct regimes. Unless otherwise stated, we hereafter use XRBs to refer specifically to low-mass BH XRBs.

As in many other classes of accreting compact object \citep[e.g. tidal disruption events, gamma-ray bursts, and active galactic nuclei, AGN;][]{livio2002,kumar2015,DeColle2020,ruiz2021,Gottlieb2023}, accretion in XRBs is often accompanied by the launching of astrophysical jets. These highly collimated plasma outflows, travelling at relativistic speeds, transport mass and energy across distances exceeding $\gtrsim10^{7}$ times the characteristic size of the accretor \citep[e.g.][]{Corbel2002Ejecta,Gallo2005Bowshock,Blandford2019AGNreview,Carotenuto2024}. This combination of dynamical range and energy output ($\sim10^{30}$--$10^{50}$ erg s$^{-1}$) makes jets key sites for high-energy particle acceleration \citep{deAngelis2008NVHEreview,Matthews2020} and major agents of mechanical and radiative feedback \citep{fabian2012,Hardcastle2020}. Within this broader jet population, XRBs are uniquely accessible because their stellar-mass accretors drive accretion and ejection cycles that unfold on humanly observable time-scales. They therefore provide access to stages of jet evolution that are effectively inaccessible in their more energetic, but much more slowly evolving, AGN analogues \citep{McHardy2006AGNXRB}.

XRBs spend most of their lifetimes in low-luminosity quiescent states \citep[typically quantified by X-ray luminosities, $L_X\,{<}\,10^{34}{\rm\,erg\,s^{-1}}$;][]{Plotkin2013QS}, punctuated by sporadic bright outbursts ($L_X{\geq}\,10^{38}{\rm\,erg\,s^{-1}}$) that can last months to years. During these outbursts, the accretion flow evolves through a sequence of `accretion states', defined primarily by the X-ray spectral and timing properties of the source. In the simplified picture most relevant here, the two canonical regimes are the \textit{hard} and \textit{soft} states, with the outburst typically beginning in the hard state, transitioning to the soft state, and later returning to the hard state \citep[although this phenomenology can be substantially more complex; see, e.g.][for reviews]{mclintock2006,Kalemci2022review,DeMarco2022states}. Jet properties are closely coupled to this accretion-state evolution, although the jet itself is usually traced at radio, rather than X-ray, wavelengths \citep[see, e.g.][for reviews]{fender2004,fender2006}.

In the hard accretion state, the X-ray spectrum is dominated by emission from a hot inner flow or corona \citep{Thorne1975,Shapiro1976}, with sub-dominant contributions from an accretion disc \citep{Done2007}. In this state, the radio jet appears to be coupled to the accretion flow, giving rise to the observed radio--X-ray luminosity correlation \citep{Hannikainen1998b,Corbel2003,gallo2003}. The radio emission is commonly interpreted in the Blandford--Königl picture \citep{blandfordkonigl1979}, in which the observed spectrum arises from a superposition of synchrotron-emitting regions whose properties vary with distance downstream from the jet base. This produces the characteristically flat or mildly inverted radio spectra of hard-state jets \citep[$\alpha\,{\gtrsim}\,0$, where $F_{R,\nu} \propto \nu^\alpha$;][]{Corbel2000,Fender2001alpha}. These steady, partially self-absorbed outflows are commonly referred to as `compact jets', reflecting the fact that their structure is generally unresolved except with very long baseline interferometry (VLBI) on milli-arcsecond scales \citep[e.g.][]{Stirling2001,Wood2024}.

In contrast, X-rays in the soft accretion state are dominated by emission from the accretion disc, while the compact jet is strongly suppressed or quenched \citep[e.g.][]{Tananbaum1972,Fender1999gx339quench,coriat2011,drussell2011,russell2019,Maccarone2020quench}. During the hard$\rightarrow$soft transition, however, many systems undergo bright radio flaring episodes. The radio spectra of these flares evolve rapidly from optically thick to optically thin synchrotron emission as the emitting plasma expands and the self-absorption frequency decreases \citep[e.g.][]{vdl1966,Hjellming1988,atetarenko2017,fender2019,CF_PLACEHOLD}.Such behaviour is often associated with the launch of transient ejecta \citep[e.g. as seen with VLBI;][]{millerjones2019,wood2021}, although radio variability alone does not uniquely imply discrete ejection events. Shock propagation within a pre-existing jet has also been considered as an origin of XRB variability and flaring \citep[e.g.][]{kaiser2000,turler2004,turler2011,Malzac2014}, similar to the shock-in-jet models more commonly applied to flaring AGN \citep{Blazar2015shockinjet}. As is common practice, we use `core' to refer to unresolved radio emission at the position of the black hole. We use `ejecta' in an observational sense, to refer to discrete radio components that are resolved as separate components from the core, without implying a specific physical origin.

When spatially resolved, these transient ejecta are typically observed as discrete, optically thin ($\alpha\,{\lesssim}\,-0.6$) components of synchrotron-emitting plasma moving away from the binary at mildly to highly relativistic speeds \citep[e.g.][]{mirabel1994,Fender1999,Brocksopp2002ejection,russel2019,bright2020,Carotenuto2021,Wood2025SWJ1717Ejecta}. Unlike the steady hard-state outflow, they separate from the core and evolve on milli-arcsecond to arcsecond angular scales. In some systems, the longest-lived ejecta remain detectable because of ongoing particle acceleration driven by interactions with the ambient interstellar medium \citep[ISM;][]{Corbel2002Ejecta,russell2019,Espinasse2020,bright2020,Carotenuto2021}. In the most extended cases, BH XRB jets can produce interaction regions tens of parsecs from the binary \citep[e.g.][]{Gallo2005Bowshock,Tetarenko2018b,Atri2025bowshock,Motta2025bowshock}, while ejecta that initially fade below the detection threshold may re-brighten months or years later as they encounter denser regions of the ISM \citep[e.g.][]{Corbel2002Ejecta,Carotenuto2021}.

The key point is that this evolution unfolds on month-to-year timescales, making it directly accessible to time-domain observations. For this work, the most relevant phases are the bright flaring episodes and subsequent launching of discrete jet components. These phases provide opportunities to infer fundamental jet properties, including speeds, energetics, geometry, and environmental coupling, through synchrotron self-absorption modelling of flaring light curves \citep[e.g.][]{fender2019,CF_PLACEHOLD}, proper-motion and deceleration measurements of spatially resolved ejecta \citep[e.g.][]{Carotenuto2024}, and, in favourable cases, direct signatures of relativistic jet dynamics such as the Lense--Thirring precession inferred in V404~Cygni \citep{millerjones2019}. This is not to say that compact jets are less useful. They offer their own routes to physical inference, including timing analyses \citep[e.g.][]{Tetarenko2021Timing}, VLBI-resolved mapping and core-shift measurements \citep{2023PrabuCoreShift,Wood2024,2025ZdziarskiJ1727steady}, and energetic arguments based on large-scale interaction regions. However, whether applied to compact jets or transient ejecta, these approaches are still built primarily on total-intensity information: flux densities, spectra, positions, and variability. Here, we instead focus on the polarisation properties of synchrotron emission, which provide a critical complement to the jet properties recoverable from total intensity alone.

Synchrotron radiation is intrinsically polarised, making polarimetry a direct probe of jet magnetic-field structure and plasma content. For a uniform magnetic field and a typical electron energy distribution, the maximum linear polarisation fraction is ${\sim}75\%$ and ${\sim}10\%$ for optically thin and optically thick emission respectively \citep{Ginzburg1969,longair2011}. In this idealised case, the observed linear polarisation angle is tied to the sky-projected magnetic-field direction, so ordered field geometries produced by shock compression or velocity shear can connect the measured polarisation angle to the projected jet position angle \citep{Laing1980,hughes1985jetpolshocks,Bell2004_magnetic}. In practice, XRB jets often deviate strongly from this picture: disordered or turbulent magnetic fields, unresolved overlapping components \citep[e.g.][]{Beck2003depol}, and opacity transitions or inhomogeneous source structure \citep{Jones1977,Jones1977_inhomo} all act to reduce the observed polarisation fraction and rotate the polarisation angle \citep[e.g.][]{Sokoloff1998,Lyutikov2005_jetpol}. Empirically, XRBs with linear-polarisation detections are typically observed at only the ${\sim}1$--$10\%$ level \citep[e.g.][]{gallo2004,Rushton2010,brocksopp2007,brocksopp2013,curran2015}, although values range from ${<}\,1\%$ in some systems \citep[e.g.][]{hughes2023,Kravtsov2025_cygx1_pol} to tens of per cent in unusually highly polarised ejecta \citep[e.g. Swift J1745--26;][]{curran2014}.

At radio frequencies, Faraday effects imprint a strong signature on linearly polarised synchrotron emission, especially toward the low-frequency end of the band \citep{Brentjens2005,rundick2023RMSynth}. Broadband radio spectropolarimetry can therefore separate the intrinsic polarisation state of the synchrotron emission from propagation-induced rotation and depolarisation. When the Faraday-rotating material is external to the emitting region, the measured rotation probes the intervening magneto-ionic medium. When it is internal to the jet, it can instead encode the density, magnetic field, and composition of the jet plasma, through the observed wavelength-dependent depolarisation/repolarisation \citep[e.g.][]{Burn1966RMSynth,Sokoloff1998,Brentjens2005,stirling2004,Osull2012,2016Anderson,Pasetto2018}. Circular polarisation, produced either intrinsically by synchrotron emission or through Faraday conversion of linear to circular polarisation in a relativistic plasma \citep{Legg1968,Pacholczyk1973,Jones1977,kennett1998}, can in principle provide complementary constraints on magnetic-field geometry and particle content \citep[e.g.][]{Wardle1998,Eblin2003,Gabuzda2008,Osull2013}. However, it is typically much weaker than linear polarisation, and has only been detected in a small number of XRBs at the few-tenths-of-a-percent level \citep{fender2000,Macquart2002,fender2003}.

In this work, we apply spectropolarimetric techniques developed primarily for AGN studies and surveys to the bright flaring phase of the 2023 outburst of Swift~J1727.8$-$1613, hereafter \src, using the dense time sampling available for this outburst. The source has been extensively studied across the electromagnetic spectrum \citep[see, for example,][for a radio-focused view]{Hughes2025_lrlx,HughesAKH2025a}, with contemporaneous X-ray behaviour discussed by \citet{Yu2024SwiftJ1727} and others \citep[e.g.][]{ingram2023ixpe,Peng2024,Cao2025SwiftJ1727}. Here, we search for transient Faraday rotation and depolarisation during the radio flaring. The remainder of this paper is structured as follows. Section~\ref{sec:obs} describes the observations, calibration, imaging, and analysis methodology. Section~\ref{sec:results} presents the spectropolarimetric results without assuming their physical origin. Section~\ref{sec:discussion} argues that the observed behaviour is intrinsic to the jet, and uses it to infer internal jet properties. Finally, Section~\ref{sec:conclusion} summarises the findings and discusses implications for applying this approach to future XRB outbursts.

\section{Observations and Analysis}
\label{sec:obs}
We observed \src as part of the MeerKAT Large Survey Programme ThunderKAT \citep{tkat}, with monitoring continuing under the subsequent ``X-KAT'' programme (Proposal ID: SCI-20230907-RF-01). In this work, we focus on 15 epochs spanning the \relto{\sim}{3}{months} from 2023 September 04 to 2023 November 25 (MJD 60191--60273). These observations cover the brightest radio flaring, but end before the core had faded substantially and the source became strongly extended by the propagation of discrete ejecta. This avoids complications in spectral extraction from comparably bright, partially overlapping components. We also include three supplementary observations obtained between 2024 February 10 and 25 (MJD 60350--60365), when \src exhibited clearly resolved, polarised components. These later observations allow us to test whether line-of-sight effects alone can explain the observed spectropolarimetric behaviour. Finally, we include one further observation from 2024 March 31 (MJD 60400), after the source had returned to the hard state and the radio core had depolarised.

MeerKAT is a radio interferometer consisting of 64 13.5-m antennas, each equipped with two orthogonal linearly polarised feeds \citep{MeerKATCitation}. We used the L-band receiver, with a central frequency of \relto{\sim}{1.28}{GHz} and a total unflagged bandwidth of 856\MHz, which at the time of observing was divided evenly into 32768 frequency channels. To reduce the data volume, we averaged the data in 32-channel increments, resulting in 1024 channels for subsequent processing. Each observation consisted of a single 15-minute scan on source, bracketed by two 2-minute scans of a nearby gain calibrator (\secondary). For flux density scale and bandpass calibration, we included either two 5-minute scans of \primary (J1939$-$6342) at the beginning and end of each observation, or a single 10-minute scan at one end. In addition, a single 10-minute scan of \polcal (J1331$+$3030) was obtained for cross-hand phase calibration for all observations except that of 2023 October 16 (MJD 60247), which inadvertently did not include a \polcal scan. For that epoch, we used the secondary calibrator to determine the cross-hand phase (see Appendix~\ref{sec:appendix_1GC_Oct16}). A full description of the MeerKAT campaign is presented in \citep{HughesAKH2025a}.

Throughout the remainder of this paper, we adopt the standard Stokes formalism, in which polarised emission is described by four flux densities: Stokes $I$, $Q$, $U$, and $V$. Stokes $I$ measures the total flux density; $Q$ and $U$ describe the linear polarisation; and $V$ measures the circularly polarised emission.

\subsection{Full Stokes calibration and imaging routine}
\label{sec:polkat}
In this subsection, we detail our semi-automated calibration and imaging routine \textsc{polkat}\footnote{Found at: \url{https://github.com/AKHughes1994/polkat}} \citep{polkatASCL}, which is a modified version of \textsc{oxkat}\footnote{Found at: \url{https://github.com/IanHeywood/oxkat}} \citep{oxkat2020}, adapted for full-Stokes processing. The routine was applied to each epoch, adopting the following workflow: 

\subsubsection{Reference Calibration (1GC)}
\label{sec:polkat_1GC}
We used \textsc{casa} \citep[v6.6;][]{Casa2022} to remove radio frequency interference (RFI) and derive calibration solutions from the calibrator fields (i.e. \textit{first-generation} or ``reference'' calibration). We began with an initial round of flagging, manually removing frequency channels known to be corrupted by persistent RFI (e.g. satellites), and applying the \textsc{casa} auto-flagging routines \textsc{tfcrop} and \textsc{rflag} to excise transient RFI. For continuum imaging, it is common to flag specific frequency ranges on short (e.g. \relto{<}{600}{m}) baselines, which are more susceptible to deterministic RFI; however, we found that this targeted strategy introduced artificial steps in the Stokes $I$ spectrum at the level of \relto{\pm}{1}\% in flux density. Although these features were sub-dominant in $Q$, $U$, and $V$ relative to the noise, we adopted a more conservative approach and flagged all baselines with \relto{<}{600}{m}. Given that the target is bright (\relto{>}{50}{mJy}) and unresolved, this has negligible impact on detectability. Even during the multi-component epochs, the angular separation of the point-like components (\relto{\lesssim}{15^{\prime\prime}}) remains well below the largest angular scale recoverable at the top of the band: \relto{\sim}{30^{\prime\prime}} for a 600\,m baseline at \relto{\sim}{1.6}{GHz}.

Following flagging, we initialised the sky model of \primary with \textsc{crystalball} \citep{crystalball}. Using \primary, we solved for the parallel-hand delay and bandpass solutions with the \textsc{casa} tasks \textsc{gaincal} (\texttt{gaintype='K'}) and \textsc{bandpass} (\texttt{gaintype='B'}), respectively. As \primary is unpolarised, we used it to correct for instrumental polarisation arising from imperfectly orthogonal feeds (i.e. \textit{leakage}) with the \textsc{casa} task \textsc{polcal} (\texttt{gaintype='Df'}).

After bootstrapping the \primary calibration solutions onto \secondary and \polcal, we derived complex gain solutions using \textsc{gaincal}. Adopting a similar approach to \citep{Taylor2024}, we solved for feed-averaged solutions (\texttt{gaintype='T'}) rather than per-feed solutions (\texttt{gaintype='G'}). The feed-averaged solutions correct for temporal drifts but, unlike per-feed solutions, do not modify the relative gain amplitudes of the feeds. These relative amplitudes are intrinsic to the polarisation properties of the calibrators and therefore require a full-Stokes calibration mode, as well as \textit{a priori} knowledge of the ionospheric rotation measure for polarised calibrators. The implicit assumption of this approach is that the relative feed amplitudes are adequately corrected during bandpass calibration (as the bandpass calibrator is unpolarised), and that atmospheric line-of-sight differences and temporal evolution between the two feeds remain sufficiently small over our observing frequencies and scan ordering. At MeerKAT's moderate-to-low observing frequencies ($\sim0.5\text{--}4{\rm\,GHz}$), this assumption appears to be well satisfied, with the calibrator properties remaining stable across the observations (see Appendix~\ref{sec:appendix_systematics}). The complex gain amplitudes were then placed on an absolute flux-density scale using the \textsc{casa} task \textsc{fluxscale}, by comparison with the reference values derived for \primary.

The final reference calibration step used the strongly polarised source \polcal to solve for the instrumental cross-hand phase ($\rho$) with \textsc{polcal} (\texttt{gaintype='Xf'}). Without this correction, the measured Stokes $U$ and $V$ flux densities are mixed into one another. For linear-feed interferometers, a linearly polarised and circularly unpolarised calibrator (e.g. \polcal) can be used to derive cross-hand phase solutions without a full-polarisation calibrator model \citep{Hugo2024}. This solution is defined only up to a ${\pm}\pi$ degeneracy arising from $\rho = \arctan(-V^{\prime}/U^{\prime})$, where the primes denote Stokes parameters measured in the feed frame after parallel-hand and leakage corrections. We found that initialising a quasi-arbitrary polarisation model with the correct Stokes quadrant provided sufficient information for \textsc{casa} to resolve this degeneracy natively. Using the \textsc{casa} task \textsc{setjy}, we defined a point-source model for \polcal with Stokes $I$, $Q$, $U$, and $V$ flux densities of 1.0, 0.0, 0.5, and 0.0 Jy, respectively (\texttt{fluxdensity=[1.0,0.0,0.5,0.0]}). We tested alternative initialisation values and found that the cross-hand phase solutions were insensitive to the precise starting model, provided that $V \equiv 0.0$ and $U\,{\gtrsim}\,Q\,{>}\,0$. These conditions are consistent with the \textit{a priori} known polarisation angle of \relto{\sim}{30}{deg} at L-band \citep{Hugo2024}, which implies a polarised spectrum dominated by positive Stokes $U$ emission. For some epochs, the ionospheric rotation measure was large enough for the measured $V^{\prime}$ and $U^{\prime}$ spectra to change sign at the lowest frequencies. In these cases, \textsc{casa} could no longer resolve the $\pm\pi$ degeneracy, producing discontinuities in the cross-hand phase spectrum at the zero crossing and introducing artificial spectropolarimetric \textit{breaks}. We corrected these cases manually by adding or subtracting $\pi$ from the \polcal phase solutions beyond the discontinuity, and flagged channels with low signal-to-noise ratio.

After solving for the cross-hand phase, each calibration table was interpolated onto \src using the nearest solution in time. We deferred the correction for parallactic angle rotation until the final stage of self-calibration, noting that this choice has a negligible effect for phase-only self-calibration. The calibrated target visibilities were then exported using the \textsc{casa} task \texttt{split}. Finally, before beginning iterative imaging, we performed an additional round of flagging with the \textsc{casa} autoflaggers and \textsc{tricolour} \citep{tricolour}.

\subsubsection{Self-Calibration and Imaging (2GC)}
\label{sec:polkat_2GC}
Following reference calibration, we performed an iterative imaging, masking, and self-calibration procedure. We first generated a deep preliminary image by combining the off-peak epochs of \src with the CLEAN-based imager \textsc{wsclean} \citep{Hogbom1974,wsclean}. From this image, we constructed a deconvolution mask with \textsc{breizorro} \citep{breizorro}, allowing subsequent deconvolution and self-calibration modelling to be restricted to genuine sky emission. For each epoch, this mask was then used for an initial shallow deconvolution of all four Stokes parameters, cleaning pixels $>10\sigma$ down to 1$\sigma$ with the \texttt{local-rms} option in \textsc{wsclean}. Here, $\sigma$ denotes the image root-mean-square noise from emission-free, noise-dominated regions of the image.

We then performed direction-independent, phase-only self-calibration (\textit{second-generation} calibration) with \textsc{quartical} \citep{quartical} on dump time-scales (8\,s). The resulting images were used to update the sky model, before repeating the imaging and self-calibration cycle (again on dump timescales) with \texttt{local-rms} enabled but a lower cleaning threshold that included all masked pixels $>3\sigma$. We then performed a final, deeper masked deconvolution with \texttt{local-rms} disabled. In total, the procedure comprised two self-calibration cycles and three imaging steps, excluding the preliminary image used to define the initial mask. After confirming that self-calibration reduced the image noise without introducing artificial spectral structure, we adopted the final self-calibrated images as the data products.

For most epochs, phase-only self-calibration was sufficient. During the brightest flaring epochs, however, the source region became dynamic-range limited, particularly at lower frequencies. This was most apparent during the \relto{\sim}{1}{Jy}-level epoch on 2023 October 14 (MJD 60231). To improve the calibration in these cases, we applied complex self-calibration during the second self-calibration cycle, solving for the full 2$\times$2 amplitude-and-phase Jones matrix, including the off-diagonal cross-hand terms, on longer $\sim4$-min solution intervals to increase the signal-to-noise ratio. We applied this step to three epochs: 2023 October 06, 14, and 22 (MJD 60223, 60231, and 60239). The first two were the brightest flaring epochs, while the third had the highest polarised signal-to-noise ratio among the otherwise spectrally simple epochs. The October 22 observation therefore served as a control case, allowing us to verify that complex self-calibration did not introduce the artificial spectropolarimetric structure that this work seeks to identify.

For the epochs treated with complex self-calibration, we verified that the resulting flux densities remained consistent with the range allowed by the dynamic-range-limited images, with no evidence for an unphysical rescaling of the target flux density. We also tested whether a bright off-axis field source at 17h24m45s $-$16d56m41s (\relto{\sim}{1}{deg} off-axis) could bias the gain solutions. Although this source was sub-dominant, it could in principle contribute residual direction-dependent structure. We therefore applied direction-dependent self-calibration (\textit{third-generation} calibration; ``peeling'') to remove the off-axis source emission, finding that this had a negligible effect on the final image properties. Together, these tests confirm that the final spectra are not driven by the additional calibration steps, but are robust features of the data.

During imaging, we adopted Briggs weighting \citep{Briggs1995} with a robustness parameter of 0 to maximise sensitivity while preserving reliable deconvolution. Although more natural weighting schemes, with robustness \relto{>}{0}, typically yield lower rms noise, the MeerKAT synthesized beam becomes increasingly non-Gaussian at higher robustness, hindering accurate deconvolution. Each imaging run produced a set of frequency-channelised images, with the number of channels chosen according to the signal-to-noise ratio and outburst phase. In practice, we generated 64--512 channels per epoch.

The three epochs on 2023 October 06, 14, and 16, obtained during or shortly after the flare peaks \citep{HughesAKH2025a}, were imaged with 512 channels because they showed the most complex spectropolarimetric behaviour. The remaining 2023 observations were imaged with 128--256 channels, while the late-time 2024 February observations, when the source had significantly faded, were imaged with 64 channels. These choices were primarily practical: imaging 512$\times$8500$\times$8500 pixels for each of the four Stokes parameters is computationally expensive and unnecessary for epochs with simple spectra. This strategy therefore balances spectral resolution, deconvolution fidelity, and computational cost, while preserving the frequency sampling needed for the spectropolarimetric analysis that follows.

\subsubsection{Flux density extraction}
\label{sec:pol_extraction}
We extracted source properties for each epoch and Stokes parameter using the \textsc{casa} task \textsc{imfit}, which models the flux distribution with a user-specified number of elliptical Gaussian components and returns their flux densities and positions. For the 2024 epochs, which exhibited resolved jet ejecta, we adopted a dual-Gaussian model; for the remaining, spatially unresolved epochs, we used a single Gaussian. As the emission was consistent with one or more overlapping point sources, we fixed the component shapes to that of the synthesized beam. Flux density uncertainties (1$\sigma$) were estimated from the local rms noise, measured with \textsc{imstat} in an emission-free 2-arcminute region southwest of the source (corresponding to hundreds of synthesized beam areas). We note that these uncertainties do not include potential systematic contributions (e.g. residual calibration errors); instead, we adopt conservative Bayes-factor thresholds in our model comparison to mitigate the impact of such effects (see Section~\ref{sec:method_nested}).

For each channel, we first fit the Stokes $I$ emission, and then used the Stokes $I$ position as the reference for fitting $Q$, $U$, and $V$. When the signal-to-noise ratio in a polarised Stokes parameter was \relto{<}{6\sigma}, we fixed the position to that measured in Stokes $I$, effectively performing forced aperture photometry. Otherwise, we allowed the position to vary to capture small frequency- or Stokes-dependent systematic offsets, such as those arising from residual phase errors, although this choice had a negligible effect in practice.

We then applied a final round of RFI excision to remove channels that passed the visibility-plane flagging but remained visibly corrupted in the extracted spectra. This was done through visual inspection, first, followed by iterative $\sigma$-clipping against fourth-order polynomial fits to the Stokes $I$, $Q$, and $U$ spectra, under the assumption that the intrinsic spectra should vary smoothly across the band. This assumption is reasonable for \src, since its small Faraday depth means that Faraday rotation does not introduce rapid oscillatory structure in $Q$ and $U$ across the band. For sources with much larger Faraday depths, the clipping could instead be applied to the polarised intensity. We conservatively flagged any frequency channel that was an outlier in any Stokes parameter across all Stokes parameters. The most affected ranges were \relto{\sim}{1-1.3}{GHz} and \relto{\sim}{1.5-1.6}{GHz}, where the RFI occupancy is known to be high, resulting in the loss of 30--60\% of the total bandwidth in each epoch. The resulting Stokes $I$, $Q$, $U$, and $V$ spectra form the basis of our polarimetric analysis. These flux-density fitting routines are available within \textsc{polkat}.

We do not detect Stokes $V$, and therefore find no significant circular polarisation in any epoch, with $|V/I|\,{\lesssim}\,0.08\%$ as a 3$\sigma$ upper limit. The remainder of this section therefore focuses on the linear polarisation analysis. The absence of detectable circular polarisation nevertheless provides tentative constraints on the emitting particle population, which we return to in Section~\ref{sec:disc_composition}.

\subsection{Spectropolarimetry}
The complex linearly polarised flux density\footnote{As our sources are unresolved, we refer to flux density rather than intensity.} can be written as
\begin{align}
P(\lambda^2) = Q(\lambda^2) + i U(\lambda^2),
\end{align}
where the dependence on wavelength squared, $\lambda^2$, is made explicit, as is standard in Faraday spectropolarimetric analyses. For brevity, we suppress this functional dependence where possible. The linearly polarised flux density and linear polarisation angle, $\psi$, are then
\begin{align}
|P| &= \sqrt{Q^2 + U^2},\\
\psi &= \frac{1}{2}\arg(P)
      = \frac{1}{2}\arctan\left(\frac{U}{Q}\right),
\end{align}
with the quadrant of the angle determined by the signs of both $Q$ and $U$. The latter quantity gives the orientation of the electric field vector of the incoming radiation, measured east of north. This is often referred to as the electric vector position angle; however, to avoid confusion with the position angle measured from VLBI imaging, we refer to it simply as the polarisation angle.

Because $|P|$ is a positive-definite quadratic combination of Gaussian random variables, it follows a Rice distribution \citep{RiceanStats}. The measured polarised flux density is therefore biased high at low signal-to-noise \citep{Vaillancourt2006}. For $|P|/\sigma_{QU}$\,\relto{>}{3}, an approximately unbiased estimator of the intrinsic polarised flux density is
\begin{align}
|P_0|^2 = |P|^2 - \sigma_{QU}^2.
\end{align}
Here, $\sigma_{QU}$ is the effective rms uncertainty in linear polarisation,
\begin{align}
\sigma_{QU}^2 = A_Q \sigma_Q^2 + (1 - A_Q)\sigma_U^2,
\end{align}
where $\sigma_Q$ and $\sigma_U$ are the rms noise values in the Stokes $Q$ and $U$ images at the source location. Following \citet{Hales2012}, we adopt $A_Q = 0.8$ when $\sigma_Q \geq \sigma_U$ and $A_Q = 0.2$ otherwise. In practice, $\sigma_Q \simeq \sigma_U$, consistent with the near-ideal case of equal noise in both Stokes parameters.

\subsubsection{Faraday depth, Rotation measure (RM) and RM Synthesis}
Radio waves commonly exhibit propagation-induced spectropolarimetric effects. The focus of this work is Faraday rotation, the rotation of the linear polarisation angle as radiation propagates through a magneto-ionic plasma threaded by a magnetic-field with a component along the line of sight. This rotation scales with wavelength squared, and its strength is quantified by the \textit{Faraday depth},
\begin{align}
\phi_f=\frac{q^3}{2\pi m_e^2 c^4}\int_{\text{source}}^{\text{observer}}n_{\ell} B_{\parallel}\,\mathrm{d}l
\end{align}
where $n_{\ell}$ is the lepton number density and $B_{\parallel}$ is the line-of-sight magnetic field strength, both of which may vary along the path length $l$. For generality, we write the charge as $q$, since Faraday rotation is dominated by the lightest charged particles. In an electron--proton plasma, this corresponds to the electrons. In a pair plasma, electrons and positrons contribute with opposite signs, cancelling exactly in the perfectly symmetric limit. Here, $m_e$ denotes the electron or positron mass.

In the simplest case of a single homogeneous Faraday-rotating screen external to the emitting region, hereafter referred to as \textit{Faraday simple}, the polarisation angle follows the canonical linear relation
\begin{align}
\psi(\lambda^2) = \psi_0 + \phi_f \lambda^2 ,
\end{align}
where $\psi_0$ is the intrinsic polarisation angle and $\phi_f$ is the Faraday depth. In this case, $\phi_f$ is what is typically referred to as the rotation measure. Departures from this linear behaviour arise when multiple unresolved components contribute at different Faraday depths, or when the emitting and rotating material are mixed along the line of sight. These effects produce non-linear spectropolarimetric structure, which we refer to generally as \textit{Faraday-complex} behaviour. When this complexity evolves with the source, it can be used to probe changes in the local magneto-ionic environment.

In such cases, it is useful to move from the wavelength-squared domain to a representation in Faraday depth. This is commonly done using \textit{rotation measure synthesis} \citep[see, e.g.][]{Burn1966RMSynth,Brentjens2005,Macquart2012,Hales2012,rundick2023RMSynth}, which treats the observed complex polarisation and its Faraday-depth distribution as a Fourier pair:
\begin{align}
P(\lambda^2)&=\int_{-\infty}^{+\infty}F(\phi_f)\, e^{2i\phi_f(\lambda^2 - \lambda_0^2)}\, \mathrm{d}\phi_f,\\
F(\phi_f)&=\int_{-\infty}^{+\infty}P(\lambda^2)\, e^{-2i\phi_f(\lambda^2 - \lambda_0^2)}\, \mathrm{d}\lambda^2.
\end{align}
We refer to this Faraday-depth representation as \textit{Faraday space}. The function $F(\phi_f)$ is the \textit{Faraday dispersion function}, referred to hereafter as the Faraday spectrum, and describes the complex linearly polarised flux density as a function of Faraday depth. This is complementary to $P(\lambda^2)$, which describes the same emission as a function of wavelength squared.

In practice, given the measured $P(\lambda^2)$, one evaluates $F(\phi_f)$ over a grid of trial Faraday depths. For each trial $\phi_f$, the data are de-rotated to a common reference wavelength squared, $\lambda_0^2$. This reference is often taken to be the $1/\sigma_{QU}^2$-weighted mean of the sampled $\lambda^2$ values; however, \citet{rundick2023RMSynth} highlighted the benefits of adopting $\lambda_0^2 = 0$ for most contexts, and we adopt that convention here. Faraday depths associated with real polarised structure then produce coherent summation of the de-rotated polarisation vectors across frequency, appearing as peaks or broad features in $|F(\phi_f)|$.

The simplest case is an unresolved component in Faraday space, which we refer to as a Faraday-thin component. In this limit, the component is described by the same linear rotation law as Equation~(7): the peak Faraday depth, $\phi_{f,\rm peak}$, corresponds to the slope that would be measured if the component were isolated. The linearly polarised flux density and polarisation angle at the reference wavelength squared are then estimated from the complex peak of the Faraday spectrum \citep[with a different bias correction;][]{Hales2012},
\begin{align}
|P_0|^2&=|F(\phi_{f,\rm peak})|^2 - 2.3\sigma^2_{f}, \qquad \text{where } \sigma_{QU}\sim\sigma_{f},\\
\psi(\lambda_0^2)&=\frac{1}{2}\arctan\left(\frac{\mathrm{Im}\left[F(\phi_{f,\rm peak})\right]}{\mathrm{Re}\left[F(\phi_{f,\rm peak})\right]}\right).
\end{align}
The corresponding intrinsic polarisation angle is then
\begin{align}
\psi_0 = \psi(\lambda_0^2) - \phi_{f,\rm peak}\lambda_0^2.
\end{align}
For the convention adopted here, $\lambda_0^2=0$, this reduces directly to $\psi_0=\psi(\lambda_0^2)$.

This simple interpretation becomes less direct when components are only marginally separated or unresolved in Faraday depth, or when the emission is intrinsically extended in Faraday space. A canonical example is the Burn slab model, in which synchrotron-emitting and Faraday-rotating material are mixed within a homogeneous plasma, so that emission from the far side of the slab undergoes more internal Faraday rotation than emission from the near side, broadening the component to a top hat in Faraday space \citep{Burn1966RMSynth}. We refer to this case, and to other intrinsically broadened structures in Faraday space, as Faraday-thick. We have also begun using imaging language, such as \textit{unresolved} and \textit{extended}, because the analogy captures how spectral sampling limits what can be recovered. Before describing our specific rotation measure synthesis approach, it is therefore useful to make this terminology explicit by comparison with synthesis imaging, where ideas such as unresolved structure, blending, and systematic sensitivity to extended emission are already familiar.

\subsubsection{Terminology and parallels to synthesis imaging}

As in synthesis imaging, finite and incomplete sampling introduces observational effects that shape what can be recovered. Here these effects are set by the sampling in $\lambda^2$, rather than by the baseline distribution, and determine both the resolution and scale sensitivity of the recovered spectrum.

The resolution analogy is the most direct. Multiplication by the sampling function in $\lambda^2$ produces a convolution in $\phi_f$, so intrinsically point-like, or Faraday-thin, components are broadened by a kernel given by the Fourier transform of the sampling function: the rotation measure spread function (RMSF). The RMSF is therefore directly analogous to the point-spread function, or \textit{dirty beam}, in synthesis imaging. In the \citet{rundick2023RMSynth} $\lambda_0^2=0$ formalism, its characteristic full width is approximately $\Phi_{\rm RMSF}\simeq2/(\lambda^2_{\rm max}+\lambda^2_{\rm min})$. For MeerKAT L-band, this gives $\Phi_{\rm RMSF}\,{\sim}\,16$~\radPerSqm, with modest epoch-to-epoch variations depending on flagging. We therefore refer to structures as unresolved when their intrinsic separation or width is small compared with this scale. This can apply either to two or more thin components with $\Delta\phi_f\equiv\phi_2-\phi_1\,{\ll}\,\Phi_{\rm RMSF}$, or to thick components with intrinsic widths $\sigma_\phi\,{\ll}\,\Phi_{\rm RMSF}$. As in synthesis imaging, the effective discrimination of sub-RMSF structure improves with signal-to-noise \citep[e.g.][]{Lobanov2005}.

The recovered spectra also suffer from scale-recoverability effects, similar to the loss of sensitivity to extended angular structure set by the shortest baselines in an interferometer. In the \citet{rundick2023RMSynth} formalism, this introduces a maximum recoverable scale, $W_{\rm max}\simeq0.67(\lambda^{-2}_{\rm min} + \lambda^{-2}_{\rm max})$ \citep[Eq.~A3 in][]{rundick2023RMSynth}. This is the FWHM of a thick component for which the power in the recoverable signal is reduced by a factor of two. Components broader than this nominal scale are not simply recovered as smooth broad structures. Instead, the interaction between the intrinsic width and the RMSF can produce apparent depolarisation and repolarisation structure across the width of the component in the observed spectrum. This is the analogue of `resolving out flux' in synthesis imaging. For MeerKAT L-band, this scale is $W_{\rm max}\,{\sim}\,25$~\radPerSqm.

The complementarity of these two scales is important. The RMSF width is driven primarily by the largest $\lambda^2$ channels, corresponding to the lowest observing frequencies, so moving to lower frequencies, such as UHF-band, improves the resolution in $\phi_f$. By contrast, $W_{\rm max}$ is set by $\lambda^2_{\rm min}$, corresponding to the highest observing frequencies, so moving to higher frequencies, such as S-band, improves sensitivity to broader structures. Improving one of these properties therefore comes at the cost of the other, unless the total frequency coverage is expanded rather than simply shifted. Thus, the analogy to synthesis imaging is two-sided: lower frequencies act like longer baselines, improving resolution, while higher frequencies act like shorter baselines, improving sensitivity to extended structure. This motivates interpreting Faraday spectra with the same caution applied to interferometric images. Compact features can be robustly identified, but the precision with which their intrinsic Faraday depths can be localised scales with the RMSF width, so poorer resolution gives poorer centroiding at fixed signal-to-noise. Broad structures, meanwhile, may be distorted or only partially recovered depending on the available $\lambda^2$ coverage.

Given the discussion above, Faraday spectra are affected by sampling-induced convolutional structure in much the same way as synthesis images. They can therefore be deconvolved using one-dimensional CLEANing. We generate and CLEAN the Faraday spectra using the \textsc{rm-tools} package \citep{rmtools}, adopting a two-tiered strategy. In the ideal thermal-noise-limited case, we would first use a blind threshold of $7\sigma_f$ to identify significant CLEAN components \citep[e.g.][]{Hales2012,Macquart2012}, and then CLEAN down to $3\sigma_f$ within masked regions centred on those initial peaks, with each mask extending by one RMSF FWHM on either side. However, \textsc{rm-tools} estimates $\sigma_f$ from the theoretical noise, which we found to be smaller than the observed noise by factors of \relto{\sim}{1.25-2}. As for our use of the \texttt{-local-rms} option in \textsc{wsclean} imaging, we therefore adopt the empirically measured noise rather than the nominal thermal value. We rescale the CLEAN thresholds by this excess factor, first identifying only high-significance peaks and then cleaning within the corresponding masks down to a lower, noise-scaled threshold.

We present these results in Section~\ref{sec:results_rmsynth}. Although this approach is computationally inexpensive and non-parametric relative to the $QU$-fitting described below, it inherits many of the limitations of (CLEAN-based) imaging. These include sensitivity to masking choices, strong dependence on the resolution and scale-recoverability limits discussed above, and systematic CLEAN biases \citep[e.g.][]{Condon1998,rundick2023RMSynth}. We therefore use the RM-synthesis results primarily as a descriptive diagnostic, both to inform the prior distributions adopted in the parametric fitting described in the following subsection and to provide a direct comparison with those fitted results.

\subsection{Parametric $QU$-fitting}
\label{sec:method_QUfit}
In contrast to RM synthesis, which describes the data in Faraday space, $QU$-fitting models the wavelength-dependent Stokes $Q$ and $U$ spectra directly. Extending the synthesis imaging analogy, it is closer to fitting the visibilities than to modelling the reconstructed image. This provides a flexible way to describe broadband spectropolarimetric behaviour using analytic models for both Faraday-thin and Faraday-complex emission \citep[although non-parametric $QU$-fitting approaches are being developed, e.g.][]{pratley_nonparametric_QU}. Standard analytic models include Faraday-thin components, external Faraday dispersion, and Burn slabs \citep{Burn1966RMSynth,Sokoloff1998}. In external Faraday dispersion, the emission passes through a separate Faraday-rotating screen with unresolved turbulent fluctuations in Faraday depth, producing wavelength-dependent depolarisation. Such models, along with combinations of multiple emitting components and propagation effects, have been widely applied to broadband spectropolarimetric observations of AGN \citep[e.g.][]{Osull2012,Pasetto2018,2026Oberhelman}, but remain comparatively unexplored in time-domain studies of transient sources such as XRB flares.

For X-ray binaries, both the appropriate Faraday-complex model(s) and the number of distinct components within an unresolved source are uncertain. This is particularly relevant during flares, when these systems can launch multiple rapidly evolving ejecta \citep[e.g.][]{millerjones2019,wood2021}, including in \src\ \citep{Wood2025SWJ1717Ejecta}, that remain unresolved in our MeerKAT observations. We therefore reduce the dimensionality by adopting the parametric Faraday-component framework of \citet{2016Anderson}.

Before fitting the spectropolarimetric models, we first model the Stokes $I$ spectrum in each epoch with a simple power law, $S_\nu\SB{\propto}\nu^\alpha$, using the \texttt{scipy.curve\_fit} implementation of non-linear least-squares fitting. This smooths over noise and residual systematics while providing a per-epoch in-band spectral index. Because the Stokes $I$ spectra are often limited by systematics rather than thermal noise, for example from secular spectral evolution of the bandpass calibrator, we rescale the Stokes $I$ uncertainties such that the fit has reduced $\chi^2 \equiv 1$. This primarily inflates the uncertainty on $\alpha$. We verified that replacing the power-law fit with a phenomenological fourth-order polynomial passing (almost) exactly through the data has a negligible effect on the final spectropolarimetric results. We then fit the fractional complex polarisation, $P/I = q + iu$, where $q = Q/I$ and $u = U/I$, modelling each spectrum as a superposition of Faraday-thin components, denoted ``S'' (for Faraday \textit{simple}), and Faraday-thick components, denoted ``T''.

\begin{figure}
    \centering
    \includegraphics[width=1.0\linewidth]{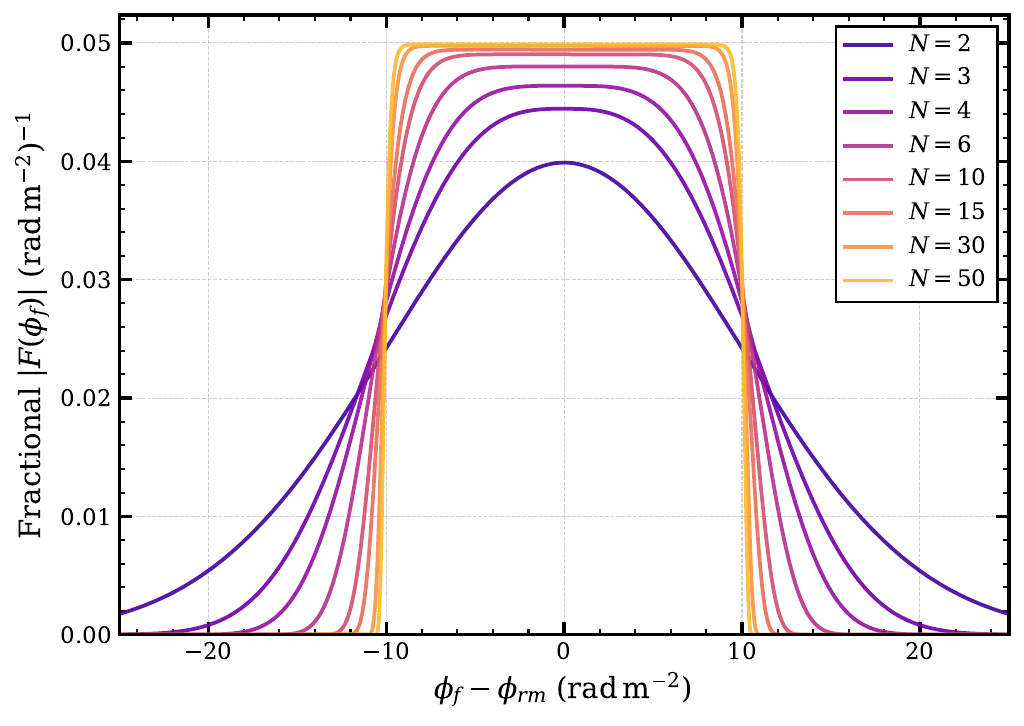}
    \caption{Schematic super-Gaussian Faraday-thick component profiles (Equation~15) for increasing shape parameter $N$. The $N=2$ case is Gaussian, while large $N$ (\relto{>}{15}) approaches a Burn-slab-like top-hat distribution.}
    \label{fig:method_supergaussian}
\end{figure}

A Faraday-thin component is represented in Faraday space as
\begin{align}
f_S(\phi_f) = p_0 \exp\left[ 2 i \psi_0 \right] \delta(\phi_f - \phi_{\rm rm}),
\end{align}
where $\phi_{\rm rm}$ is the Faraday depth of the rotating screen. This gives the $\psi$-linear-in-$\lambda^2$-space form
\begin{align}
p_S(\lambda^2) = p_0 \exp\left[ 2 i (\psi_0 + \phi_{\rm rm}\lambda^2) \right].
\end{align}
Following \citet{2016Anderson}, we model a Faraday-thick component as a super-Gaussian \citep{SuperGaussian} in Faraday space,
\begin{align}
f_T(\phi_f) = \frac{p_0}{C_N} \exp\left[ -\frac{|\phi_f - \phi_{\rm rm}|^{N}}{2\sigma_\phi^{N}} \right] \exp\left[ 2 i \psi_0 \right],
\end{align}
where $\sigma_\phi$ characterises width of the component in Faraday space, $N$ controls the profile shape, and $C_N = 2^{\,1 + 1/N}\,\sigma_\phi\,\Gamma(1/N)/N$ normalises the profile such that $\int |f_T(\phi_f)|\, d\phi_f = p_0$. Thus, $p_0$ represents the integrated fractional polarisation of the component. This formalism connects commonly used depolarisation models. For $N = 2$, the profile reduces to the Gaussian form associated with external Faraday dispersion, while for $N \rightarrow \infty$ (in practice \relto[N]{>}{15}) it approaches the top-hat Faraday-depth distribution corresponding to the Burn slab \citep{Burn1966RMSynth,Sokoloff1998,Brentjens2005}. A comparison of these limiting cases and intermediate profiles is shown in Figure~\ref{fig:method_supergaussian}. Because no general closed-form expression exists for the Fourier transform of the super-Gaussian profile for arbitrary $N$, we evaluate $p_T(\lambda^2)$ numerically during fitting.

Although the models are written in polar form in terms of $(p_0, \psi_0)$, we perform the fitting in Cartesian coordinates by defining,
\begin{align}
q_0 = p_0 \cos(2\psi_0), \qquad u_0 = p_0 \sin(2\psi_0),
\end{align}
We found that fitting in $(q_0, u_0)$ rather than $(p_0, \psi_0)$ is numerically more stable, particularly near $p_0 \rightarrow 0$. The recovered polar parameters are then obtained via $p_0 = \sqrt{q_0^2 + u_0^2}$ and $\psi_0 = \tfrac{1}{2}\arctan(u_0/q_0)$.

\begin{figure}
    \centering
    \includegraphics[width=1.0\linewidth]{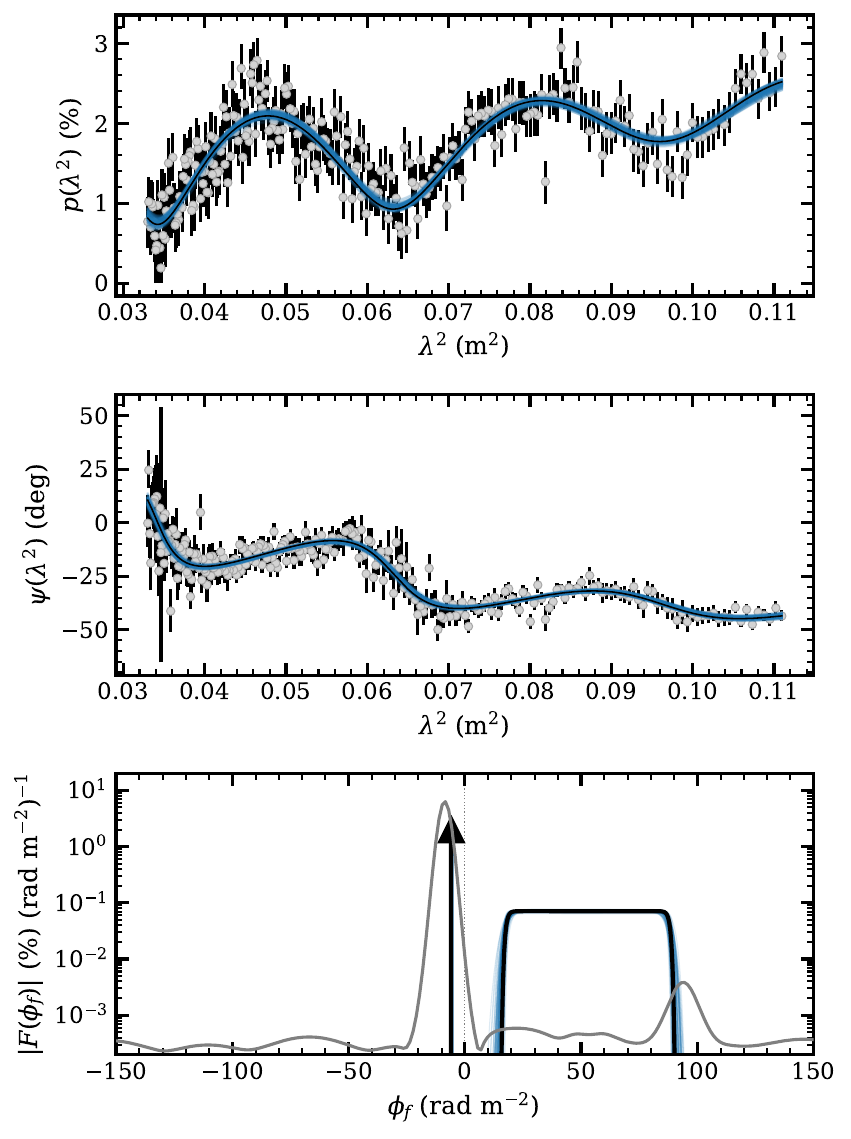}
    \caption{Simulated two-component ST model used to test the fitting procedure. We draw 250 synthetic data points from a known input model, inject Gaussian noise, and refit the data. The true model is shown in black, with posterior samples from the fit shown in blue. The top and middle panels show the linear polarisation fraction and polarisation angle as functions of $\lambda^2$, respectively. The bottom panel shows the corresponding Faraday spectrum, comparing the parametric $QU$-fitting model with the CLEANed RM-synthesis spectrum. For the Faraday-thin component, the posterior spread is narrower than the width of the model line in Faraday space, so the blue samples are hidden beneath the black curve. The injected parameters are accurately recovered, while the RM-synthesis spectrum shows little evidence for the broad Faraday-thick component.}
    \label{fig:method_simulation_example}
\end{figure}

The total fractional polarisation is modelled as the coherent complex sum of all components, $p(\lambda^2)=\sum_k p_k(\lambda^2)$. Figure~\ref{fig:method_simulation_example} illustrates the fitting procedure with a simulated two-component ST model, consisting of a Faraday-thin component and a Faraday-thick component at different Faraday depths. In the Faraday-spectrum panel, arrows mark Faraday-thin S components, which are $\delta$-functions in Faraday space and are plotted at a finite height only for visual clarity. As a consistency check, we generate mock data from the known input model, inject Gaussian noise, and apply the same fitting routine used for the observations. The model is accurately recovered, as shown by the agreement between the posterior samples and the injected truth.

The same example also illustrates why the Faraday-spectrum representation can make multiple polarised components more visually intuitive than their combined behaviour in polarisation fraction and angle alone. However, the number of components in this space should not be interpreted as the number of distinct emitting regions. This is particularly important for Faraday-thin emission, where several unresolved ejecta or compact emitting regions viewed through the same external screen can add coherently and appear as a single thin component. Conversely, they would appear as multiple sources in Faraday space only if the integrated emission contained components at distinct Faraday depths.

Comparison with the CLEANed RM-synthesis spectrum also highlights how broad Faraday structure can be suppressed or distorted by scale-filtering effects. In this example, the parametric model recovers the injected thick component, whereas RM synthesis primarily returns narrow features. This increased sensitivity to broad structure comes at the cost of model dependence: if the adopted functional form is inappropriate, the fit can return a precise but misleading description of the data. For example, our thick components assume a symmetric super-Gaussian profile in $\phi$ with a constant intrinsic polarisation angle, $\psi_0$, across the component. This cannot capture strongly asymmetric Faraday-depth distributions, or cases in which the complex phase varies across the emitting region, as may occur in more complex geometries \citep[e.g.][]{Sokoloff1998}. Future work should therefore explore both more physically motivated models and more flexible parametric descriptions, including profiles that allow skewness or internal phase structure.

Implicit in this approach, and in Faraday spectra more generally, is the assumption that all components share the same total-intensity spectral index across the observing band. Under this assumption, frequency-dependent structure in $q$ and $u$ is attributed to internal rotation and depolarisation, rather than to distinct intrinsic spectra. To test whether the data could instead be described by components with different intrinsic spectral shapes but a common rotation measure, we also fit spectral power-law, or ``P'', components of the form
\begin{align}
p(\lambda^2) = (q_{\lambda_0^2} + i u_{\lambda_0^2}) \left( \frac{\lambda^2}{\lambda_0^2} \right)^{-\beta/2} \exp\left[ 2 i \phi_{\rm rm}\lambda^2 \right].
\end{align}
Here, $\beta$ is defined such that the fractional polarisation scales as $p \propto \nu^{\beta}$ in frequency space, using $\lambda^2 \propto \nu^{-2}$. The parameters $q_{\lambda_0^2}$ and $u_{\lambda_0^2}$ are the fractional Stokes parameters evaluated at the reference wavelength $\lambda_0^2$, so that $p_{\lambda_0^2} = \sqrt{q_{\lambda_0^2}^2 + u_{\lambda_0^2}^2}$ is the fractional polarisation at $\lambda_0^2$. This differs from the Faraday-thin and Faraday-thick models above, where $p_0$ represents the integrated fractional polarisation in Faraday-depth space, equivalent to the intrinsic fractional polarisation at $\lambda^2 = 0$.

For multi-component P models, we tie the Faraday depths of all components, so that each component shares the same $\phi_{\rm rm}$ while retaining an independent spectral slope $\beta$. This represents multiple emitting regions with distinct intrinsic spectral behaviour that pass through a common Faraday-thin rotating screen. Curvature in $q(\lambda^2)$ and $u(\lambda^2)$ can therefore arise from intrinsic spectral structure rather than Faraday thickness. We discuss these models, and subsequently disfavour them, in Section~\ref{sec:result_pcomps}. More general models could allow each component to have both an independent Faraday depth and an independent spectral shape, but this would substantially increase the dimensionality and degeneracy of the parameter space. For complex combinations, per-component properties may be impossible to recover without additional \textit{a priori} constraints on the total-intensity spectra, for example from spatially resolved VLBI observations. We therefore defer such extensions to future work.

\begin{figure*}
\centering
\includegraphics[width=1.0\linewidth]{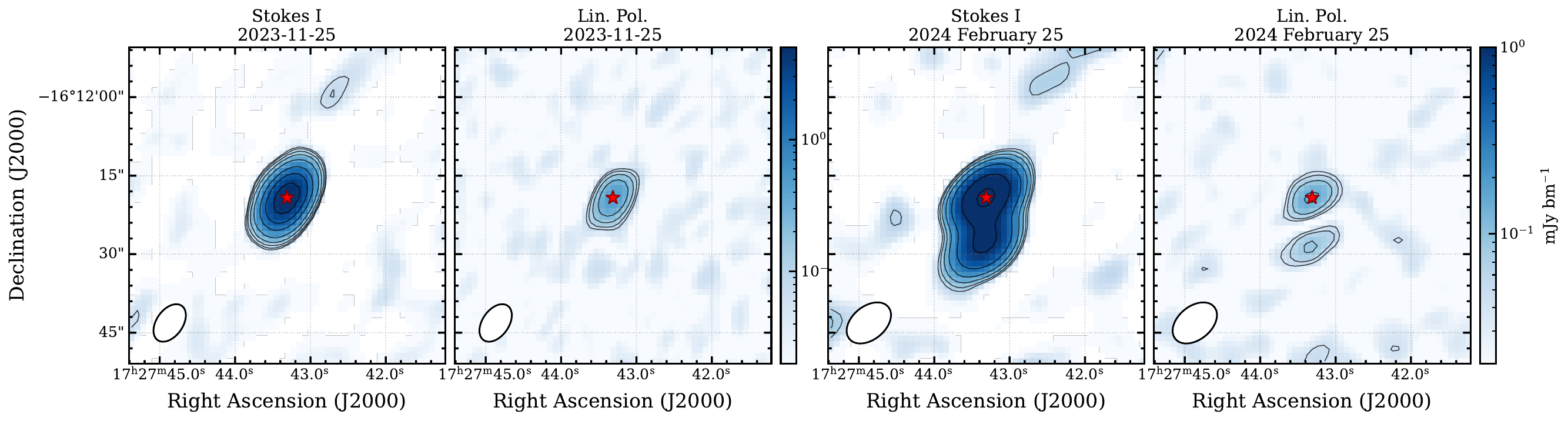}
\caption{A subset of MeerKAT images of \src\ during its outburst. The left-most panels correspond to final observations around the flaring interval on 2023 November 25 (MJD 60273), following the cessation of bright flaring, but while the emission was still point-like. The right-most panels correspond to 2024 February 25 (MJD 60365) and highlight the later multi-component morphology, including polarised ejecta at an angular separation of \relto{\sim}{5^{\prime\prime}}. The red star corresponds to the source position from \citep{millerjones2023a}.}
\label{fig:results_raster_images}
\end{figure*}

\subsubsection{Nested sampling}
\label{sec:method_nested}
To perform the fitting we utilise the \textsc{dynesty} implementation \citep{speagle_dynesty_code,koposov_dynesty_code} of the nested sampling algorithm \citep{2004_skilling_Nested,2006_skilling_nested}, which provides a Bayesian inference framework that is relatively insensitive to initial conditions compared to commonly used Markov-chain Monte Carlo methods and naturally yields high-fidelity evidence estimates required for model comparison. A detailed description of the fitting procedure and prior choices is provided in Appendix~\ref{sec:appendix_nested_sampling}. For each epoch we fit combinations of Faraday-thin and Faraday-thick components (e.g. S, SS, ST) as well as models composed of spectral power-law components (P, PP, \ldots), increasing the number of components until the further additions is disfavoured by the Bayesian evidence ($\mathcal{Z}$). 

Model comparison is performed using the Bayes factor derived from the evidence estimates returned by \textsc{dynesty}. In common practice \citep{Kass_bayes,Trotta2008_bayes_factor}, the evidence difference $\Delta \ln \mathcal{Z}$ is interpreted on a qualitative scale in which values of $\Delta \ln \mathcal{Z}$\relto[]{\sim}{1} indicate weak preference, \relto{\sim}{2.5} moderate preference, and \relto{>}{5} strong preference for one model over another. However, because our modelling does not explicitly include potential systematic effects in the spectra (e.g. residual calibration errors), we adopt a more conservative threshold of $\Delta \ln \mathcal{Z}$\relto[]{>}{10} before favouring a more complex model over a simpler one. 

For each epoch, we report the model preferred under this criterion, the next-closest model of greater complexity, the single Faraday-thin fit, and the preferred P-type model for comparison. When two models have near-identical evidence, we apply a simple Occam-like tie-breaking rule that favours Faraday-thin components over Faraday-thick components, since these represent the simpler physical parameterisation. The resulting $QU$-fitting results are presented in Section~\ref{sec:results_qu} and Table~\ref{tab:qu_fitting_summary}\&\ref{tab:qu_fitting_appendix}. Where relevant, we report the posterior mode and 68\% highest-density interval (HDI) for each parameter, since the posteriors are generally unimodal but can be skewed.

\begin{figure}
\centering
\includegraphics[width=1.05\linewidth]{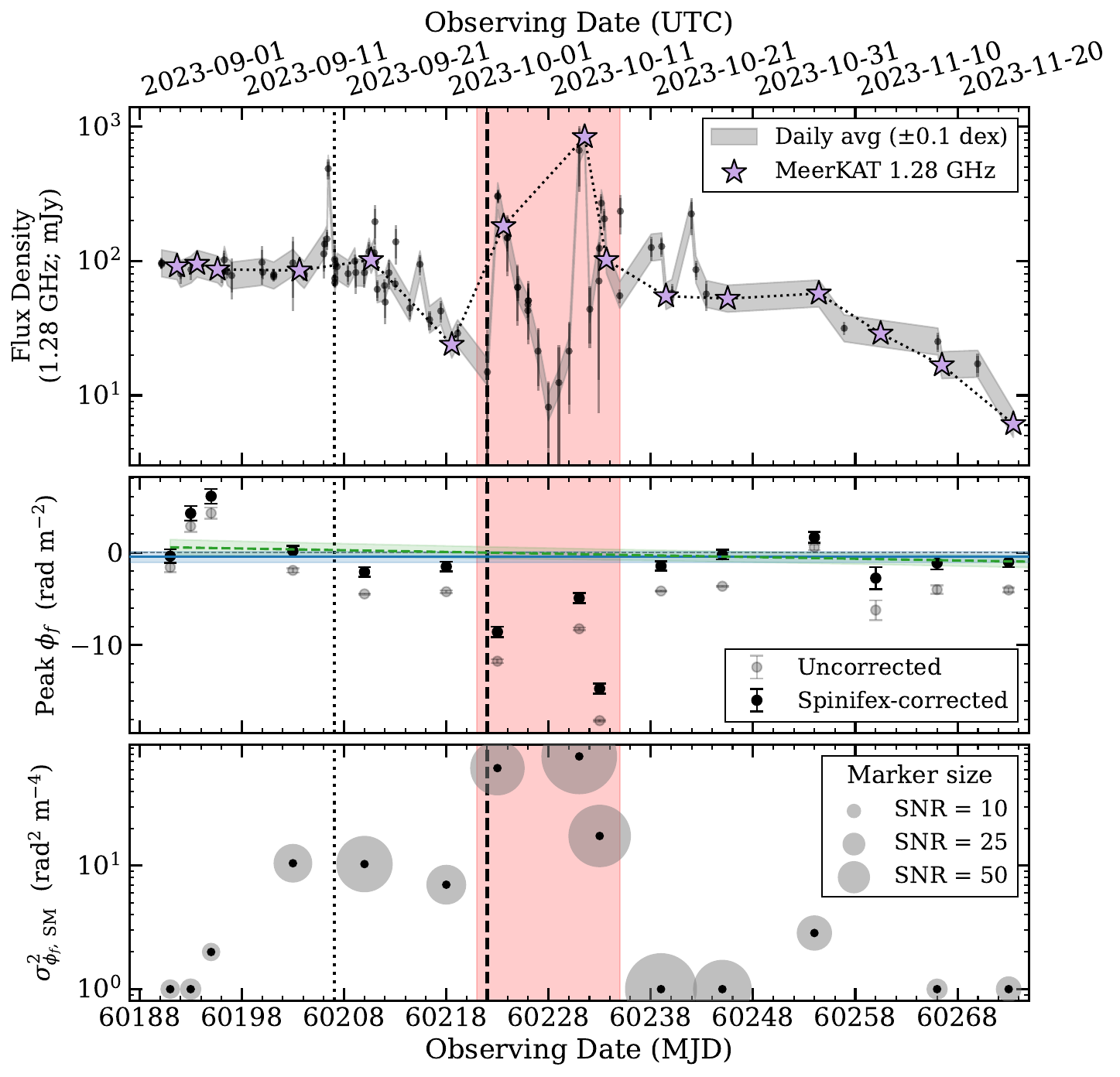}
    \caption{Time evolution of the Stokes $I$ flux density and Faraday-spectrum diagnostics during the 2023 flaring interval. (\textit{top}) 1.28\GHz\ flux-density evolution, with daily-averaged monitoring shown in grey/black and the MeerKAT spectropolarimetric epochs marked by purple stars. The dashed vertical line marks the reported hard$\rightarrow$soft state transition on 2023 October 5 \citep[MJD 60222;][]{Bollemeijer2023a,Bollemeijer2023b}, while the dotted vertical line marks the first reported launch of discrete jet components \citep{Wood2025SWJ1717Ejecta}. (\textit{middle}) Peak Faraday depth of each CLEANed spectrum, before and after ionospheric correction, shown in grey and black, respectively. The blue and green curves show constant and linearly evolving descriptions of the ISM contribution. (\textit{bottom}) Second moment of the CLEAN component distribution, with grey marker size scaling with the peak signal-to-noise of the spectrum. The red shaded interval marks the brightest flaring period, during which the peak depths become offset from the surrounding epochs and the spectra become more complex.}
\label{fig:results_faraday_moments}
\end{figure}

\section{Results}
\label{sec:results}
In this section, we present our time-domain spectropolarimetric results, focusing on the observations obtained immediately before, during, and after the brightest flaring epochs of \src's 2023 outburst. In addition to these bright flaring epochs, we include several later-time observations, taken approximately \relto{\sim}{3}{months} after the flaring, that allow us to measure the spectropolarimetry of both the unresolved core and the angularly separated ejecta (see Figure~\ref{fig:results_raster_images}).

To place the spectropolarimetric behaviour in the broader context of the flare evolution, the top panel of Figure~\ref{fig:results_faraday_moments} shows the Stokes $I$ flux densities around the main flaring period, taken from the comprehensive radio light curves compiled in \citet{HughesAKH2025a}. We include 1--15\GHz\ observations from facilities with simultaneous multi-band coverage, allowing us to derive inter-band spectral indices and estimate the flux density at 1.28\GHz (grey band), the central frequency of the MeerKAT observations (purple stars). The dotted vertical line marks the epoch at which VLBI observations confirmed a morphological change in the jet, corresponding to the emergence of multiple distinct components \citep{Wood2025SWJ1717Ejecta}. The dashed vertical line marks the X-ray inferred start of the hard$\rightarrow$soft state transition \citep[MJD 60222;][]{Bollemeijer2023a,Bollemeijer2023b}, which also coincides with the onset of the brightest radio flaring, likely associated with repeated transitions and thus ejection episodes \citep{Wenfei2023}. To account for secular evolution of the primary calibrators \citep[e.g.][]{2011MNRAS.415.1597M,2012MNRAS.422.1527M}, and for differences in the flux-density scales used by each observatory, we add a 3\% systematic uncertainty in quadrature to the Stokes $I$ flux densities.

Throughout this section, we show a subset of representative plots to illustrate the evolution of the best-fitting models, the Faraday spectra, and the underlying data. However, given the number of epochs, Stokes parameters, and trialled models, it is not practical to include the full set of diagnostic plots in the paper. We therefore treat the subset shown here as representative of the full analysis, and make all associated resources available on GitHub\footnote{\url{https://github.com/AKHughes1994/SwJ1727_2023_Outburst}}. In addition, by comparing \src\ with our polarised calibrators, \polcal\ and \secondary, we show that the spectropolarimetric evolution is intrinsic to the source, and use the calibrator behaviour to quantify the systematic precision of the data. We present this systematic analysis in full in Appendix~\ref{sec:appendix_systematics} and refer to those results as necessary in the sections that follow.

\begin{figure*}
    \centering
    \includegraphics[width=1.0\linewidth]{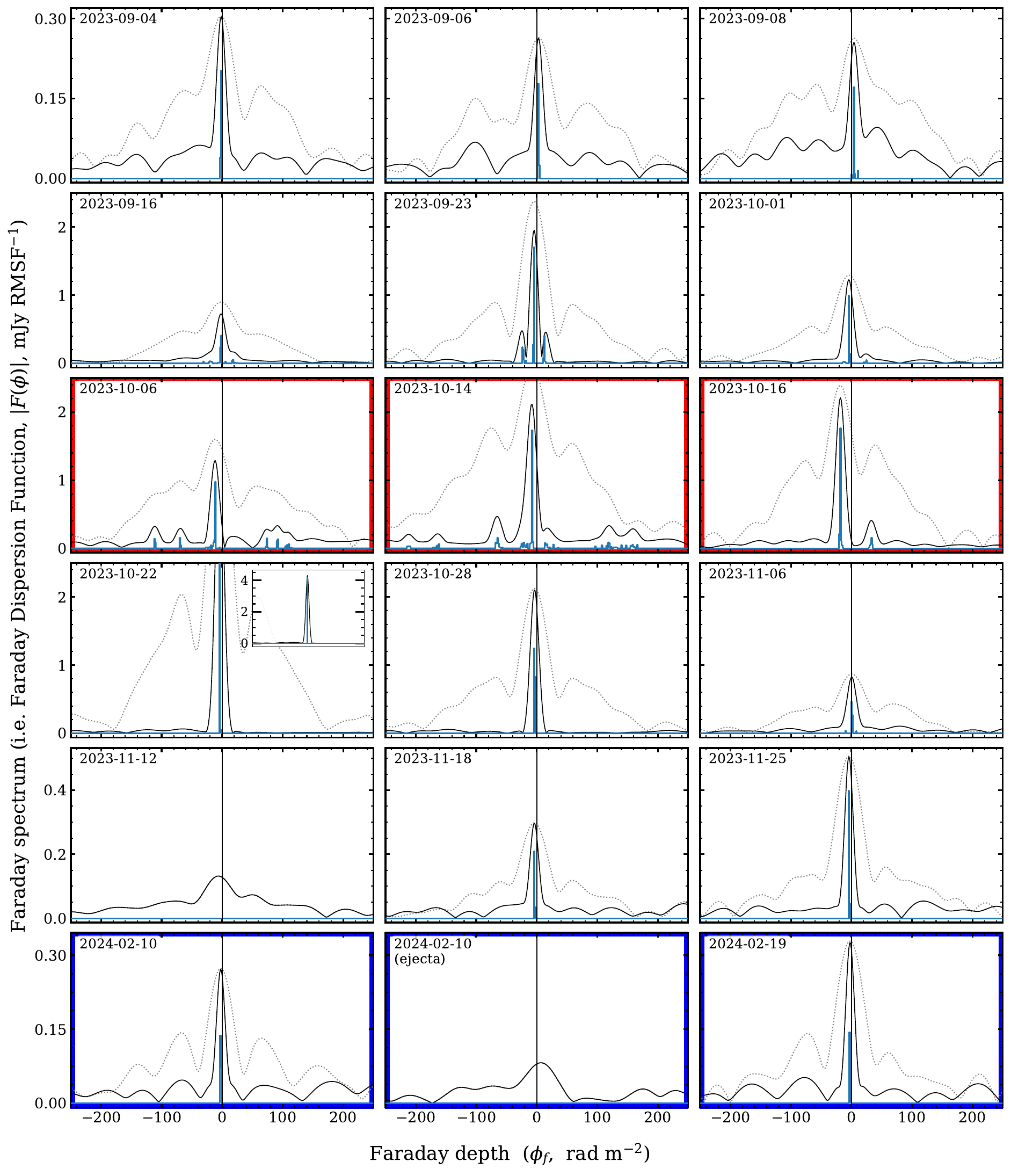}
    \caption{Faraday spectra of \src. The dotted grey curve, solid black curve, and solid blue step function show the observed, CLEANed, and model Faraday spectra derived using \textsc{rm-tools}. Red borders highlight the epochs with the largest changes during the brightest flaring interval of the outburst, while blue borders mark epochs outside the range shown in Figure~\ref{fig:results_faraday_moments}. For ease of comparison, the second through fourth panels share the same y-axis scale. The 2023 October 22 peak is truncated on this scale, so an inset shows the full range. The remaining panels use independent scaling because of the weakly polarised compact jet at early times and the fading source at later times.}
    \label{fig:results_fdf_panels}
\end{figure*}

\subsection{Rotation Measure Synthesis}
\label{sec:results_rmsynth}
In Figure~\ref{fig:results_fdf_panels}, we show the CLEANed Faraday spectra of the \textit{core} radio emission component, by which we mean the compact central component present in both epochs shown in Figure~\ref{fig:results_raster_images}. We again emphasize, that this is an observational definition and does not require the emission to arise from a stationary compact jet. For context, we also include one late-time ejecta epoch, which shows an excess at a depth consistent with the core emission, although it is too faint to meet the CLEANing threshold. We highlight the most complex epochs, from 2023 October 06 to 2023 October 16 (MJD 60223--60233), in red. These correspond to the period when MeerKAT captured the brightest flaring, shown by the red shaded region in Figure~\ref{fig:results_faraday_moments}. We note, however, that the epoch on 2023 September 23 also shows clear structure in the Faraday spectrum. Epochs outside the time range shown in Figure~\ref{fig:results_faraday_moments} are indicated with blue borders.

We first consider the peak Faraday depth in each epoch, shown in the middle panel of Figure~\ref{fig:results_faraday_moments}. We plot both the raw measurements in grey and the ionospheric-corrected values derived using \textsc{spinifex} in black \citep{spinifex}. We also apply an additional correction of $-0.3$\radPerSqm~to account for a known systematic bias in the \textsc{spinifex} solutions \citep[Appendix~\ref{sec:appendix_systematics}; see also][]{perley_ionosphere}. To the corrected measurements, we add a conservative systematic uncertainty of \relto{\pm}{0.5}\radPerSqm~in quadrature, motivated by the observed spectropolarimetric fluctuations of \polcal\ (Fig.~\ref{fig:3c286_systematics}). Excluding the red shaded flaring interval in Figure~\ref{fig:results_faraday_moments}, we find a weighted mean peak depth of $\phi_{\rm rm} = -0.5 \pm 0.2$\radPerSqm. If the uncertainties are instead inflated such that the weighted mean has a reduced $\chi^2 \relto{\equiv}{1}$, this becomes $\phi_{\rm rm} = -0.5 \pm 0.6$\radPerSqm. This approximately constant component, shown by the horizontal blue line in the middle panel, can naturally be identified with a time-independent interstellar contribution, $\phi_{\rm rm, ISM}$. Even if we relax the assumption of strict temporal stability, allowing for linear-in-time deviations, the implied secular evolution remains modest, as shown by the green region in the same panel.

Even allowing for a time-evolving foreground contribution, the flaring epochs excluded from the mean calculation are inconsistent with plausible foreground evolution alone. Inspection of the corresponding red-bordered Faraday spectra reveals the transient emergence of additional structure, and a clear departure from a single Faraday-thin component. While it is tempting to associate these features with multiple thin components, several resemble the signatures expected from (partially resolved out) Faraday-thick structure \citep[e.g. Fig.~\ref{fig:method_simulation_example}; see also][]{rundick2023RMSynth}. In particular, on 2023 October 06 (MJD 60223), the excess at \relto{\sim}{100}\radPerSqm resembles the RMSF-convolved response of a top-hat-like distribution in $\phi_f$. The pair of apparent peaks and central depression near \relto{\sim}{-100}\radPerSqm are also consistent with a thick component broader than the nominal maximum recoverable scale for MeerKAT L-band (i.e. \relto{\sim}{25}\radPerSqm). We therefore caution against direct component counting in Faraday space, particularly when the inferred structures may be broad $\phi_f$-distributions. A similar interpretation may apply to 2023 October 14 (MJD 60231). However, this epoch occurs at the peak of the brightest flare, and shows additional features relative to 2023 October 06, which may indicate a larger number of transient thick components.

As a complementary means of quantifying the emergent Faraday complexity, we calculate the second moment of the CLEAN component distribution, effectively the signal-weighted variance of the CLEAN components:
\begin{align}
\mu_{\phi_f} &= \frac{\sum_{n=0}^N \phi_{f,n}\lvert F(\phi_{f,n})\rvert}{\sum_{n=0}^N \lvert F(\phi_{f,n})\rvert}, \\
\sigma^2_{\phi_f, \mathrm{SM}} &= \frac{\sum_{n=0}^N (\phi_{f,n} - \mu_{\phi_f})^2 \lvert F(\phi_{f,n})\rvert}{\sum_{n=0}^N \lvert F(\phi_{f,n})\rvert}.
\end{align}
Here, $\mu_{\phi_f}$ and $\sigma^2_{\phi_f, \mathrm{SM}}$ are the weighted mean and second moment of the $N$ CLEAN components, where the weights are given by the polarised flux density of each component. We present these second moments in the bottom panel of Figure~\ref{fig:results_faraday_moments}, where larger values indicate broader structure in Faraday space \citep[e.g.][]{RMSecondMomement,2016Anderson}; for plotting purposes, any value of $\sigma^2_{\phi_f, \mathrm{SM}}$\relto{<}{1} was fixed at that value.

Brighter epochs, particularly during flaring, are naturally more favourable for detecting Faraday complexity that might otherwise be masked by noise. However, the emergence of this structure does not appear to be driven by data quality alone. To test this, we scale the second-moment marker sizes by the peak significance of each Faraday spectrum, which acts as a proxy for the strength of the linear-polarisation detection given the similar noise levels across epochs. High-significance epochs can appear either Faraday-simple or Faraday-complex, suggesting that the complexity observed during flaring reflects intrinsic evolution. This is particularly clear for the epochs on 2023 October 22 and 28 (MJD 60239 and 60245), which have comparably high signal-to-noise to the flaring epochs but remain Faraday-simple.

\subsection{Parametric $QU$-fitting}
\label{sec:results_qu}
Motivated by the RM-synthesis signatures, we model the spectropolarimetric behaviour using the $QU$-fitting framework described in Section~\ref{sec:method_QUfit}. Since \src\ was already known to have launched multiple discrete jet components by 2023 September 20 \citep{Wood2025SWJ1717Ejecta}, we test whether more than one polarised component is required to reproduce the time-dependent spectropolarimetry and what the inferred component properties imply about the jet.

We explored two fitting strategies. The first is deliberately model-agnostic, testing combinations of Faraday-thin, or \textit{simple} (S), Faraday-thick (T), and power-law (P) components with minimal physical assumptions, and allowing each component to occupy any region of Faraday space permitted by the observed CLEAN component distribution. The second is more physically guided: given the prominent peaked features seen throughout Figure~\ref{fig:results_fdf_panels}, and the predominantly simple behaviour before and after the flaring interval, we assume that each epoch contains at least some thin contribution associated with the foreground component identified in Section~\ref{sec:results_rmsynth}. We refer to this component as the ISM contribution.

We adopt the second approach because it is better motivated physically and more tractable numerically. X-ray binary ejecta expand as they propagate \citep[e.g.][]{MJ_2011_expansion,rushton2017,atetarenko2017,Wood2025SWJ1717Ejecta}, becoming fainter, less dense, and likely threaded by weaker magnetic fields. These changes should reduce both the amplitude and width of the internal Faraday-depth distribution, so initially Faraday-thick material may evolve towards effectively Faraday-thin behaviour on relatively short time-scales. Repeated ejections \citep{Wood2025SWJ1717Ejecta} can therefore produce a spatially unresolved mixture of thick and thin components within the MeerKAT beam. Since many epochs are clearly dominated by thin emission, and even the Faraday-complex epochs show pronounced peaks in their Faraday spectra, it is reasonable to assume that a thin contribution remains present throughout the observations. From a practical perspective, anchoring one component to a narrower region of parameter space also improves convergence, particularly when T components approach the S-component limit as $\sigma_{\phi} \rightarrow 0$. We therefore adopt a weakly informative prior on the first S-component in the trialled models, using a truncated Gaussian centred on the mean depth derived in Section~\ref{sec:results_rmsynth}, with the inflated uncertainty on that mean as the standard deviation ($-0.5 \pm 0.6$\radPerSqm). Since the fitting is performed on the observed values, we then reintroduce the systematic offset of $-0.3$\radPerSqm, together with the ionospheric contribution predicted by \textsc{spinifex}, and truncate the Gaussian at \relto{\pm}{2}\radPerSqm around the resulting observed value. In the following sections, we present only this physically guided approach; for completeness, the more model-agnostic results are presented in Appendix~\ref{sec:appendix_systematics}. For most epochs, the two approaches recover nearly identical component properties within the posterior credible intervals, and therefore lead to the same conclusion: the data favour the transient emergence of Faraday-thick components during the flaring interval.

\subsubsection{Transient Faraday complexity}
\begin{figure*}
    \centering
    \includegraphics[width=0.90\linewidth]{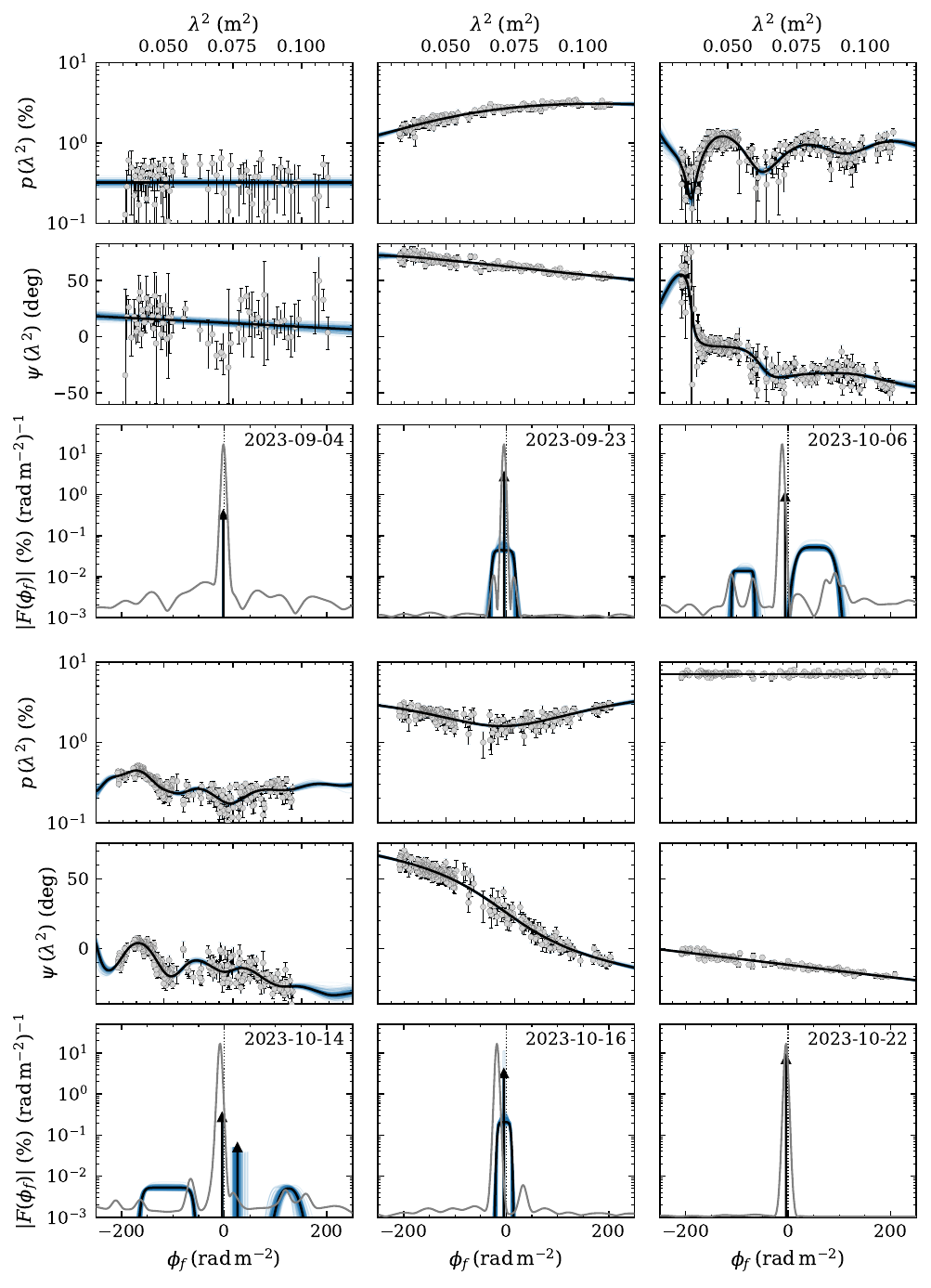}
    \caption{Representative spectropolarimetric behaviour of \src\ across six epochs before, during, and after the bright flaring period. For each epoch, the upper and middle panels show the observed $p(\lambda^2)$ and $\psi(\lambda^2)$, respectively, while the lower panel shows the corresponding model amplitudes, $|F(\phi_f)|$. Black points show the data, blue curves show posterior draws from the preferred $QU$-fitting model, and grey curves show the CLEANed spectra from Figure~\ref{fig:results_fdf_panels}. The flaring epochs show depolarisation/repolarisation and non-linear $\psi(\lambda^2)$ structure than the simple pre- and post-flare epochs: note, that the Faraday simple epochs may have different $\psi$-slopes due to changes in the ionosphere.}
    \label{fig:results_QU_sample}
\end{figure*}

In Figure~\ref{fig:results_QU_sample}, we highlight a subset of the spectropolarimetric behaviour across six epochs before, during, and after the brightest flaring. Although the modelling was performed directly on our observables, $q(\lambda^2)$ and $u(\lambda^2)$, these quantities are not ideal for qualitatively illustrating Faraday complexity, as even a single Faraday-thin screen can produce strong sinusoidal structure across the band. We therefore show the linear polarisation fraction, $p(\lambda^2)$, the observed linear polarisation angle, $\psi(\lambda^2)$, and the model Faraday spectrum, with the CLEANed versions from Figure~\ref{fig:results_fdf_panels} overlaid in grey. A more comprehensive diagnostic plot for the 2023 October 06 epoch, including both the inferred and directly observed quantities, as well as its corner plot, is shown in Appendix~\ref{sec:appendix_systematics}. Because the ISM contribution lies close to $\phi_{\rm rm}=0$, $\psi(\lambda^2)$ does not show strong wrapping across the band, making departures from simple linear Faraday rotation straightforward to identify by eye.

The spectropolarimetric properties of \src\ show a temporary but significant departure from a Faraday-simple spectrum, i.e. constant $p(\lambda^2)$ and linear $\psi(\lambda^2)$. The surrounding Faraday-simple epochs imply that this complexity emerges and disappears within the approximately month-long flaring period. This is clearest when comparing the two most Faraday-complex epochs, 2023 October 06 and 14, with 2023 October 22. Despite having the highest polarised signal-to-noise in the sample, the 2023 October 22 epoch is remarkably Faraday simple, whereas the October 06 and 14 epochs each require multiple components, including two Faraday-thick components. The markedly different Faraday spectra observed on 2023 October 14 and 16 further suggest that the thick components can evolve or disappear on timescales of \relto{<}{2}{d}, consistent with the rapidly evolving VLBI morphology, where ejecta are seen to appear/disappear over \relto{\sim}{24}{h} intervals (Wood et al., submitted). Individual components therefore may not persist across multiple epochs, highlighting how easily this transient phenomenon could be missed without suitable cadence and frequency coverage. We searched for intra-observation variability by splitting the brightest epochs in half, but found no significant evidence for such behaviour across our \relto{\sim}{15}{min} observations. If the evolution is gradual rather than impulsive, continuous observations on $\gtrsim4\,{\rm h}$ timescales should be sufficient to track the evolution of these components.

\begin{figure*}
    \centering
    \includegraphics[width=0.9\linewidth]{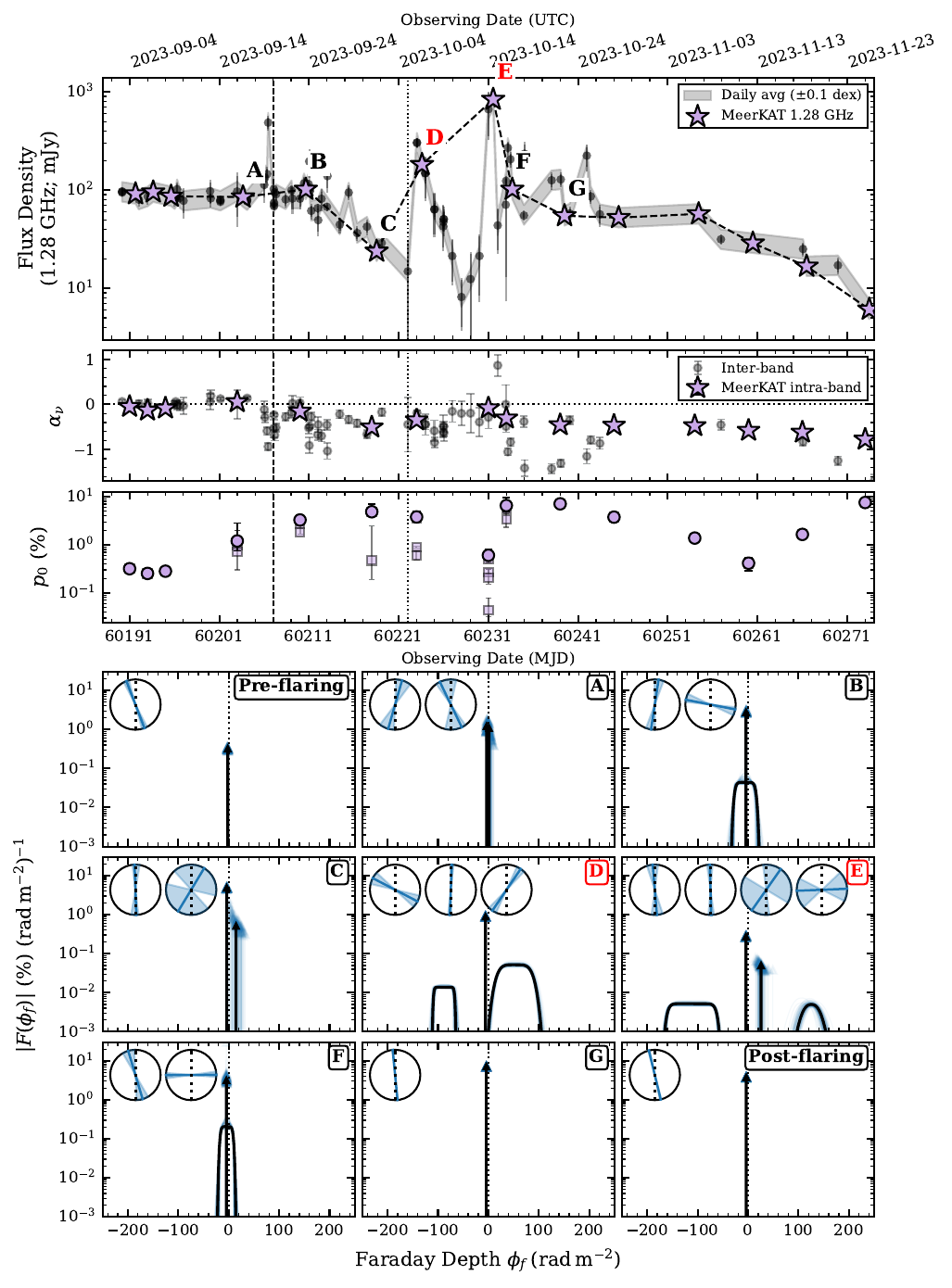}
    \caption{
    Summary of the radio and spectropolarimetric evolution of \src\ during the 2023 flaring period. The upper panels show the 1.28\,GHz MeerKAT light curve, the radio spectral index, and the inferred intrinsic polarisation fractions, $p_0$, including both the total values (solid markers) and per-component values (translucent markers). The lower panels show the preferred models for representative epochs, ordered and alphabetically labelled by their position in the flaring timeline. Model amplitudes are shown as fractional polarisation percentages, while the circular dials indicate the intrinsic linear polarisation angles and their 68\% credible intervals, ordered from the most negative to the most positive Faraday-depth component. The Faraday-complex models are confined to the bright flaring interval, while the pre- and post-flaring epochs are consistent with comparatively simple Faraday structure.}
    \label{fig:results_ILoveThisPlot}
\end{figure*}

Figure~\ref{fig:results_QU_sample} compares the parametric $QU$-fitting results with the non-parametric RM-synthesis/RM-CLEAN reconstruction. The broad fitted components generally coincide with excess structure in the CLEANed spectra, supporting the interpretation that they trace Faraday-thick emission. For example, on 2023 September 23 the fitted thick component spans the pronounced wings around the Faraday-thin ISM component, while on 2023 October 06 the negatively shifted component aligns with the bi-peaked structure separated from the main thin peak. The correspondence is not one-to-one, however. Several inferred thick components, particularly the positively shifted component on 2023 October 06 and much of the non-ISM structure on 2023 October 14, align only approximately with features in the CLEANed spectra. Their behaviour nevertheless resembles the simulated example in Figure~\ref{fig:method_simulation_example}, and they would be difficult to identify or classify reliably by eye. Moreover, the strongest CLEANed peak does not necessarily trace the $\phi_{\rm rm}$ of any thin component, since interference between multiple components can shift the dominant Faraday-spectrum peak away from the $QU$-inferred ISM component.

The main caveat is therefore not whether complex structure is present, but how literally the individual fitted components should be interpreted. Our parametric models are simplified descriptions of structure that may be broader or more irregular than our adopted symmetric super-Gaussian components can represent. This uncertainty limits component-level interpretation, but does not favour an intrinsically thinner interpretation, since the observational limitations discussed above preferentially suppress or fragment broad distributions in $\phi_f$. The fitted thick components may therefore be conservative approximations to broader or more complex underlying structure. For example, the oppositely signed components could in principle originate from a single broad but asymmetric distribution, implying intrinsic variation in the magneto-ionic or emissivity properties of the emitting plasma along the line of sight. We discuss possible configurations in Section~\ref{sec:discu_results_origin} and Figure~\ref{fig:disc_schematic}.

Having established that the emergent Faraday complexity is robust and transient, we next place it in the context of the Stokes $I$ flaring timeline. Figure~\ref{fig:results_ILoveThisPlot} summarises the outburst evolution, showing the light curve alongside the total and per-component intrinsic polarisation fractions, $p_0$. The lower panels show the best-fitting model Faraday spectra selected using the Bayesian evidence criterion described in Section~\ref{sec:method_nested}, demonstrating that the most pronounced Faraday complexity occurs near the peaks of the major flares. There is one exception to this model-selection procedure. On 2023 October 16, the ST model falls slightly below the $\Delta\ln\mathcal{Z}\,{\sim}\,10$ threshold, but the next-simplest model, SS, is strongly seed-dependent and shows high covariance between the inferred per-component $p_0$ values. The two amplitudes can increase or decrease together while reproducing a similar spectrum, but with less favourable evidence and a worse $\chi^2$ than the ST model. We therefore show the ST model for this epoch, noting that this choice does not affect the main result, since the two preceding epochs already require Faraday-thick components.

A subset of the epoch fits is presented in Table~\ref{tab:qu_fitting_summary}, highlighting cases where the data strongly favour Faraday-thick components. In both peak-flaring epochs, $\Delta\ln\mathcal{Z}\,{\gtrsim}100$ relative to the highest-evidence models containing only $S$ and $P$ components, despite testing pure $S$ and $P$ models with up to six and four components, respectively. This conclusively favours Faraday-thick structure. Within each model Faraday-spectrum panel, the fractional amplitudes are shown as percentages. The circular dials mark the intrinsic linear polarisation angles and their 68\% credible intervals, ordered from the most negative to the most positive $\phi_{\rm rm}$ component. We use this figure to contextualise the remaining results and discussion.

\subsection{Evolution of the linear polarisation properties}

\begin{figure}
    \centering
    \includegraphics[width=1.0\linewidth]{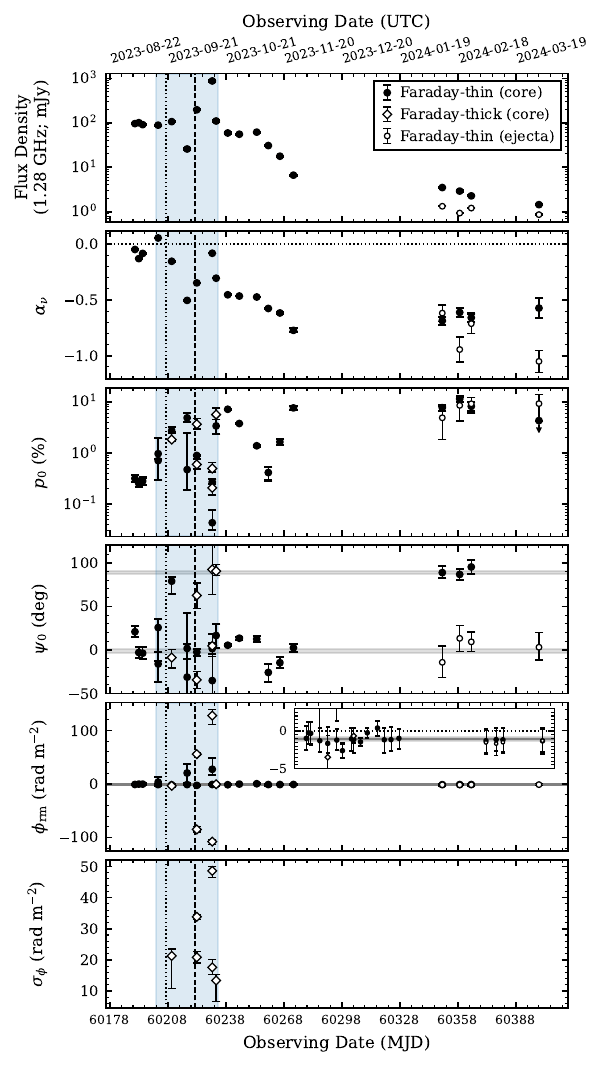}
    \caption{Time evolution of the radio and linear polarisation properties of \src. From top to bottom, the panels show the 1.28\,GHz Stokes $I$ flux density, radio spectral index, intrinsic polarisation fraction, $p_0$, intrinsic polarisation angle, $\psi_0$, Faraday depth, $\phi_{\rm rm}$, and, where relevant, the Faraday width of the super-Gaussian components, $\sigma_\phi$. Filled circles show Faraday-thin core components, diamonds show Faraday-thick core components, and open circles show Faraday-thin ejecta components. The shaded blue region marks the bright flaring interval, while the horizontal grey bands in the $\psi_0$ panel indicate the canonical jet-aligned orientations, parallel and orthogonal to the VLBI jet axis.}
    \label{fig:results_pol}
\end{figure}

Figure~\ref{fig:results_ILoveThisPlot} showed the visual evolution of the $QU$-fitting models in Faraday space, with the largest departures from a single Faraday-thin component occurring during the flaring epochs D and E. We now examine the corresponding fitted parameters as time series. Figure~\ref{fig:results_pol} summarises the Stokes $I$, spectral-index, and linear-polarisation evolution, including the post-flaring epochs from 2024 February 10 to 2024 March 31 (MJD 60350--60400), when the L-band emission is well described by an unresolved core and an angularly separated, southward-moving jet ejection.

The blue shaded region marks the flaring interval over which the emission is Faraday-complex, where Faraday-complex denotes epochs requiring a multi-component parametric model. As discussed in \citet{HughesAKH2025a}, the contemporaneous spectral-index evolution is consistent with changing optical depths in expanding knots of plasma \citep{fender2019,CF_PLACEHOLD}, and with the milliarcsecond-scale morphological changes observed at the start of this interval \citep{Wood2025SWJ1717Ejecta}. This is also when we observe the most rapidly varying spectropolarimetric behaviour. Outside this interval, the polarisation properties evolve more smoothly on week-to-week time-scales.

\subsubsection{Linear polarisation fraction}

The intrinsic polarisation fraction is steady but low, at \relto{\sim}{0.3\%}, near the start of the outburst, when the radio emission is dominated by the compact jet. This is broadly consistent with previous detections of compact jets in X-ray binaries, which typically report \relto{\lesssim}{1\%} linear polarisation \citep[e.g.][]{han1992,Corbel2000,Russell2015maxij1836LrLx,Kravtsov2025_cygx1_pol}. Similarly low fractions are also observed in the cores of radio-bright AGN \citep[e.g.][]{Hodge2018}. Following the onset of flaring, the inferred component polarisation fractions vary substantially, spanning \relto{\sim}{0.1-8\%} from week to week. The post-flaring evolution also shows significant changes, most notably a \relto{\sim}{5}{week} depolarisation and re-polarisation trend beginning immediately after the end of the flaring interval, between 2023 October 22 and 2023 November 25.

Even when the source is clearly optically thin ($\alpha_\nu$\relto{\simeq}{-0.6}), the polarisation fraction remains far below the theoretical maximum for optically thin synchrotron emission in a uniform magnetic field \citep[\relto{\sim}{75\%};][]{longair2011}. The maximum value we infer is only \relto{\sim}{8\%}, even for the resolved ejecta, where there is less ambiguity from component blending than in the unresolved core. The low observed fractions therefore suggest that the magnetic-field coherence length is smaller than the angular resolution of MeerKAT, so each component contains partially tangled field structure and appears depolarised after spatial averaging. A useful comparison is provided by M87, where unresolved ALMA observations show much lower polarisation than high-angular-resolution Event Horizon Telescope observations ($\sim2$ versus $\sim20\%$), with the resolved maps revealing the underlying tangled field structure \citep{m87_pol1}.


\subsubsection{Linear polarisation angle}
\label{sec:results_angle}
The intrinsic polarisation angle also shows its strongest variability during and immediately after the flaring interval. In Figure~\ref{fig:results_pol}, we highlight the position-angle ranges parallel and orthogonal to the jet axis, which lies $\in[-2,1]\,{\rm deg}$ east of north based on VLBI observations of both the compact jet and discrete ejecta \citep{Wood2024,Wood2025SWJ1717Ejecta}. These angles provide a useful reference because simple magnetic-field ordering mechanisms, such as shock compression and velocity shear \citep{Laing1980,hughes1985jetpolshocks,Hughes1989jetpola,Hughes1989jetpolb,Bell2004_magnetic}, tend to align the magnetic field either parallel or perpendicular to the jet axis. The observed linear polarisation angle should follow the same basic division, being parallel to the projected magnetic field for an ideal homogeneous optically thick synchrotron source, and perpendicular for optically thin or inhomogeneous synchrotron emission \citep[although the inhomogeneous case has a small region of parameter space where it remains parallel;][]{Jones1977,Jones1977_inhomo,longair2011,CF_PLACEHOLD}. In the following sections, we use this parallel/orthogonal relation between the intrinsic polarisation angle and the jet axis as the canonical jet-aligned orientation, and highlight departures from it as obliquity.

This simple picture can be complicated by polarisation-dependent relativistic beaming \citep{Lyutikov2005_jetpol}, shocks, and complex magnetic-field structures \citep[e.g.][]{Cawthorne1966Obliqueshocks,Gomez2001Shocks,Jorstad2007_agnpoll,Gomez2008oblique}. Nevertheless, unresolved XRB and AGN jets often show approximate clustering around the canonical orientations \citep[e.g.][]{Lister1998jetpol,gallo2004,stirling2004,2005Lister_mojav_pol,Lyutikov2005_jetpol,rushton2017,Hodge2018,Pushkarev_2023_mojave_pol,Kravtsov2025_cygx1_pol}, although oblique orientations are also observed in some cases \citep[e.g.][]{Gomez2001Shocks,Dulwhich2009_oblique,fender2003,hughes2023}. While \src\ shows clear clustering around the parallel and orthogonal bands, the inferred angles also show transient obliquity. The obliquity is much larger than the epoch-to-epoch changes in the \polcal\ polarisation angles, where $\Delta\psi_0$\relto{\sim}{0.3}{deg}, indicating that it is source-specific and physical.

\begin{figure}
    \centering
    \includegraphics[width=1.0\linewidth]{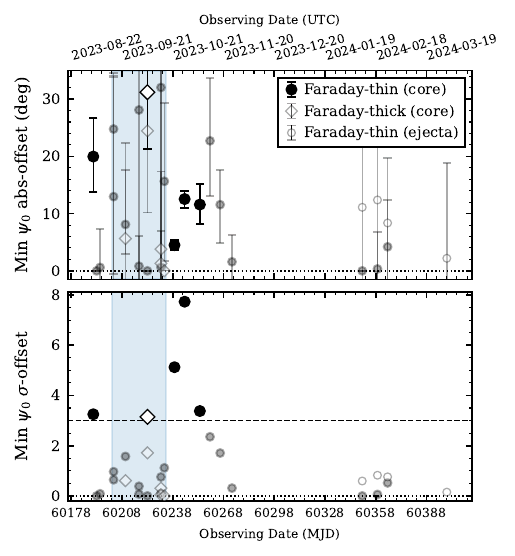}
    \caption{
    Deviation of the intrinsic polarisation angle from the nearest canonical jet-aligned orientation. The upper panel shows the minimum absolute offset between $\psi_0$ and the position angles parallel or orthogonal to the VLBI jet axis, while the lower panel shows the same offset scaled by the posterior uncertainty. The horizontal dashed line marks the \relto{>}{3\sigma} threshold used to identify significantly oblique orientations.
    }
    \label{fig:psi0_offset_subfig}
\end{figure}

We quantify this behaviour in Figure~\ref{fig:psi0_offset_subfig} by plotting the absolute angular offset from the nearest canonical orientation, either parallel or perpendicular to the jet axis, in the top panel, and the same offset scaled by its posterior uncertainty in the bottom panel. We highlight epochs where the inferred angle differs from both canonical jet-aligned orientations at \relto{>}{3\sigma}. The system shows one apparently misaligned point before the flare, during one of the hard-state epochs. However, at polarisation fractions of only \relto{\sim}{0.3\%}, systematic uncertainties become more relevant. Although these measurements are likely still noise dominated, the formal posterior uncertainties may underestimate the true errors and therefore overestimate the significance of this obliquity.

During the flaring interval, the clearest oblique configuration occurs on 2023 October 06. The two Faraday-thick components have intrinsic polarisation angles that are approximately perpendicular to one another, while one is also significantly offset from the nearest jet-aligned orientation. The first two post-flaring epochs show the most statistically significant departures, including 2023 October 22, which has the highest polarised signal-to-noise ratio in the sample ($p_0$\relto{\gtrsim}{5\%}). These offsets are more modest in absolute angle than during the flaring epoch, but are better constrained. The subsequent post-flare, lower-significance epochs appear to wobble around the expected jet-aligned orientations. This behaviour in $\psi_0$ coincides with the slow depolarisation and re-polarisation trend seen in the fractional polarisation.

Finally, between the post-flaring epochs (i.e., November 2023) and the late-time 2024 observations, the unresolved core shows an apparent 90\,deg inversion in polarisation angle. By contrast, the late-time resolved ejecta remain consistent with the post-flaring alignment relative to the jet position angle, and are approximately orthogonal to the simultaneous core detections. This occurs despite the core and ejecta having statistically consistent polarisation fractions of \relto{\sim}{8\%} and both being optically thin, with $\alpha$~\relto{\sim}{-0.6}. This is notable because the late-time epochs occur well into the soft state, when the core emission is likely dominated by unresolved ejecta, since jet activity from close to the black hole should have long ceased \citep[e.g.][]{fender2010}. The inversion therefore requires either intrinsic differences between the emitting components, or an observational effect that changes the apparent angle. The former would be particularly interesting, since it is unclear why otherwise empirically similar jets, launched by the same system and presumably through the same mechanism, would have orthogonal polarisation angles. Overall, even setting aside the transient Faraday complexity, these more classical polarisation diagnostics show atypical and strongly time-dependent behaviour.

\subsubsection{Faraday properties}
\label{sec:results_QU_faraday}
Here we define the ``Faraday properties'' as the peak Faraday depth, $\phi_{\rm rm}$, the super-Gaussian width, $\sigma_\phi$, and the shape parameter, $N$. Each epoch is consistent with at least some polarised emission passing through a time-invariant Faraday-thin screen. Applying the same ionospheric and systematic corrections as above, and now including the flaring epochs and the resolved-ejecta fits, we derive an average rotation measure of $\phi_{\rm rm,ISM}=-1.0\pm0.3$~\radPerSqm. We interpret this as the ISM contribution along the line of sight to \src. It is consistent with the spread of rotation measures from nearby pulsars \citep[from the Australian Telescope National Facility pulsar catalogue][]{ATNF}, and lies well within the distribution of Galactic sources \citep{Taylor2009RMGal}. The stability of these solutions gives $\chi_{\rm red}^2\relto{\sim}{0.5}$, shown in the inset of Figure~\ref{fig:results_pol}, suggesting that our systematic error treatment is likely conservative. This value is consistent with the tentative prior estimate of $-0.5\pm0.6$~\radPerSqm, but is offset from the prior peak. Since the posteriors are unimodal in all epochs, the inferred mean does not appear to be inherited from the prior shape.

\def\arraystretch{1.55}
\begin{table*}
    \centering
    \small
    \setlength{\tabcolsep}{1.0pt}
    \begin{tabular}{lccllcccccccc}
        \Xhline{5\arrayrulewidth}
        \multicolumn{1}{c}{Date} &
        \multicolumn{1}{c}{\begin{tabular}[c]{@{}c@{}}$I_0$ (mJy)\\$\alpha_\nu$\end{tabular}} &
        \multicolumn{1}{c}{Ejecta} &
        \multicolumn{1}{c}{Model} &
        \multicolumn{1}{c}{Comp.} &
        \multicolumn{1}{c}{$p_0$ (\%)} &
        \multicolumn{1}{c}{$\psi_0$ (deg)} &
        \multicolumn{1}{c}{$\phi_f$ ($\rad\,\mathrm{m}^{-2}$)} &
        \multicolumn{1}{c}{$\sigma_\phi$ ($\rad\,\mathrm{m}^{-2}$)} &
        \multicolumn{1}{c}{$N$} &
        \multicolumn{1}{c}{$\beta$} &
        \multicolumn{1}{c}{$\chi^2$ (dof)} &
        \multicolumn{1}{c}{$\ln \mathcal{Z}$} \\
        \Xhline{5\arrayrulewidth}
        \multirow[t]{4}{*}{\begin{tabular}[t]{@{}l@{}}2023-09-04 15:55\\60191.664\end{tabular}} & \multirow[t]{4}{*}{\begin{tabular}[t]{@{}c@{}}$95.98 \pm 0.07$\\$-0.047 \pm 0.005$\end{tabular}} & \multirow[t]{4}{*}{$\times$} & ${\rm \textbf{S}}$ & ${\rm S}_1$ & $0.318_{-0.016}^{+0.019}$ & $21_{-2}^{+3}$ & $-2.2_{-0.6}^{+0.6}$ & $\cdots$ & $\cdots$ & $\cdots$ & $141\,(135)$ & $688$ \\
        \cline{4-13}
         &  &  & \multirow{2}{*}{${\rm SS}$} & ${\rm S}_1$ & $0.30_{-0.18}^{+0.08}$ & $20_{-9}^{+8}$ & $-2.2_{-0.6}^{+0.7}$ & $\cdots$ & $\cdots$ & $\cdots$ & $140\,(132)$ & $682$ \\
         &  &  &  & ${\rm S}_2$ & $0.05_{-0.05}^{+0.15}$ & $14_{-26}^{+12}$ & $-1_{-2}^{+7}$ & $\cdots$ & $\cdots$ & $\cdots$ & $140\,(132)$ & $682$ \\
        \cline{4-13}
         &  &  & ${\rm P}$ & ${\rm P}_1$ & $0.30_{-0.03}^{+0.03}$ & $21_{-2}^{+3}$ & $-2.2_{-0.6}^{+0.6}$ & $\cdots$ & $\cdots$ & $0.3_{-0.4}^{+0.3}$ & $140\,(134)$ & $686$ \\
        \Xhline{3\arrayrulewidth}
        \multirow[t]{8}{*}{\begin{tabular}[t]{@{}l@{}}2023-09-23 15:30\\60210.646\end{tabular}} & \multirow[t]{8}{*}{\begin{tabular}[t]{@{}c@{}}$106.32 \pm 0.05$\\$-0.153 \pm 0.003$\end{tabular}} & \multirow[t]{8}{*}{$\times$} & ${\rm S}$ & ${\rm S}_1$ & $2.229_{-0.012}^{+0.015}$ & $80.7_{-0.6}^{+0.5}$ & $-4.46_{-0.15}^{+0.12}$ & $\cdots$ & $\cdots$ & $\cdots$ & $1928\,(325)$ & $801$ \\
        \cline{4-13}
         &  &  & \multirow{3}{*}{${\rm SSS}$} & ${\rm S}_1$ & $6.8_{-1.1}^{+1.7}$ & $-79_{-3}^{+2}$ & $-4.7_{-0.6}^{+0.5}$ & $\cdots$ & $\cdots$ & $\cdots$ & $329\,(319)$ & $1577$ \\
         &  &  &  & ${\rm S}_2$ & $8_{-2}^{+1}$ & $-2_{-4}^{+3}$ & $-1.0_{-0.4}^{+1.1}$ & $\cdots$ & $\cdots$ & $\cdots$ & $329\,(319)$ & $1577$ \\
         &  &  &  & ${\rm S}_3$ & $1.4_{-0.2}^{+1.9}$ & $61_{-5}^{+4}$ & $4_{-1}^{+3}$ & $\cdots$ & $\cdots$ & $\cdots$ & $329\,(319)$ & $1577$ \\
        \cline{4-13}
         &  &  & \multirow{2}{*}{${\rm \textbf{ST}}$} & ${\rm S}_1$ & $2.72_{-0.08}^{+0.13}$ & $79_{-3}^{+3}$ & $-4.0_{-0.5}^{+0.6}$ & $\cdots$ & $\cdots$ & $\cdots$ & $327\,(320)$ & $1585$ \\
         &  &  &  & ${\rm T}_1$ & $1.82_{-0.09}^{+0.15}$ & $-9_{-3}^{+3}$ & $-5.9_{-1.3}^{+1.1}$ & $21_{-3}^{+1}$ & $4_{-2}^{+12}$ & $\cdots$ & $327\,(320)$ & $1585$ \\
        \cline{4-13}
         &  &  & \multirow{2}{*}{${\rm PP}$} & ${\rm P}_1$ & $4.5_{-0.6}^{+1.4}$ & $-6.2_{-1.3}^{+1.3}$ & $-2.9_{-0.5}^{+0.3}$ & $\cdots$ & $\cdots$ & $-3.8_{-0.3}^{+0.4}$ & $325\,(321)$ & $1591$ \\
         &  &  &  & ${\rm P}_2$ & $7.3_{-0.5}^{+1.5}$ & $79.2_{-0.6}^{+0.7}$ & $-2.9_{-0.5}^{+0.3}$ & $\cdots$ & $\cdots$ & $-2.56_{-0.16}^{+0.19}$ & $325\,(321)$ & $1591$ \\
        \Xhline{3\arrayrulewidth}
        \multirow[t]{17}{*}{\begin{tabular}[t]{@{}l@{}}2023-10-06 14:21\\60223.599\end{tabular}} & \multirow[t]{17}{*}{\begin{tabular}[t]{@{}c@{}}$196.40 \pm 0.04$\\$-0.3466 \pm 0.0014$\end{tabular}} & \multirow[t]{17}{*}{$\times$} & ${\rm S}$ & ${\rm S}_1$ & $0.788_{-0.007}^{+0.009}$ & $6.5_{-0.3}^{+0.3}$ & $-6.539_{-0.004}^{+0.013}$ & $\cdots$ & $\cdots$ & $\cdots$ & $3215\,(549)$ & $1486$ \\
        \cline{4-13}
         &  &  & \multirow{6}{*}{${\rm SSSSSS}$} & ${\rm S}_1$ & $3.0_{-0.1}^{+0.2}$ & $-26.3_{-1.7}^{+1.7}$ & $-6.0_{-0.3}^{+0.7}$ & $\cdots$ & $\cdots$ & $\cdots$ & $841\,(534)$ & $2602$ \\
         &  &  &  & ${\rm S}_2$ & $0.1_{-0.1}^{+0.8}$ & $-20_{-6}^{+13}$ & $10_{-3}^{+1}$ & $\cdots$ & $\cdots$ & $\cdots$ & $841\,(534)$ & $2602$ \\
         &  &  &  & ${\rm S}_3$ & $23_{-3}^{+3}$ & $-21_{-3}^{+3}$ & $11.6_{-0.6}^{+1.2}$ & $\cdots$ & $\cdots$ & $\cdots$ & $841\,(534)$ & $2602$ \\
         &  &  &  & ${\rm S}_4$ & $29_{-3}^{+4}$ & $47_{-3}^{+4}$ & $16.8_{-1.1}^{+0.9}$ & $\cdots$ & $\cdots$ & $\cdots$ & $841\,(534)$ & $2602$ \\
         &  &  &  & ${\rm S}_5$ & $8.5_{-0.9}^{+1.2}$ & $88_{-5}^{+3}$ & $28.5_{-1.1}^{+1.2}$ & $\cdots$ & $\cdots$ & $\cdots$ & $841\,(534)$ & $2602$ \\
         &  &  &  & ${\rm S}_6$ & $0.89_{-0.10}^{+0.09}$ & $42_{-8}^{+5}$ & $60.4_{-1.9}^{+1.2}$ & $\cdots$ & $\cdots$ & $\cdots$ & $841\,(534)$ & $2602$ \\
        \cline{4-13}
         &  &  & \multirow{3}{*}{${\rm \textbf{STT}}$} & ${\rm S}_1$ & $0.893_{-0.012}^{+0.013}$ & $-2.3_{-1.8}^{+1.3}$ & $-5.8_{-0.3}^{+0.3}$ & $\cdots$ & $\cdots$ & $\cdots$ & $689\,(539)$ & $2709$ \\
         &  &  &  & ${\rm T}_1$ & $0.60_{-0.05}^{+0.04}$ & $63_{-6}^{+5}$ & $-88_{-2}^{+2}$ & $20.8_{-0.7}^{+0.7}$ & $8_{-4}^{+16}$ & $\cdots$ & $689\,(539)$ & $2709$ \\
         &  &  &  & ${\rm T}_2$ & $3.6_{-0.3}^{+0.4}$ & $-34_{-4}^{+4}$ & $52.6_{-1.2}^{+1.1}$ & $33.8_{-0.3}^{+0.3}$ & $4.6_{-0.4}^{+0.6}$ & $\cdots$ & $689\,(539)$ & $2709$ \\
        \cline{4-13}
         &  &  & \multirow{4}{*}{${\rm STTT}$} & ${\rm S}_1$ & $0.86_{-0.02}^{+0.02}$ & $-10_{-3}^{+1}$ & $-4.2_{-0.3}^{+0.3}$ & $\cdots$ & $\cdots$ & $\cdots$ & $658\,(534)$ & $2709$ \\
         &  &  &  & ${\rm T}_1$ & $0.67_{-0.08}^{+0.08}$ & $60_{-8}^{+9}$ & $-87_{-3}^{+3}$ & $20.5_{-0.7}^{+0.7}$ & $6_{-3}^{+9}$ & $\cdots$ & $658\,(534)$ & $2709$ \\
         &  &  &  & ${\rm T}_2$ & $2.0_{-0.2}^{+0.5}$ & $-37_{-15}^{+9}$ & $54_{-2}^{+2}$ & $18.2_{-1.4}^{+1.3}$ & $3_{-1}^{+8}$ & $\cdots$ & $658\,(534)$ & $2709$ \\
         &  &  &  & ${\rm T}_3$ & $3.9_{-1.1}^{+2.0}$ & $-25_{-35}^{+8}$ & $50_{-1}^{+2}$ & $33.7_{-1.0}^{+1.2}$ & $4.8_{-0.5}^{+0.6}$ & $\cdots$ & $658\,(534)$ & $2709$ \\
        \cline{4-13}
         &  &  & \multirow{3}{*}{${\rm PPP}$} & ${\rm P}_1$ & $0.74_{-0.07}^{+0.09}$ & $-54.0_{-1.8}^{+1.9}$ & $-2.1_{-0.3}^{+0.1}$ & $\cdots$ & $\cdots$ & $-0.6_{-0.5}^{+0.6}$ & $1292\,(542)$ & $2413$ \\
         &  &  &  & ${\rm P}_2$ & $4.2_{-0.6}^{+0.9}$ & $2_{-2}^{+2}$ & $-2.1_{-0.3}^{+0.1}$ & $\cdots$ & $\cdots$ & $4.62_{-0.13}^{+0.05}$ & $1292\,(542)$ & $2413$ \\
         &  &  &  & ${\rm P}_3$ & $3.4_{-0.7}^{+0.9}$ & $-89.1_{-1.7}^{+1.8}$ & $-2.1_{-0.3}^{+0.1}$ & $\cdots$ & $\cdots$ & $4.99_{-0.03}^{+0.01}$ & $1292\,(542)$ & $2413$ \\
        \Xhline{5\arrayrulewidth}
    \end{tabular}
    \caption{Summary of the $QU$-fitting results for all epochs. Parameter values are posterior modes with 68 per cent highest-density intervals. The Stokes~$I$ column gives the flux density at 1.28 GHz and the spectral index $\alpha_\nu$. The $\phi_f$ column gives $\phi_{\rm rm}$ for Faraday-thin and power-law components, and $\phi_{\rm peak}$ for Faraday-thick components. Bold entries mark the favoured models. We also show a subset of comparison models, including the best all-S or all-P model and the next-best S/T-type mixture, to indicate the relative support for the favoured model.}
    \label{tab:qu_fitting_summary}
\end{table*}

\def\arraystretch{1.55}
\begingroup
\renewcommand{\thetable}{\ref{tab:qu_fitting_summary}}
\begin{table*}
    \centering
    \small
    \setlength{\tabcolsep}{1.0pt}
    \begin{tabular}{lccllcccccccc}
        \Xhline{5\arrayrulewidth}
        \multicolumn{1}{c}{Date} &
        \multicolumn{1}{c}{\begin{tabular}[c]{@{}c@{}}$I_0$ (mJy)\\$\alpha_\nu$\end{tabular}} &
        \multicolumn{1}{c}{Ejecta} &
        \multicolumn{1}{c}{Mod.} &
        \multicolumn{1}{c}{Comp.} &
        \multicolumn{1}{c}{$p_0$ (\%)} &
        \multicolumn{1}{c}{$\psi_0$ (deg)} &
        \multicolumn{1}{c}{$\phi$ ($\rad\,\mathrm{m}^{-2}$)} &
        \multicolumn{1}{c}{$\sigma_\phi$ ($\rad\,\mathrm{m}^{-2}$)} &
        \multicolumn{1}{c}{$N$} &
        \multicolumn{1}{c}{$\beta$} &
        \multicolumn{1}{c}{$\chi^2$ (dof)} &
        \multicolumn{1}{c}{$\ln \mathcal{Z}$} \\
        \Xhline{5\arrayrulewidth}
        \multirow[t]{17}{*}{\begin{tabular}[t]{@{}l@{}}2023-10-14 12:47\\60231.533\end{tabular}} & \multirow[t]{17}{*}{\begin{tabular}[t]{@{}c@{}}$875.03 \pm 0.09$\\$-0.0809 \pm 0.0007$\end{tabular}} & \multirow[t]{17}{*}{$\times$} & ${\rm S}$ & ${\rm S}_1$ & $0.304_{-0.002}^{+0.002}$ & $12.4_{-0.2}^{+0.2}$ & $-6.68_{-0.01}^{+0.04}$ & $\cdots$ & $\cdots$ & $\cdots$ & $4525\,(537)$ & $1583$ \\
        \cline{4-13}
         &  &  & \multirow{4}{*}{${\rm SSSS}$} & ${\rm S}_1$ & $0.65_{-0.12}^{+0.13}$ & $-18_{-2}^{+3}$ & $-5.2_{-0.5}^{+0.6}$ & $\cdots$ & $\cdots$ & $\cdots$ & $1865\,(528)$ & $2859$ \\
         &  &  &  & ${\rm S}_2$ & $0.50_{-0.13}^{+0.14}$ & $49_{-4}^{+2}$ & $-1.9_{-1.0}^{+1.3}$ & $\cdots$ & $\cdots$ & $\cdots$ & $1865\,(528)$ & $2859$ \\
         &  &  &  & ${\rm S}_3$ & $0.045_{-0.003}^{+0.003}$ & $-77_{-13}^{+8}$ & $119_{-2}^{+3}$ & $\cdots$ & $\cdots$ & $\cdots$ & $1865\,(528)$ & $2859$ \\
         &  &  &  & ${\rm S}_4$ & $0.044_{-0.002}^{+0.003}$ & $21_{-9}^{+9}$ & $156_{-2}^{+2}$ & $\cdots$ & $\cdots$ & $\cdots$ & $1865\,(528)$ & $2859$ \\
        \cline{4-13}
         &  &  & \multirow{4}{*}{${\rm \textbf{SSTT}}$} & ${\rm S}_1$ & $0.270_{-0.008}^{+0.012}$ & $2_{-3}^{+3}$ & $-4.4_{-0.6}^{+0.5}$ & $\cdots$ & $\cdots$ & $\cdots$ & $922\,(524)$ & $3326$ \\
         &  &  &  & ${\rm S}_2$ & $0.043_{-0.005}^{+0.014}$ & $-35_{-23}^{+21}$ & $24_{-5}^{+6}$ & $\cdots$ & $\cdots$ & $\cdots$ & $922\,(524)$ & $3326$ \\
         &  &  &  & ${\rm T}_1$ & $0.50_{-0.03}^{+0.05}$ & $5_{-4}^{+5}$ & $-111.2_{-1.8}^{+1.5}$ & $48.6_{-0.7}^{+0.8}$ & $10_{-2}^{+4}$ & $\cdots$ & $922\,(524)$ & $3326$ \\
         &  &  &  & ${\rm T}_2$ & $0.21_{-0.03}^{+0.03}$ & $-88_{-15}^{+8}$ & $125_{-3}^{+6}$ & $17.6_{-0.9}^{+0.9}$ & $2.3_{-0.3}^{+0.3}$ & $\cdots$ & $922\,(524)$ & $3326$ \\
        \cline{4-13}
         &  &  & \multirow{4}{*}{${\rm STTT}$} & ${\rm S}_1$ & $0.267_{-0.007}^{+0.012}$ & $2_{-3}^{+2}$ & $-4.6_{-0.5}^{+0.6}$ & $\cdots$ & $\cdots$ & $\cdots$ & $922\,(522)$ & $3332$ \\
         &  &  &  & ${\rm T}_1$ & $0.50_{-0.05}^{+0.04}$ & $5_{-5}^{+5}$ & $-111.3_{-1.8}^{+1.5}$ & $48.7_{-0.8}^{+0.8}$ & $10_{-3}^{+4}$ & $\cdots$ & $922\,(522)$ & $3332$ \\
         &  &  &  & ${\rm T}_2$ & $0.047_{-0.009}^{+0.013}$ & $-36_{-24}^{+21}$ & $24_{-5}^{+6}$ & $0_{-1}^{+3}$ & $4_{-2}^{+15}$ & $\cdots$ & $922\,(522)$ & $3332$ \\
         &  &  &  & ${\rm T}_3$ & $0.20_{-0.03}^{+0.04}$ & $90_{-10}^{+13}$ & $125_{-5}^{+4}$ & $17.7_{-0.9}^{+0.8}$ & $2.3_{-0.3}^{+0.3}$ & $\cdots$ & $922\,(522)$ & $3332$ \\
        \cline{4-13}
         &  &  & \multirow{4}{*}{${\rm PPPP}$} & ${\rm P}_1$ & $5.1_{-0.7}^{+0.5}$ & $31.8_{-0.9}^{+0.9}$ & $-6.66_{-0.02}^{+0.09}$ & $\cdots$ & $\cdots$ & $1.03_{-0.12}^{+0.12}$ & $1899\,(527)$ & $2782$ \\
         &  &  &  & ${\rm P}_2$ & $10.3_{-1.0}^{+1.0}$ & $-57.0_{-0.9}^{+1.0}$ & $-6.66_{-0.02}^{+0.09}$ & $\cdots$ & $\cdots$ & $2.68_{-0.07}^{+0.06}$ & $1899\,(527)$ & $2782$ \\
         &  &  &  & ${\rm P}_3$ & $13.3_{-0.3}^{+0.3}$ & $33.4_{-0.9}^{+0.9}$ & $-6.66_{-0.02}^{+0.09}$ & $\cdots$ & $\cdots$ & $4.47_{-0.05}^{+0.04}$ & $1899\,(527)$ & $2782$ \\
         &  &  &  & ${\rm P}_4$ & $7.6_{-0.4}^{+0.4}$ & $-56.4_{-0.8}^{+1.0}$ & $-6.66_{-0.02}^{+0.09}$ & $\cdots$ & $\cdots$ & $4.997_{-0.011}^{+0.003}$ & $1899\,(527)$ & $2782$ \\
        \Xhline{3\arrayrulewidth}
        \multirow[t]{7}{*}{\begin{tabular}[t]{@{}l@{}}2023-10-16 15:57\\60233.665\end{tabular}} & \multirow[t]{7}{*}{\begin{tabular}[t]{@{}c@{}}$109.89 \pm 0.03$\\$-0.3047 \pm 0.0017$\end{tabular}} & \multirow[t]{7}{*}{$\times$} & ${\rm S}$ & ${\rm S}_1$ & $1.883_{-0.016}^{+0.014}$ & $64.8_{-0.2}^{+0.2}$ & $-6.820_{-0.001}^{+0.003}$ & $\cdots$ & $\cdots$ & $\cdots$ & $5859\,(537)$ & $-216$ \\
        \cline{4-13}
         &  &  & \multirow{2}{*}{${\rm SS}$} & ${\rm S}_1$ & $10_{-2}^{+5}$ & $49_{-2}^{+2}$ & $-5.1_{-0.5}^{+0.4}$ & $\cdots$ & $\cdots$ & $\cdots$ & $545\,(534)$ & $2434$ \\
         &  &  &  & ${\rm S}_2$ & $9_{-2}^{+5}$ & $-51_{-3}^{+2}$ & $-2.6_{-0.6}^{+0.7}$ & $\cdots$ & $\cdots$ & $\cdots$ & $545\,(534)$ & $2434$ \\
        \cline{4-13}
         &  &  & \multirow{2}{*}{${\rm \textbf{ST}}$} & ${\rm S}_1$ & $3.4_{-0.6}^{+0.4}$ & $17_{-4}^{+5}$ & $-4.6_{-0.5}^{+0.7}$ & $\cdots$ & $\cdots$ & $\cdots$ & $532\,(532)$ & $2442$ \\
         &  &  &  & ${\rm T}_1$ & $5.6_{-0.6}^{+0.5}$ & $-89.6_{-1.9}^{+2.0}$ & $-4.1_{-0.3}^{+0.4}$ & $13_{-2}^{+1}$ & $5_{-3}^{+8}$ & $\cdots$ & $532\,(532)$ & $2442$ \\
        \cline{4-13}
         &  &  & \multirow{2}{*}{${\rm PP}$} & ${\rm P}_1$ & $6.0_{-2.0}^{+0.9}$ & $-3_{-2}^{+3}$ & $-4.2_{-0.3}^{+0.4}$ & $\cdots$ & $\cdots$ & $-2.1_{-0.4}^{+0.3}$ & $539\,(533)$ & $2438$ \\
         &  &  &  & ${\rm P}_2$ & $5_{-2}^{+1}$ & $78.4_{-1.6}^{+1.2}$ & $-4.2_{-0.3}^{+0.4}$ & $\cdots$ & $\cdots$ & $-0.3_{-0.2}^{+0.3}$ & $539\,(533)$ & $2438$ \\
        \Xhline{3\arrayrulewidth}
        \multirow[t]{4}{*}{\begin{tabular}[t]{@{}l@{}}2023-10-22 11:32\\60239.481\end{tabular}} & \multirow[t]{4}{*}{\begin{tabular}[t]{@{}c@{}}$59.05 \pm 0.03$\\$-0.453 \pm 0.004$\end{tabular}} & \multirow[t]{4}{*}{$\times$} & ${\rm \textbf{S}}$ & ${\rm S}_1$ & $7.11_{-0.03}^{+0.03}$ & $5.5_{-0.4}^{+0.3}$ & $-4.15_{-0.09}^{+0.09}$ & $\cdots$ & $\cdots$ & $\cdots$ & $294\,(277)$ & $1172$ \\
        \cline{4-13}
         &  &  & \multirow{2}{*}{${\rm SS}$} & ${\rm S}_1$ & $7.3_{-1.7}^{+1.9}$ & $6_{-3}^{+3}$ & $-4.1_{-0.1}^{+0.4}$ & $\cdots$ & $\cdots$ & $\cdots$ & $291\,(274)$ & $1168$ \\
         &  &  &  & ${\rm S}_2$ & $0.4_{-0.4}^{+1.5}$ & $-79_{-38}^{+7}$ & $-4_{-1}^{+2}$ & $\cdots$ & $\cdots$ & $\cdots$ & $291\,(274)$ & $1168$ \\
        \cline{4-13}
         &  &  & ${\rm P}$ & ${\rm P}_1$ & $7.15_{-0.04}^{+0.05}$ & $5.5_{-0.3}^{+0.3}$ & $-4.16_{-0.09}^{+0.09}$ & $\cdots$ & $\cdots$ & $-0.02_{-0.02}^{+0.03}$ & $293\,(276)$ & $1168$ \\
        \Xhline{5\arrayrulewidth}
    \end{tabular}
    \caption[]{Continued.}
\end{table*}
\endgroup

The four epochs that favour Faraday-thick component(s) span a range of widths, shapes, and offsets from the ISM-like component; full parameter constraints are listed in Table~\ref{tab:qu_fitting_summary}. The two off-peak epochs, 2023 September 23 and 2023 October 16 (B and F in Figure~\ref{fig:results_ILoveThisPlot}), favour relatively narrow Faraday-thick components centred close to the Faraday-thin emission, with $\sigma_\phi\,{\sim}\,20$ and ${\sim}\,13$~\radPerSqm, respectively. In both cases, the shape parameter is only weakly constrained, with $N\,{\sim}\,9$ but 99\% credible intervals extending approximately to the prior bounds.

By contrast, the flare-peak epochs require broader Faraday-thick components that are significantly offset from the stable ISM-like component. On 2023 October 06 (D in Figure~\ref{fig:results_ILoveThisPlot}), we infer two such components centred at $\phi_{\rm rm}\,{\sim}\,-88$ and ${\sim}\,53$~\radPerSqm, with widths of $\sigma_\phi\,{\sim}\,21$ and ${\sim}\,34$~\radPerSqm. On 2023 October 14 (E in Figure~\ref{fig:results_ILoveThisPlot}), the corresponding components are centred at $\phi_{\rm rm}\,{\sim}\,-111$ and ${\sim}\,125$~\radPerSqm, with widths of $\sigma_\phi\,{\sim}\,49$ and ${\sim}\,18$~\radPerSqm; this epoch also requires an additional, poorly constrained Faraday-thin component at $\phi_{\rm rm}\,{\sim}\,25$~\radPerSqm. Unlike the off-peak thick components, some of the flare-peak components have shape parameters that are meaningfully constrained away from the prior bounds. This indicates that the data contain information about the Faraday-depth profile, not only its characteristic width. Some of these constraints favour values below the $N\relto{>}{15}$ regime associated with a canonical Burn slab, suggesting that the Faraday-rotating material, or at least our preferred parametric description of it, is not well described by a simple uniform emitting and rotating slab.

The thick-component profile is described by both a scale parameter, $\sigma_\phi$, and a shape parameter, $N$. These two parameters are strongly degenerate: different combinations of $\sigma_\phi$ and $N$ can produce profiles with similar effective widths but different detailed shapes. As a result, $\sigma_\phi$ alone is not a shape-independent measure of the Faraday thickness. For example, in the $N\rightarrow\infty$ limit, the component approaches a top-hat-like profile with a full width of approximately $2\sigma_\phi$, whereas for a Gaussian-like component ($N\,{\sim}\,{2}$), $\sigma_\phi$ corresponds to the standard 68\% interval. To compare widths consistently across components, we therefore convert each posterior sample of $(\sigma_\phi,N)$ into a direct containment width by calculating the Faraday-depth interval that contains 99\% of the integrated component profile. This yields a posterior distribution for the 99\% highest-density containment width, $W_{\rm rm,99}$. The inferred $W_{\rm rm,99}$ ranges are 40--80~\radPerSqm\ for 2023 September 23, 40--70 and 90--130~\radPerSqm\ for the negative and positive components on 2023 October 06, 100--130 and 60--120~\radPerSqm\ for the negative and positive components on 2023 October 14, and 25--50~\radPerSqm\ for 2023 October 16. Motivated by the flare-peak components, which are most directly comparable to synchrotron self-absorption analyses of flaring ejecta \citep{fender2019,CF_PLACEHOLD}, we adopt $W_{\rm rm,99}\,{\sim}\,100$~\radPerSqm\ as a characteristic Faraday thickness for the physical interpretation in Section~\ref{sec:disc_thick_mass}.

\begin{figure}
    \centering
    \includegraphics[width=1.0\linewidth]{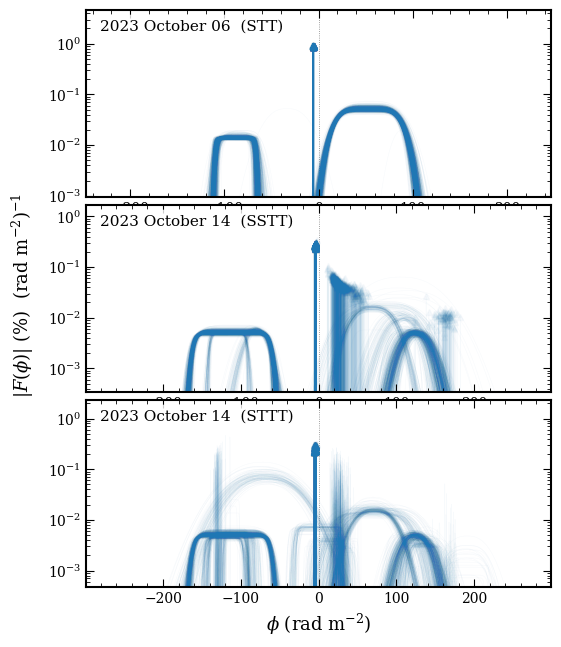}
    \caption{Seed-dependence test for the peak-flaring $QU$ fits. Each panel shows 100 posterior samples from each of 10 independent random seeds. The 2023 October 06 STT solution is stable across trials, while the 2023 October 14 SSTT and STTT fits show larger seed-to-seed variation, including solutions with very broad Faraday-thick components.}
    \label{fig:results_seed_check}
\end{figure}

Since the peak-flaring epochs favour several broad components that are susceptible to scale-filtering effects from the finite $\lambda^2$ coverage, we tested the robustness of these models against the sampler initialisation. For each peak epoch, we repeated the fitting with 10 independent random seeds and drew 100 posterior samples from each trial, as shown in Figure~\ref{fig:results_seed_check}. The 2023 October 06 fits are stable across trials, with the best-fitting STT model consistently recovered. By contrast, the 2023 October 14 fits show substantial seed-to-seed variation for both the preferred SSTT model and the next-best STTT model, although the fiducial solution shown in Figure~\ref{fig:results_ILoveThisPlot} corresponds to the highest-density posterior region. Of particular interest are the trials that favour extremely broad Faraday-thick components, where smaller $N$ values produce extended wings and increase the width enclosing 99\% of the integrated profile. These solutions suggest that the intrinsic Faraday-thick medium may be broader or more structured than our fiducial fits imply. Qualitatively, however, they still favour strong Faraday-thick components on either side of the ISM-like component. Given the greater stability of the 2023 October 06 model, we focus on that epoch in the synchrotron self-absorption analysis in Section~\ref{sec:disc_thick_mass}.

\subsection{Plausibility of P-component models}
\label{sec:result_pcomps}
We included the additional parametric P-type models to test whether component-specific spectral slopes could explain the observed departures from Faraday simplicity without invoking Faraday-thick structure. For the simpler deviations, this is at least plausible. The epochs on 2023 September 16 and 2023 October 01 (A and C in Figure~\ref{fig:results_ILoveThisPlot}) can be adequately described with modest power-law indices, $\beta\relto{\sim}{1}$, where $p(\nu)\propto\nu^\beta$. The Bayesian evidence also favours these models, with $\Delta\ln\mathcal{Z}\lesssim6$, although this remains below our adopted threshold for a strong preference.

However, this explanation becomes less satisfactory for the more complex epochs. For example, the 2023 September 23 epoch requires a negative $\beta$, such that the polarisation fraction decreases with increasing frequency. This is difficult to motivate in the usual picture for jetted sources, where lower frequencies are thought to preferentially probe larger distances downstream and more disordered magnetic fields \citep[see the discussion of Mrk~501 in][]{Liodakis2022Mrk501}. In that framework, intrinsic spectral structure should more naturally produce positive polarisation spectral indices, as seen empirically in X-ray binary flares \citep[e.g.][]{han1992,hannikainen2001,brocksopp2007,brocksopp2013,curran2014,hughes2023}, although simultaneous comparisons are complicated by frequency-dependent time delays in expanding ejecta \citep[see][]{vdl1966,atetarenko2017,fender2019,CF_PLACEHOLD}. A cleaner comparison comes from AGN, where the time-domain ambiguity is less severe \citep[e.g.][]{Taylor2024}; even there, negative polarised spectral indices, or apparent ``repolarisation'', are commonly attributed to Faraday effects rather than intrinsic spectral stratification \citep[e.g.][]{Farnes_polcatalog2014}.

More severely, the flaring epochs (D--F in Figure~\ref{fig:results_ILoveThisPlot}) require multiple polarised power-law components with dramatically different $\beta$ values, despite the Stokes $I$ spectrum itself remaining well described by a single power law (see, e.g. Figure~\ref{fig:J1727_systematics}). Some components require $\beta\relto{>}{4}$ (see Table~\ref{tab:qu_fitting_summary}), which would imply polarisation fractions \relto{>}{100\%} just outside our observing bandwidth, while still giving worse model-comparison ($\ln\mathcal{Z}$) and model-assessment ($\chi^2$) statistics. One could invoke spectral turnovers, or relax the assumption that the P-components share tied $\phi_{\rm rm}$ values, but at that point the model is no longer a simple intrinsic spectral-slope explanation. It would instead require additional frequency-dependent structure whose physical origin would itself need explaining, effectively reintroducing Faraday complexity by another route. We therefore do not consider the P-models plausible for the complex epochs, and focus the rest of this work on the implications of transient Faraday-thick structure.

\section{Discussion}
\label{sec:discussion}
In this section, we present our main physical interpretation. We argue that the transient Faraday complexity is most naturally intrinsic to the jets, and therefore provides a probe of their internal plasma conditions. In particular, we show that this complexity can distinguish between electron-positron and electron-proton plasma compositions, constrain the mass contained within the Faraday-rotating material when combined with Stokes $I$ flaring analyses \citep[using the methods detailed in][]{CF_PLACEHOLD}, and provide insight into the underlying particle energy spectrum. We conclude by considering possible physical origins for the obliquity and temporal evolution of the linear polarisation angles.

\subsection{The origin of the transient Faraday complexity}
\label{sec:discu_results_origin}
We can separate the mechanisms responsible for the observed spectropolarimetric evolution into external and internal Faraday rotation. External rotation may arise either in the Galactic ISM or in material local to the outburst. A variable Galactic ISM contribution is disfavoured by the stable Faraday-thin component measured before and after the flaring interval, $\phi_{f,{\rm ISM}}\relto{\sim}{-1.0}\radPerSqm$, and by the consistent Faraday depths measured toward the angularly separated core and ejecta. Together, these show that the foreground Faraday depth is stable in time and across angular scales of \relto{\lesssim}{5^{\prime\prime}}, consistent with the modest Galactic structure function expected on such small scales along this line of sight \citep{MinterStructureFunc1996}. We therefore see no Faraday-rotation analogue of the line-of-sight H\,{\sc i} variations observed across the ejecta of GRS~1915+105 \citep{Dhawan1915HI}. A purely foreground ISM screen is also unlikely to explain the Faraday-thick structure, since producing a broadened component through external Faraday dispersion would require strong unresolved fluctuations in $\phi_f$ across the beam, which are difficult to reconcile with the observed foreground stability.

A transient local external screen is more plausible than the Galactic ISM, since an outflow could in principle introduce local rotation, and turbulence within that material could produce approximately Gaussian broadening in Faraday space. This explanation would still struggle to reproduce the broader $N>3$ components, or the presence of multiple distinct components, but the most natural candidate would be an accretion-disc wind. Strong disc winds in XRBs are typically associated with softer accretion states \citep[see, e.g.][for reviews]{Ponti2012discwinds,DiazTrigo2016Winds,Teo2026Winds}, although optical line emission indicates that a wind was likely present in \src\ close to the start of the outburst, with inferred velocities of \relto{\lesssim}{1000}\kms\ in the lead-up to the flaring epochs \citep{Castro2026J1727winds}. However, \src\ then faces the same difficulty identified for the 2015 outburst of V404~Cygni \citep{hughes2023}. If the wind is isotropic, the required mass-loss rate is high, orders of magnitude larger than the $10^{-9}{\rm\,M_\odot\,yr^{-1}}$ inferred by \citet{Castro2026J1727winds}. Alternatively, the wind would need to be highly directional, and would have to favour, by chance, the line of sight connecting the jets to Earth.

\begin{figure*}
    \centering
    \includegraphics[width=1.0\linewidth]{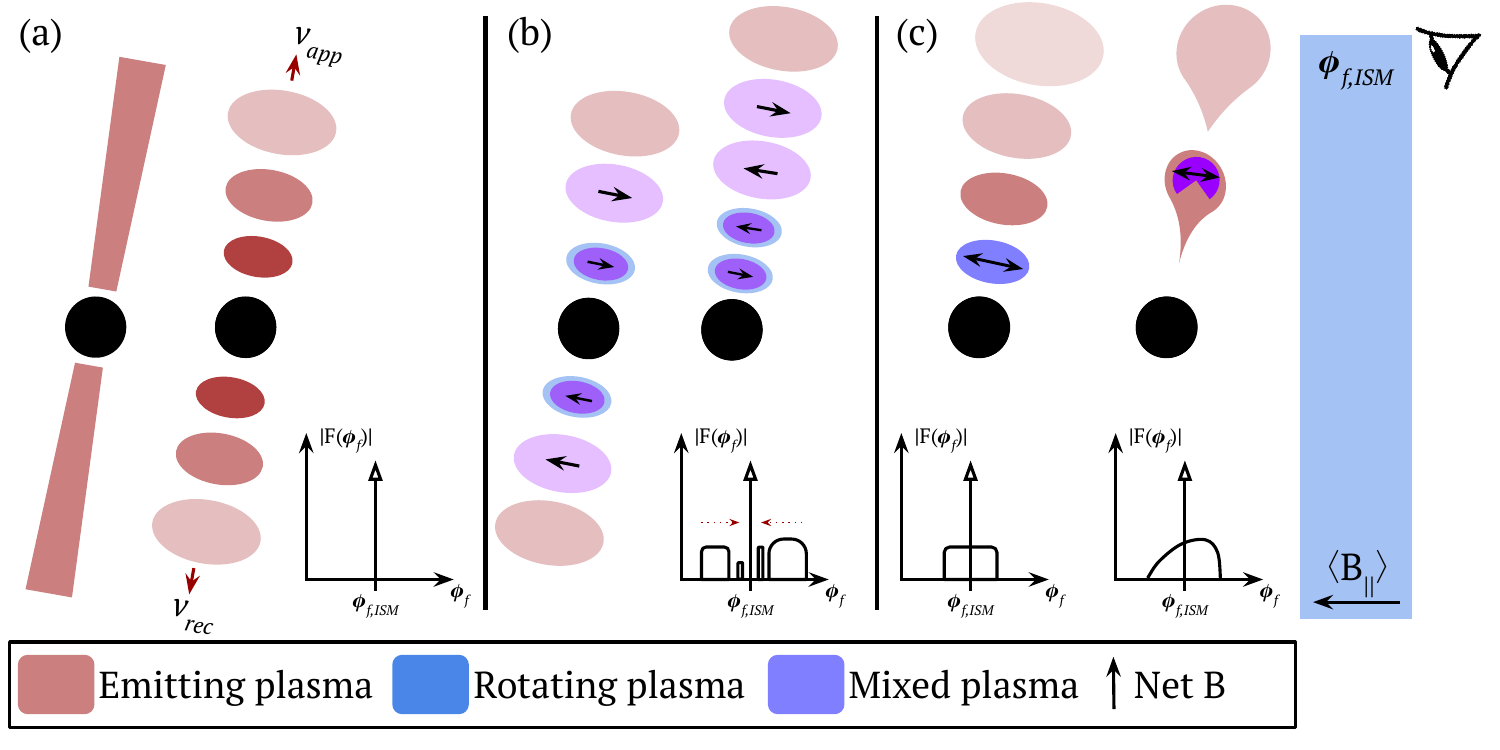}
\caption{
Schematic examples of Faraday-space signatures expected from different jet and foreground configurations. Red denotes synchrotron-emitting plasma, blue denotes Faraday-rotating plasma, and purple denotes plasma that is both emitting and rotating. Dimmer colours indicate lower emissivity or a weaker Faraday-rotating contribution. The rightmost blue screen represents the Galactic ISM, the eye marks the observer's line of sight, and black arrows indicate the magnetic-field direction. For each case, we show the corresponding schematic Faraday spectrum. The magnetic fields are drawn perpendicular to the propagation direction for clarity, but parallel or oblique fields would give the same qualitative interpretation. (a) For an unresolved source containing only emitting plasma behind the ISM, any jet or ejecta geometry collapses to a single Faraday-thin component at $\phi_{f,{\rm ISM}}$. (b) Local rotating plasma surrounding or coincident with the emitting regions can shift components away from $\phi_{f,{\rm ISM}}$ and give them finite width. This can occur in two-sided jets, or when only the approaching side is detected, for example due to beaming. As the ejecta expand and dilute, these components should tend back toward the ISM value in both centroid and width. (c) Mixed emitting and rotating plasma with magnetic-field reversals and/or more complex geometry can produce broad or skewed Faraday-depth distributions, including both positive and negative offsets relative to $\phi_{f,{\rm ISM}}$.}
\label{fig:disc_schematic}
\end{figure*}

A more serious problem with a disc wind explanation is the flow timescale. The optically thin photospheric distance of the compact jet at 1.28\GHz\ in \src\ is of order $10^{15}$\cms\ \citep[e.g.][]{2025ZdziarskiJ1727steady}. If this is the minimum relevant emission radius, a wind moving at \relto{\lesssim}{1000}\kms\ and maintaining a constant velocity would travel only ${\sim}\,10^{13}$\cms\ per day. It would therefore require $\sim100$\,d to reach $10^{15}$\cms, far longer than the interval between the outburst onset and the first Faraday-thick epoch. A substantially faster wind could reduce the travel time, but for a fixed mass-loss rate it would also be less dense, and so would exacerbate the mass-budget problem. More generally, an external screen provides little explanatory value for the rapid emergence and disappearance of the thick components. These shortcomings point instead to a simpler interpretation in which the transient rotating plasma is internal to the jets themselves. Figure~\ref{fig:disc_schematic} shows possible internal-jet configurations capable of producing the observed Faraday spectra, which we discuss further below.

\subsubsection{Faraday-simple configurations}
\label{sec:disc_simple}
Panel (a) shows Faraday-thin-only configurations, illustrating why the number of components in Faraday space should not be equated with the number of discrete jet components. In principle, the unresolved radio emission could contain a compact jet, a single ejecta, several unresolved ejecta, or a simultaneous compact jet and ejecta. However, if these components are viewed through the same foreground screen, such as the ISM, or otherwise share the same effective Faraday depth, they collapse to a single component in Faraday space. This is likely the case before and after the flare, when the source transitions from a compact jet to multiple unresolved ejecta without a corresponding increase in Faraday complexity.

\subsubsection{Multiple Faraday-thick components}
\label{sec:disc_multi_thick}
Panel (b) shows configurations that can produce multiple Faraday-thick components shifted to positive and negative depths relative to the Faraday-thin ISM emission, as inferred during the flare-peak epochs on 2023 October 06 and 14. In this case, the additional components may map more closely onto individual ejecta, or other shock-based jet features. The offsets of the thick-component centroids from the ISM-rotated emission could be enhanced if each emitting feature also lies behind additional Faraday-rotating material that is local to the jet but separate from the emitting plasma (schematic interpretation shown in Figure~\ref{fig:disc_schematic}b, e.g. the inner-most ejecta). The left-hand configuration illustrates one possible geometry, in which two thick components arise from an approaching--receding pair of ejecta, alongside one or more thin components. This is qualitatively appealing, since the mirror symmetry of a two-sided outflow could naturally reverse the line-of-sight magnetic-field contribution and produce components on either side of the ISM value.

The idea is also consistent with the polarisation fractions of the thick components. Within a given epoch, each $p_0$ is measured relative to the same unresolved Stokes $I$ flux density, integrated over all contributing components. Ratios of $p_0$ therefore cancel this common denominator and trace ratios of linearly polarised flux density, not ratios of intrinsic polarisation fraction. For the two thick components in each flaring epoch, we find
\begin{align*}
    R_F = \frac{p_{0,a}}{p_{0,r}} \sim 2{\rm-}6,
\end{align*}
where the subscripts $a$ and $r$ denote the approaching and receding components. We can assess whether such ratios are feasible for an approaching--receding ejecta pair using the geometric constraints on \src, including its distance, inclination, and proper motion. We adopt the formalism of \citet{millerjones2004}, based on \citet{mirabel1994}, to calculate the flux-density ratio of approaching and receding components, assuming they are intrinsically symmetric. We evaluate this ratio at a fixed observer time, rather than at a fixed angular separation from the core, because this is the relevant comparison for a single observing epoch. In this case, light-travel-time effects mean that the approaching and receding ejecta are observed at different intrinsic ages. In this framework, the measured flux-density ratio is
\begin{align*}
    R_F = \left(\frac{1 + \beta\cos i}{1 - \beta\cos i}\right)^{\kappa},
\end{align*}
where $\beta$ is the intrinsic ejecta speed in units of $c$, and $i$ is the inclination angle to the line of sight. We keep the exponent as a general parameter, $\kappa$, because it depends on the source geometry, homogeneity, expansion history, and intrinsic evolution. For optically thin, linearly expanding homogeneous ejecta, $\kappa=k-p$, where $k=3$ for discrete ejecta \citep[although empirical measurements find $k\,{<}\,3$ for ejecta]{mirabel1994,Fender1999}, and $p$ is the power-law index of the non-thermal electron energy distribution, $n_{\rm rel}(\gamma)\propto\gamma^{-p}$ \citep{millerjones2004,longair2011}.

\begin{figure}
    \centering
    \includegraphics[width=1.0\linewidth]{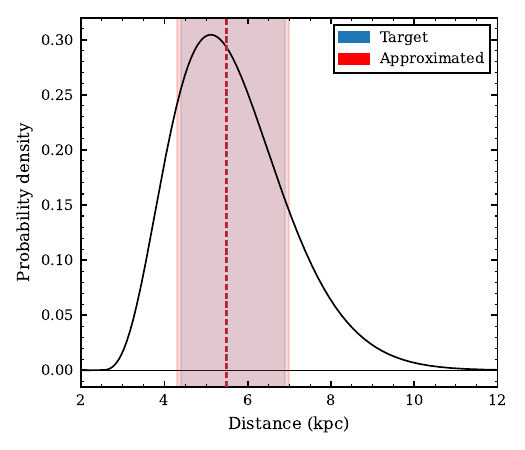}
    \caption{
    Bounded beta-distribution approximation to the distance posterior for \src. The blue and red intervals compare the approximated and target medians and 68\% equal-tailed intervals from \citet{Burridge2025distance1727}. The distribution is parametrised using the \texttt{scipy.stats.beta} convention, with $a=4.71696$, $b=30.00000$, $\mathrm{loc}=2.41143$, and $\mathrm{scale}=23.79047$.
    }
    \label{fig:disc_distance}
\end{figure}

We then use a Monte Carlo approach to sample the relevant intrinsic properties. For the distance to \src, we approximate the distribution inferred by \citet{Burridge2025distance1727} with a bounded beta distribution. We optimise its parameters to reproduce their 68\% equal-tailed interval, $5.5_{-1.1}^{+1.4}$\,kpc, and the qualitatively identified mode near 5.1\,kpc, while keeping ${>}\,99\%$ of the probability density within the approximate range 2.5--10\,kpc. The resulting approximation is shown in Figure~\ref{fig:disc_distance}, demonstrating that this parameterisation captures the main features of their Figure~5.

For the inclination, X-ray spectral modelling generally favours $20\,{\lesssim}\,i\,{\lesssim}\,60$\,deg \citep{Peng2024,Svoboda2024,Liu2024,Burridge2025distance1727}. The absence of ellipsoidal modulations gives a more model-independent upper limit of $i\,{<}\,74$\,deg \citep{Mata2025distance1727}, while recent optical spectroscopic modelling of wind-line profiles favours a moderate inclination of $45\,{\lesssim}\,i\,{\lesssim}\,55$\,deg \citep{Castro2026J1727winds}. The proper motion of the discrete ejecta adds a distance-dependent constraint, since the intrinsic velocity must remain subluminal. Adopting the largest angular velocity reported by \citet{Wood2025SWJ1717Ejecta}, $\mu_{\rm max}=3.18\pm0.1{\rm\,mas\,hr^{-1}}$, the corresponding maximum inclination is
\begin{align}
    i_{\rm max} = 2\tan^{-1}\left(\frac{c}{\mu_{\rm max}D}\right).
\end{align}
For each distance draw, we sample inclinations uniformly in $\cos i$ over the broad interval $i\in[20,\min(74,i_{\rm max}(D))]$\,deg. We then draw $\beta\in[0.1,0.99]$. Although all VLBI-detected ejecta from \src\ have $\beta>0.1$ (Wood et al., submitted), we adopt this as a deliberately conservative lower bound, since slower ejecta reduce the maximum approaching--receding flux-density contrast. Finally, we allow $\kappa\in[0.1,1]$, with a conservative emphasis on lower beaming exponents that make large flux-density ratios less favoured. The purpose of this calculation is not to infer a unique jet speed or inclination, but to test whether the observed polarised-flux ratios are plausible for a paired approaching--receding ejecta interpretation, given the currently weak constraints on the geometry and dynamics of \src. Across the sampled parameter space, they are. Despite this conservative setup, the model produces $R_F\relto{\gtrsim}{2}$ in approximately 20\% of samples, and still produces $R_F\relto{\gtrsim}{6}$ in 3\% of samples. The fitted amplitudes therefore do not prove the approaching--receding interpretation, but they are quantitatively consistent with it.

This interpretation raises an immediate observational caveat: would a receding component be consistent with the available VLBI imaging? Existing VLBI observations do not provide clear evidence for one. They either show a bright stationary core with only approaching components \citep[e.g.][]{Wood2025SWJ1717Ejecta}, or contain a feature whose identification is ambiguous between a stationary core and receding ejecta blended with the core emission; in the latter case, the authors favour the stationary-core interpretation (Wood et al., submitted). Thus, a receding component is not directly supported by the VLBI imaging, although it also cannot be ruled out if it is faint or blended at the relevant epochs. Without simultaneous VLBI spectropolarimetry, we cannot uniquely associate the unresolved Faraday components with distinct Stokes $I$ components. As a crude check, we therefore treat the fitted $p_0$ values as proxies for the relative Stokes $I$ contributions of the different components,
\begin{align*}
    I_i = \frac{p_{0,i}}{\sum_{j=0}^N p_{0,j}} I_{\rm tot}, \quad {\rm for~component}~i\in[0,N].
\end{align*}
This assumes comparable intrinsic fractional polarisations and depolarisation behaviour, and should therefore be interpreted only as an order-of-magnitude estimate. Under this assumption, the putative receding components on 2023 October 06 and 14 would have flux densities of order \relto{\sim}{50-200}\mJy, far above the nominal detection thresholds of the relevant interferometers. However, the observations were not simultaneous. The VLBI campaign bookended our October 06 observation by ${\pm}\,12\,$hr; the earlier epoch showed a bright flare (\relto{\gtrsim}{100}{mJy} at \relto{\sim}{8}{GHz}; Wood et al., submitted), while the later epoch showed a single propagating ejecta at the $\sim$\,mJy level. The timing mismatch was worse for the brighter October 14 flare, with several days separating the VLBI and MeerKAT observations. Given the rapid evolution and the ambiguity in the VLBI morphology, we cannot rule out the approaching--receding interpretation. This highlights the need for simultaneous VLBI and spectropolarimetric observations. If no compact VLBI core were present when the Faraday components appear, a receding component could not easily be hidden by core blending, and its non-detection would strongly disfavour the receding-ejecta interpretation.

Regardless, this caveat does not rule out an ejecta-internal origin for the observed Faraday complexity. If the thick components do not correspond to an approaching--receding pair, they could instead arise from multiple approaching ejecta, several of which have already been seen in VLBI imaging of \src\ \citep{Wood2025SWJ1717Ejecta}, and later in MeerKAT imaging (Carotenuto et al., in prep.). In that case, the positive and negative $\phi_f$ offsets would not trace opposite jet sidedness, but stochastic changes in the sign of the line-of-sight magnetic field within the rotating plasma, as illustrated in the right-hand schematic of Figure~\ref{fig:disc_schematic}b. This need not imply distinct intrinsic polarisation angles, because linear polarisation is insensitive to 180-degree reversals in the projected magnetic-field direction. A common ordering process, such as shock compression \citep{Laing1980,hughes1985jetpolshocks,Hughes1989jetpola,Hughes1989jetpolb,Bell2004_magnetic}, could therefore produce similar $\psi_0$ values in different ejecta, while stochastic reversals in $B_\parallel$ set the sign of the internal rotation. A intuitive way to distinguish these scenarios would be to test whether future XRB outbursts repeatedly produce thick components in positive--negative pairs around the foreground ISM contribution. Persistent pairing would support an approaching--receding interpretation, while a more stochastic distribution of $\phi$ offsets, including multiple ejecta on the same side of the ISM contribution, would favour signs set by local magnetic-field realisations within individual ejecta.

In either scenario, the expected evolution is similar: as the ejecta expand and propagate, their internal Faraday depths should decrease, causing the thick components to narrow and move back toward the foreground ISM value. This is illustrated by the red arrows in the schematic Faraday spectra of Figure~\ref{fig:disc_schematic}, which mark the expected `inward motion' (in Faraday space) with time. In this picture, features that are thinner and closer to the ISM component would correspond to a later evolutionary stage. This may explain the additional Faraday-thin component required in the 2023 October 14 model, as well as the more modest complexity seen on 2023 September 16 and October 01, where the data are sufficiently described by an additional thin component with $\phi_f - \phi_{f,\rm ISM} \ll \Phi_{RMSF}$. In these lower signal-to-noise cases, the data may no longer identify sub-RMSF-scale width, so a weak or narrowing thick component can instead appear as an offset thin feature. The detailed properties of these components remain poorly constrained, particularly given the seed dependence of the 2023 October 14 fit, but the robust point is that the additional structure lies close to the ISM component. Long-track, many-hour observations would provide a powerful test of this interpretation by directly tracking this inward evolution as the source returns toward Faraday simplicity.

\subsubsection{Single Faraday-thick components}
\label{sec:disc_single_thick}
In the final panel, Figure~\ref{fig:disc_schematic}c, we show alternative Faraday-thick configurations. In these examples, the rotating plasma contains reversals in the line-of-sight magnetic field, distributing the internal depth to both positive and negative values relative to the foreground ISM contribution. The resulting thick component can therefore remain centred close to the ISM value, rather than being strongly offset from it. This provides a possible interpretation for the off-peak but complex epochs on 2023 September 23 and October 16, which do not require a large displacement in $\phi_{\rm rm}$ from the ISM contribution.

We emphasise, however, that although our L-band observations strongly favour thick structure, they also push against the fundamental limits of the observing setup. The resolution in $\phi_f$, set by the FWHM of the RMSF, together with the modest polarised signal-to-noise of the off-peak epochs, makes it difficult to distinguish unambiguously between the configurations shown in Figures~\ref{fig:disc_schematic}b and \ref{fig:disc_schematic}c. For example, an apparently single, centrally located thick component could instead be two oppositely signed components that are unresolved in Faraday space. This degeneracy could be reduced by increasing the $\lambda^2$ bandwidth and narrowing the RMSF, for example through simultaneous MeerKAT UHF-band observations. The converse ambiguity is also possible, since the strongly offset thick components inferred during the 2023 October 06 and 14 flare-peak epochs may themselves be partially resolved pieces of broader structure in $\phi_f$. Some inferred widths reach $\sigma_\phi$\relto{\gtrsim}{40}\radPerSqm, already larger than the scale at which broad components begin to be significantly resolved out at L band (\relto{\sim}{25}\radPerSqm). Our parametric modelling, together with the high signal-to-noise of the flare-peak epochs, allows us to push against this limit, but doing so increases the sensitivity to systematic errors if the true depth distribution deviates from the assumed parameterisation. The right-hand example in Figure~\ref{fig:disc_schematic}c illustrates this possibility, where a more complex ejecta geometry produces a strongly skewed distribution and the ejecta contribute to both the thin and broad structure. Such complexity is plausible given that simulations of propagating ejecta can show highly structured morphologies \citep[e.g.][]{Savard2025}. In this sense, we may be fitting patch-wise components to an intrinsically broader or skewed distribution. Testing this directly would require greater sensitivity to broad emission in Faraday depth, which in turn requires decreasing $\lambda^2_{\rm min}$, for example through simultaneous MeerKAT S-band observations.

Finally, we caution that the configurations in Figure~\ref{fig:disc_schematic} are necessarily non-exhaustive. The magnetic field is drawn perpendicular to the direction of propagation as a visual convention, although parallel or oblique field geometries would lead to the same broad interpretations, and may be expected given the polarisation-angle evolution. The schematic is intended to build intuition for how internal rotation within jet components can map onto the observed Faraday spectra, rather than to provide a complete catalogue of possible geometries or a unique solution for \src. Real jets may contain gradients in density, magnetic-field strength, field direction, and emitting-particle distribution, all of which could broaden, skew, or fragment the Faraday-depth distribution in ways not captured by our simple parametric forms. Discriminating between these possibilities will likely require broadband spectropolarimetry together with simultaneous VLBI imaging, to associate the Faraday components recovered here with individual ejecta. This comparison is not straightforward because the relevant arrays can have very different angular-scale sensitivity from MeerKAT. For example, the VLBI observations used by Wood et al. (submitted) are insensitive to scales \relto{\gtrsim}{40}\,mas at \relto{\sim}{8}\GHz, comparable to only about one thirtieth of a MeerKAT image pixel in our analysis. Spatial filtering, unresolved substructure, and different observing frequencies could therefore complicate the mapping between VLBI components, MeerKAT Stokes $I$ emission, and the polarimetric components recovered here, but such observations will be necessary to connect Faraday-space structure with the physical structure of XRB jets.

Nevertheless, these configuration-dependent uncertainties do not affect the central claim. The transient complexity is best explained by internal Faraday rotation within the jet ejecta themselves, enabling the broader, geometry-independent inferences that follow.

\subsection{Evidence for electron--proton plasma in the ejecta}
Perhaps the simplest physical consequence of internal Faraday rotation is that the ejecta must contain an electron--ion component, most naturally an electron--proton plasma. This follows directly from the leading dependence of the Faraday-depth equation, $\phi_f\propto q^3m^{-2}$ (Eq.~6). The $m^{-2}$ term means that the rotation is dominated by the lightest charged particles, while the odd power of $q$ reflects the underlying Lorentz-force symmetry, since particles of opposite charge gyrate with opposite handedness, so their Faraday rotation contributions have opposite signs. In a charge-symmetric electron--positron plasma, the electron and positron contributions therefore cancel. By contrast, in an electron--proton plasma, the proton contribution is suppressed by a factor of $(m_e/m_p)^2$, leaving the electrons to dominate the net rotation. The detection of internal rotation therefore rules out a purely charge-symmetric pair plasma as the Faraday-rotating material, and instead requires either an electron--ion component within the jet or a substantial lepton charge asymmetry. Such charge asymmetries can arise in some pair plasmas. In pulsar cascades, curvature radiation from a single primary electron seeds pair production, leaving a residual charge imbalance, but empirical estimates typically imply that this imbalance is very small, at the level of one part in $10^5$--$10^7$, although the extrema remain debated \citep{deJager2007PWN,Bucciantini2011Multiplicity,Timokhin2015multiplicty,Timokhin2019multiplicit}. This analogy does not naturally carry over to accreting black-hole systems. In XRBs and AGN, pair production is generally expected to be seeded by high-energy Compton up-scattered photons, which create electrons and positrons in pairs and therefore favour charge symmetry \citep{Henri1991pairprodAGN,Sikora2020AGNpair,Zdziarski2021pairprod,Zdziarski2022cygx1ep}.

There is complementary evidence for electron--proton plasma in jet ejecta. One route comes from ejecta masses inferred through deceleration modelling \citep[e.g.][]{Carotenuto2024,Savard2025}, which can exceed the pair-production yield expected from the accretion flow by orders of magnitude over plausible launching timescales. For the previously studied XRB MAXI~J1348--630, for example, the pair-production rate required to supply the inferred ejecta mass is about five orders of magnitude larger than expected \citep[e.g. see][]{Zdziarski2023J1348}. If the flares in systems such as MAXI~J1348--630 were also associated with Faraday-thick ejecta, the required charge asymmetry would make a pure pair-plasma interpretation even more difficult. Even an extreme 10\% lepton charge imbalance would require a pair-production rate a further order of magnitude larger than that implied by the mass argument alone, because the Faraday rotation depends on the imbalance between charges, not just the total number of pairs. A more modest imbalance, such as that corresponding to a pair multiplicity of $10^{5}$, would increase the required pair-production rate by the same factor. 

A separate but consistent argument was put forward by \citet{Zdziarski2024epsdistance}, who suggested that the larger propagation distances of discrete ejecta, compared with compact jets, may reflect different plasma compositions: electron--proton loading in the former, and a more electron--positron-dominated plasma in the latter \citep[consistent with timing analyses of compact jets; e.g.][]{Tetarenko2021Timing}. In this picture, the greater inertia of electron--proton ejecta allows them to propagate further. Thus, while our data do not require the ejecta to be pair-free, they do require the Faraday-rotating material to contain more than a pure, charge-symmetric pair plasma. Combined with previous mass and propagation arguments, the most natural interpretation is that the ejecta contain a substantial, and likely dominant, electron--proton component by the time they reach the scales probed here.

The fact that our Faraday-rotation measurements probe jet composition makes them relevant to jet launching and mass loading. In broad terms, pair-dominated jets are usually associated with Blandford--Znajek-type mechanisms \citep{Blandford1977}, whereas baryon-loaded outflows are more naturally supplied by the accretion flow, as in Blandford--Payne-type disc-driven scenarios \citep{Blandford1982}. However, it is perhaps more likely that we are tracing baryonic mass loading than identifying the launch mechanism directly, since the plasma composition can change after launch through entrainment from the local medium or disc winds \citep{Kantzas2023}. In the following sections, we compare our results with much later stages of jet evolution, where ejecta are studied tens of days after launch \citep[e.g.][]{Carotenuto2024,Savard2025,Cowie2026CirX1shocks}. Although the flaring observations analysed here probe material much closer in time to launch, within $<1$~day, they do not probe the launch region directly. The characteristic length scale, $\sim10^{15}\,{\rm cm}$, close to the $\sim1$~GHz photosphere, corresponds to $\sim10^8$--$10^9R_g$ for a $10\,M_\odot$ black hole, many orders of magnitude beyond the horizon or the expected jet-launching region \citep[e.g.][]{Blandford1977,Blandford1982,2025ZdziarskiJ1727steady}. The internal Faraday rotation could therefore reflect the plasma composition set at launch, and thus retain information about the launching mechanism; alternatively, it could mark the stage of downstream propagation at which baryonic mass loading becomes dominant. In either case, the presence of baryonic material on these scales is a robust conclusion.

\subsection{Faraday thickness as a constraint on ejecta mass}
\label{sec:disc_thick_mass}
Having argued above that the rotating material must include an electron--ion component, we use the characteristic width inferred in Section~\ref{sec:results_QU_faraday}, $W_{\rm rm,99}\sim100$~\radPerSqm, as an order-of-magnitude density constraint. We approximate the line-of-sight integral in Equation~(6) with characteristic averages, $\int n_e B_\parallel\,dl \sim n_{e,{\rm rot}}\langle B_\parallel\rangle l$, giving
\begin{align}
    W_{\rm rm,99} \sim 810
    \left(\frac{n_{e,{\rm rot}}}{1{{\rm\,cm^{-3}}}}\right)
    \left(\frac{\langle B_\parallel\rangle}{1\,{\rm mG}}\right)
    \left(\frac{l}{1\,{\rm pc}}\right)
    {\rm rad\,m^{-2}},
\end{align}
where $n_{e,{\rm rot}}$ is the electron number density of the rotating plasma, $\langle B_\parallel\rangle$ is the effective line-of-sight magnetic field through that region, and $l$ is the path length through the jet. This expression constrains only the product $n_e\langle B_\parallel\rangle l$, so estimating $n_{e,{\rm rot}}$ requires external constraints on the field strength and path length. Since the rotating and emitting plasmas are co-located, we anchor $\langle B_\parallel\rangle$ and $l$ using analytic modelling of the synchrotron-emitting population \citep[e.g.][]{fender2019,Zdziarski2022,CF_PLACEHOLD}, together with spatially resolved VLBI observations of propagating ejecta \citep{bright2020,Chauhan2021ejecta}.

Equivalently,
\begin{align}
    n_{e,{\rm rot}} \sim 3.8\times10^3
    \left(\frac{W_{\rm rm,99}}{100\,{\rm rad\,m^{-2}}}\right)
    \left(\frac{\langle B_\parallel\rangle}{1\,{\rm mG}}\right)^{-1}
    \left(\frac{l}{10^{14}{\rm\,cm}}\right)^{-1}
    {\rm cm^{-3}}.
\end{align}
For an electron--proton plasma, this density corresponds to a characteristic Faraday-rotating mass $M_{\rm rot}\sim m_p n_e V$, where $V$ is the volume occupied by the rotating material. Taking $V\sim l^3$ gives
\begin{align}
    M_{\rm rot} \sim 6.4\times10^{21}
    \left(\frac{W_{\rm rm,99}}{100\,{\rm rad\,m^{-2}}}\right)
    \left(\frac{\langle B_\parallel\rangle}{1\,{\rm mG}}\right)^{-1}
    \left(\frac{l}{10^{14}{\rm\,cm}}\right)^2
    {\rm g}.
\end{align}

These estimates assume an idealised, approximately homogeneous Faraday-rotating plasma with a coherent characteristic field, path length, and filling factor. In reality, the ejecta are likely structured, expanding, and inhomogeneous, while our L-band data may also filter out part of any broader Faraday-depth distribution. The resulting masses should therefore be treated as exploratory order-of-magnitude estimates, and potentially lower limits, rather than precise measurements. Their purpose is to show how the observed Faraday thickness can be folded into the broader ejecta picture. More detailed modelling will require less ambiguous observing campaigns, such as simultaneous VLBI and spectropolarimetry, long-track monitoring, or broader frequency coverage.

\subsubsection{Flaring and SSA modelling}
As our primary route to estimating $\langle B_\parallel\rangle$ and $l$, we adopt techniques that model radio flares as the time-domain evolution of the optical depth within an expanding jet plasma. In this picture, the observing frequency evolves through the synchrotron self-absorption (SSA) turnover: the flare rise is driven by the increasing projected surface area of optically thick plasma, while the decay is driven by adiabatic expansion once the ejecta become optically thin \citep[see][for detailed derivations]{fender2019,CF_PLACEHOLD}. These models provide estimates of the magnetic field strength perpendicular to the line of sight, $B_\perp$, and the characteristic radius of the emitting region, $R$. We use $B_\perp$ as a characteristic field scale from which to estimate $\langle B_\parallel\rangle$, as discussed below. Assuming a quasi-spherical emitting volume, we take the Faraday path length to be $l\sim2R$.

Specifically, we use the codebase\footnote{\url{https://github.com/Fraserjcowie/SSA_estimates}} accompanying \citet{CF_PLACEHOLD}, which propagates uncertainties in the main observables, including distance, flux density, and optically thin spectral index, through a Monte Carlo framework with $10^6$ samples. We use the Stokes $I$ flux density from the 2023 October 06 flare, \relto{\sim}{200}\,mJy, since this epoch gives seed-stable fits and therefore a robust inferred $\sigma_{\phi}$. The code also samples uncertain properties of the relativistic emitting-electron energy distribution, including the power-law index $p\in[2,3]$, where $dn_{e,\rm rel}/d\gamma\propto\gamma^{-p}$ and $p=1-2\alpha_{\rm thin}$ for $S_\nu\propto\nu^{\alpha_{\rm thin}}$, together with the lower Lorentz-factor cut-off $\gamma_{\rm min}\in[3,30]$ and the less influential upper cut-off $\gamma_{\rm max}\in[10^3,10^5]$. These assumptions are then propagated into the inferred distributions of $B_\perp$ and $R$. For the distance, we adopt the same beta-distribution reconstruction of the \citet{Burridge2025distance1727} estimate as in Section~\ref{sec:disc_multi_thick}.

These estimates are usually made close to equipartition, with the internal energy density in emitting electrons, $E_e$, comparable to the magnetic energy density, $E_B$, to within factors of order unity. The codebase nevertheless allows departures from strict equipartition by drawing $E_e/E_B$ from a log-normal distribution centred on equipartition, with a central 3$\sigma$ range spanning deviations of up to a factor of 50 in either direction. This flexibility is important because, while there is some evidence for equipartition in XRB jet ejecta \citep[e.g.][]{Chauhan2021ejecta}, observational and simulation-based studies of XRB and AGN jets give mixed results \citep[e.g.][]{Readhed1994Equip,Orienti2008equipartiongood,Zdziarski2014euipartCygX1,Sironi2015Equip,2025Perezequipart}. We make the \src-tailored version of the code, including the adopted input distributions, available in the GitHub repository accompanying this study.

From the SSA analysis, we find a characteristic length scale of $l=1.3_{-0.3}^{+0.3}\times10^{14}$\cms. As a consistency check, we compare this with the expected transverse size of the compact jet at the 1.28\GHz\ photosphere. \citet{2025ZdziarskiJ1727steady} used the 8.3\GHz\ VLBA observations of \src\ from \citet{Wood2024}, which imply an 8.3\GHz\ photospheric distance of $(0.9$--$1.8)\times10^{14}$\cms\ for an assumed source distance of 4\kpc. Applying the $z\propto\nu^{-1}$ scaling expected for a conical, self-absorbed jet \citep{blandfordkonigl1979}, rescaling to our nominal distance of 5.5\kpc, and adopting a full jet opening angle $\phi_j$, the corresponding compact-jet diameter at the 1.28\GHz\ photosphere is
\begin{align}
    l_{\rm cj}
    &\sim (0.7\text{--}1.4)\times10^{14}
    \left(\frac{D}{5.5{\rm\,kpc}}\right)
    \left(\frac{\phi_j}{5{\rm\,deg}}\right)
    {\rm cm}.
\end{align}
Sampling over the source distance and adopting a uniform opening-angle prior, $\phi_j\in[1,10]\,$deg \citep{JMJ2006OpeningAngles}, broadens the allowed range to $l_{\rm cj}=1.4_{-0.8}^{+0.8}\times10^{14}$\cms. This compact-jet scale is consistent with the adopted SSA-derived length scale, and this agreement is particularly useful if the preferred geometry is a shock propagating downstream through the compact jet, rather than a detached discrete knot. As expected, these length scales are also significantly smaller than the radii inferred for resolved jet knots observed days to months after launch, which can reach $\gtrsim10^{16}$\cms\ \citep[e.g., for MAXI~J1820+070;][]{bright2020, Savard2025}.

From the SSA analysis, we infer a characteristic magnetic field strength of $B_\perp=200_{-30}^{+40}$\,mG. This is consistent with the limited set of previous magnetic-field estimates from XRB ejecta in which the low-frequency spectral break has been directly constrained, although such measurements remain rare and often carry substantial uncertainties \citep[e.g., for MAXI~J1535--571, $B_\perp\simeq100\pm80$\,mG;][]{Chauhan2021ejecta}. However, this SSA-derived field is not directly equivalent to the quantity required for the mass estimate. The synchrotron calculation is sensitive to the characteristic field strength in the radiating plasma, whereas Faraday rotation depends on the signed, density-weighted line-of-sight field averaged over the path length $l$. For an inhomogeneous or disordered magnetic field, reversals along the line of sight can therefore reduce the effective $\langle B_\parallel\rangle$ relative to the local field strength inferred from the synchrotron emission. Thus, a simple geometric projection from $B_\perp$ to $B_\parallel$ is insufficient; we must also allow for cancellation by tangled or incoherent field structure within the jet component.

To do this, we adopt a simple depolarisation model in which the magnetic field is decomposed into an ordered component, $\bar{B}$, and a turbulent component, $b$, such that
\begin{align}
    B^2 = \bar{B}^2 + \langle b^2\rangle
    \quad \longrightarrow \quad
    B_\perp^2 = \bar{B}_\perp^2 + \frac{2}{3}\langle b^2\rangle .
\end{align}
Here, the factor of two-thirds reflects the contribution of two transverse components for an isotropic turbulent field \citep[see][]{Burn1966RMSynth,Sokoloff1998,Beck2003depol}. We define the depolarisation fraction as
\begin{align}
    \mathcal{D} = \frac{p_{\rm obs}}{p_0},
\end{align}
where $p_{\rm obs}$ is the observed fractional polarisation (corrected for Faraday effects; i.e. evaluated at $\lambda^2=0$) and $p_0$ is the intrinsic fractional polarisation expected for an ordered optically thin synchrotron source. In this simple ordered-plus-turbulent model,
\begin{align}
    \mathcal{D}
    = \frac{\bar{B}_\perp^2}{B_\perp^2}
    \quad \longrightarrow \quad
    \bar{B}_\perp = \sqrt{\mathcal{D}}\,B_\perp .
\end{align}
We then approximate the characteristic ordered line-of-sight field as
\begin{align}
    \langle B_\parallel\rangle \sim \sqrt{\mathcal{D}}\,B_\perp \cot{\theta_B},
\end{align}
where $\theta_B$ is the angle between the ordered magnetic-field direction and the line of sight. This angle is not necessarily tied directly to either the jet inclination or the projected jet position angle measured by VLBI, particularly if the observed obliquity of the linear polarisation angle reflects a misalignment between the projected jet axis and the ordered magnetic-field geometry. We therefore adopt a broad prior that is flat in $\cos{\theta_B}$ over $\theta_B\in[10,80]$\,deg, excluding the limiting cases $\theta_B=0$\,deg and $\theta_B=90\,$deg: the former would suppress the ordered synchrotron polarised emission, while the latter would suppress the line-of-sight field required for Faraday rotation.

The remaining question is the appropriate range of $\mathcal{D}$. We set an upper limit of $\mathcal{D}=0.1$ based on the polarisation properties of the resolved ejecta. Although these components are isolated, and therefore less affected by blending between multiple unresolved polarised components, their observed optically thin fractional polarisations peak at only $p_{\rm obs}\sim8\%$, roughly an order of magnitude below the theoretical maximum for an ordered optically thin synchrotron source \citep[\relto{\sim}{75\%};][]{longair2011}. The lower limit is harder to define. During the flaring epochs, some fitted components have fractional polarisations of only $\sim0.2\%$; however, these values are measured relative to the total Stokes $I$ flux density, so the intrinsic fractional polarisation of the individual component could plausibly be higher by a factor of a few. We therefore adopt a deliberately broad uniform range, $\mathcal{D}\in[0.01,0.1]$, corresponding to effective depolarisation by roughly one to two orders of magnitude relative to the theoretical optically thin limit. Propagating this range together with the SSA-derived field strength and the adopted geometric prior, we estimate a characteristic line-of-sight magnetic field strength of $\langle B_\parallel\rangle\sim30_{-20}^{+40}$\,mG.

\subsubsection{Mass estimate and the mass accretion rate}
\begin{figure}
    \centering
    \includegraphics[width=1.0\linewidth]{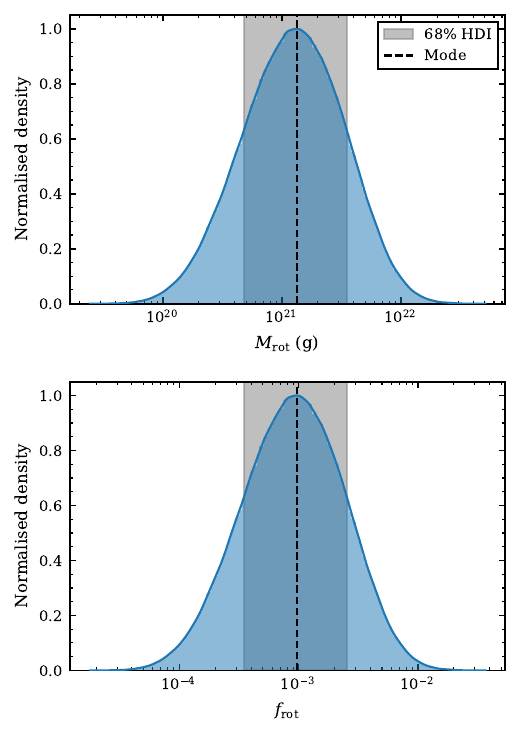}
    \caption{Posterior distributions for the inferred Faraday-rotating mass, $M_{\rm rot}$, and its fraction of the accreted mass available during the flare, $f_{\rm rot}$. The blue distributions show the normalised posterior densities, the shaded regions mark the 68 per cent highest-density intervals, and the dashed vertical lines indicate the posterior modes. The fiducial solution requires $M_{\rm rot}\sim10^{21}{\rm\,g}$, corresponding to only a small fraction, $f_{\rm rot}\sim10^{-3}$, of the available accreted mass.}    
    \label{fig:disc_mass}
\end{figure}

Equipped with these results, we can now estimate the mass contained in the Faraday-rotating plasma. The posterior distribution for this mass is shown in the top panel of Figure~\ref{fig:disc_mass}, from which we obtain a mode of $M_{\rm rot}=1.3_{-0.7}^{+2.3}\times10^{21}$\,g. In isolation, this number is difficult to interpret. However, if the jet is fed by the accretion flow, we can compare it with the mass accreted during the flaring interval, as inferred from the contemporaneous X-ray properties.

During the brightest of the radio flaring (from October 05 to 15), MAXI/GSC measured an average 2--20\,keV count rate of ${\sim}\,13\,{\rm photons\,cm^{-2}\,s^{-1}}$, varying by only a factor of ${\sim}\,1.3$ across the interval considered here. Using WebPIMMS, we convert this count rate to a 2--20\,keV X-ray flux assuming an absorbed power-law spectrum, ${\rm d}N/{\rm d}E \propto E^{-\Gamma}$, with photon index $\Gamma=2$ and neutral hydrogen column density $N_{\rm H}=2.68\times10^{21}\,{\rm cm^{-2}}$ \citep{Hughes2025_lrlx}. We then adopt a bolometric correction factor of 2 to approximately convert the in-band X-ray flux to the bolometric flux of the accretion flow. The corresponding accreted mass available over a flare duration $t_{\rm flare}$ is
\begin{align}
    M_{\rm acc}
    &= \frac{L_{\rm bol}t_{\rm flare}}{\eta c^2} \\
    &\sim 1.39\times10^{24}
    \left(\frac{D}{5.5{\rm\,kpc}}\right)^2
    \left(\frac{\eta}{0.05}\right)^{-1}
    \left(\frac{t_{\rm flare}}{12{\rm\,hr}}\right)
    {\rm g},
\end{align}
where $\eta$ is the radiative efficiency of the accretion flow. Sampling over the distance distribution and comparing the available accreted mass with the inferred Faraday-rotating mass, we obtain a modal mass fraction of
\begin{align}
    f_{\rm rot}
    \equiv \frac{M_{\rm rot}}{M_{\rm acc}}
    \sim 10^{-3}
    \left(\frac{\eta}{0.05}\right)
    \left(\frac{t_{\rm flare}}{12{\rm\,hr}}\right)^{-1}.
\end{align}

This distribution is shown in the bottom panel of Figure~\ref{fig:disc_mass} for the adopted values $\eta=0.05$ and $t_{\rm flare}=12$\,hr. We adopt a sub-10 per cent radiative efficiency because X-ray spectral modelling \citep[e.g.][]{Peng2024,Stiele2024SwJ1717,Cao2025SwiftJ1727} suggests both that the disc does not necessarily extend to the ISCO during this interval, and that a geometrically thick or coronal component may contribute significantly to the accretion flow. Both effects would reduce the radiative efficiency relative to the $\eta\sim0.1$ value commonly associated with a classic thin disc \citep{Shakura1973,Done2007}. From the radio light curve, the relevant flaring timescale is not uniquely defined, but is constrained to lie between $\sim1$ and $\sim24$\,hr \citep{HughesAKH2025a}. The fiducial choice of $t_{\rm flare}=12$\,hr therefore sits near the middle of the allowed range, and in consistent with the rapid evolution seen in VLBI (Wood et al., submitted).

For the fiducial case, the Faraday-thick components require only a small fraction of the accreted mass available during the flare, with a mode of $f_{\rm rot}\,{\sim}10^{-3}$ and a 99 per cent upper limit of $f_{\rm rot}\,{<}\,8\,{\times}\,10^{-3}$. This first provides a useful mass-budget check, since the inferred Faraday-rotating mass remains comfortably below the available reservoir. Having passed this check, it is informative to compare with other XRB ejecta mass estimates. In MAXI~J1820$+$070, modelling of the ejecta at \relto{\sim}{90}{d} post-launch requires a mass comparable to the amount accreted during the \relto{\sim}{6}{hr} flare rise associated with the launch event (i.e. $f_{\rm rot}\,{\sim}\,1$) in order to reproduce the observed deceleration trajectory \citep{bright2020,Savard2025}. Although our assumed flare duration is longer, this cannot account for the order-of-magnitude difference in inferred mass fraction, since rescaling the fiducial estimate from $t_{\rm flare}=12$\,hr to $t_{\rm flare}=6$\,hr would increase $f_{\rm rot}$ by only a factor of two.

The comparison is therefore useful because it probes two different regimes. Our analysis is sensitive to the Faraday-rotating plasma during the unresolved flaring phase, whereas the MAXI~J1820$+$070 estimate probes ejecta after they have propagated and decelerated on much larger spatial scales. The contrast may therefore carry physical information, rather than simply reflecting inconsistent mass estimates. In particular, it may point to substantial mass growth during downstream propagation, either through entrainment or interaction with the surrounding medium \citep[e.g.][]{Bowman1996Entrainment}. In this picture, the material probed here is a lighter, earlier phase of the outflow, before the ejecta acquire substantially more mass at larger radii.

That said, the inferred mass remains method-dependent. First, the SSA model treats the source as homogeneous, which is almost certainly an approximation. Most X-ray binary flares, including those from \src, do not reach the canonical optically thick spectral index of $\alpha=+2.5$ during the flare rise \citep[e.g.][]{Fender2023,HughesAKH2025a}, providing strong evidence for inhomogeneous structure \citep{Jones1977_inhomo,CF_PLACEHOLD}. The strong depolarisation points in the same direction, since it requires either substantial magnetic-field disorder, unresolved substructure, or both. Second, our mass estimate is incomplete by construction, because we calculate it only for the single broadest Faraday-thick component in the flaring epoch. Accounting for both fitted thick components, local Faraday-thin screens external to the emitting plasma, or broader Faraday-thick structure partly filtered by the L-band data could increase the total rotating mass, plausibly by a factor of a few. However, this uncertainty does not erase the scale of the comparison. Even a tenfold increase would not close the many-order-of-magnitude gap between the flaring-phase estimate and the masses inferred for large-scale decelerating ejecta. Moreover, not all systematics act in the same direction. If the SSA peak occurred before our observations, then the magnetic field during the actual launch phase may have been stronger than the value inferred at the observed epoch \citep{CF_PLACEHOLD}; a stronger field would reduce the density, and hence mass, required to produce the observed Faraday depth. A lower assumed radiative efficiency in the accretion flow would likewise increase the available accreted mass budget and reduce the inferred fractional mass. The robust conclusion is therefore not the exact value of $f_{\rm rot}$, but that the Faraday-rotating material detected during the unresolved flare can be supplied by the local accretion mass budget while remaining substantially lighter than the material inferred on much larger downstream scales.

A useful consequence is that these early-time mass constraints are naturally complementary to kinematic modelling of decelerating jets \citep[e.g.][]{Steiner2012H17distance,Carotenuto2022,Zdziarski2023J1348,Carotenuto2024,Cooper2025jets}, which is primarily sensitive to the initial kinetic energy. Because the flares occur much closer in time to the launch events of the components later seen as resolved ejecta, combining Faraday-rotation mass constraints from the unresolved flaring phase with later-time dynamical constraints could help separate the mass and velocity contributions to the initial ejecta energetics.

\subsection{Does Faraday rotation trace the emitting plasma or a cold component?}
\label{sec:disc_composition}
The SSA analysis also provides an estimate of the number density of the radiating, relativistic electron population. For the fiducial case, we find $n_{e,\rm rel}=33^{+57}_{-13}\,{\rm cm^{-3}}$, consistent with the Faraday-rotating particle density inferred above, $n_{e,\rm rot}=50^{+70}_{-30}\,{\rm cm^{-3}}$, although both distributions are wide. Taken at face value, this overlap might suggest that the synchrotron-emitting plasma alone can produce the observed internal Faraday rotation.

However, the comparison is not direct, because relativistic electrons are much less efficient Faraday rotators than non-relativistic electrons. We therefore need to convert the relativistic electron population inferred from the synchrotron emission into the equivalent density of cold electrons that would produce the same Faraday rotation. In the polarised radiative-transfer formalism of \citet{Jones1977}, the relevant transfer coefficient is the \textit{rotativity}, $\xi_V$. For a power-law distribution of relativistic electrons, \citet{Jones1977} give the rotativity relative to that of a non-relativistic, or `cold', electron population \citep[Eq.~10 of][]{Jones1977}; see also the application to AGN polarimetry by \citet{Wardle1977}. Enforcing equal rotativities then gives the approximate equivalence
\begin{align}
    \frac{n_{e,\rm rel}}{n_{e,\rm cold}}
    =
    \frac{1-\alpha}{-2\alpha(1.5-\alpha)}
    \frac{\gamma_{\rm min}^2}{\ln{\gamma_{\rm min}}},
\end{align}
where we have re-written the expression using the convention $S_\nu\propto\nu^\alpha$, and denote the low-energy cut-off of the non-thermal electron distribution by $\gamma_{\rm min}$. For the 2023 October 06 epoch, adopting the observed $\alpha\simeq -0.35$ and extending to a canonical optically thin value of $\alpha\simeq-0.6$, a deliberately conservative $\gamma_{\rm min}=3$ gives
\begin{align}
    \frac{n_{e,\rm rel}}{n_{e,\rm cold}} \simeq 5{-}10.
\end{align}
Thus, even under assumptions favourable to the relativistic population, the synchrotron-emitting electrons alone would need $n_{e,\rm rel}\simeq(5{-}10)n_{e,\rm rot}\,{\simeq}\,250{-}500\,{\rm cm^{-3}}$ to reproduce the observed Faraday rotation. The rotation therefore already requires an additional colder, weakly emitting electron population. This requirement becomes rapidly more prohibitive for larger low-energy cut-offs; for example, $\gamma_{\rm min}=10$ gives $n_{e,\rm rel}/n_{e,\rm cold}\simeq30{-}45$, requiring $n_{e,\rm rel}\,{\gtrsim}\,(1.5{-}5)\times10^{3}{\rm\,cm^{-3}}$.

The absence of detectable circular polarisation provides additional, although tentative, evidence against a large mildly relativistic electron population. During the brightest epochs, we find $|V/I|<0.08\%$ at $3\sigma$, comparable to our estimated systematic floor and 15--20 times fainter than the peak linear polarisation fraction across the band on 2023 October 06. This limit cannot be assigned cleanly to any single unresolved component, nor does it provide a direct lower limit on $\gamma_{\rm min}$, because the expected Stokes $V$ signal depends on field geometry, optical depth, inhomogeneity, and cancellation between unresolved components. Nevertheless, the non-detection is relevant because low-energy relativistic electrons can increase both intrinsic synchrotron circular polarisation, for which the circular-to-linear ratio scales approximately as $\gamma^{-1}$ in the ideal uniform-field limit \citep{Legg1968,longair2011}, and Faraday conversion between Stokes $U$ and $V$ \citep{Jones1977,Jones1977_inhomo,kennett1998}. The latter mechanism has been invoked to explain circular polarisation in both AGN and XRB jets \citep[e.g.][]{Wardle1998,fender2000,fender2003}. The absence of detected circular polarisation is therefore more consistent with $\gamma_{\rm min}>3$ than with the large mildly relativistic population that would be required for the synchrotron-emitting plasma itself to act as an efficient Faraday rotator.

Increasing $\gamma_{\rm min}$ also makes the SSA-inferred relativistic electron population less numerous, because it removes the lowest-energy electrons from the distribution. For a power-law electron population, written as $dn_{e,\rm rel}/dE=N_0E^{-p}$ with $E=m_ec^2\gamma$, the integrated relativistic electron density is
\begin{align}
    n_{e,\rm rel}
    =
    \frac{N_0(m_ec^2)^{1-p}}{p-1}
    \left(
    \gamma_{\rm min}^{1-p}
    -
    \gamma_{\rm max}^{1-p}
    \right),
\end{align}
for $p\neq1$. In the canonical synchrotron regime, $p\in[2,3]$ and $\gamma_{\rm min}\ll\gamma_{\rm max}$, this reduces to $n_{e,\rm rel}\propto N_0\gamma_{\rm min}^{1-p}$, so increasing $\gamma_{\rm min}$ directly reduces the available number of relativistic electrons. The same intuition holds for any $p>1$ in the same limit.

In the SSA calculation, $N_0$ is not held fixed. The inferred normalisation can increase mildly as the low-energy cutoff is raised, because the synchrotron observables must still be reproduced. However, over the sampled range of $p$, this increase is not sufficient to compensate for the loss of the low-energy part of the density integral \citep{CF_PLACEHOLD}. In the full Monte Carlo calculation, where the distance, flux density, spectral index, magnetic field, and size scale are sampled jointly, the practical effect is to reduce the upper envelope of allowed $n_{e,\rm rel}$ values, rather than to shift the modal density by orders of magnitude. Larger $\gamma_{\rm min}$ therefore strengthens the Faraday-rotation discrepancy, because the relativistic electron population becomes both less numerous and less efficient as a rotator.

The dominant uncertainty in the modal densities remains the broad distance range. This could be reduced in future XRB outbursts with parallax distances \citep{v404cyg2009dist,2020atri2020maxij820dist}, with constraining H{\sc i} distances \citep{Chauhan2019}, or, for \src\ itself, if a better distance estimate becomes available. However, the corresponding changes in the SSA-derived magnetic field and size scale are modest, so the Faraday-based mass estimates above are not substantially affected. Increasing $\gamma_{\rm min}$ therefore mainly worsens the relativistic-electron rotativity problem, rather than changing the inferred mass. We therefore interpret the behaviour above as evidence that a cold electron component is required to explain the observed Faraday rotation.

\subsubsection{Constraints on acceleration efficiency}

Given that the relativistic emitting and Faraday-rotating particle populations are co-spatial, their relative number densities provide an empirical constraint on the fraction of electrons accelerated into the synchrotron-emitting population. We therefore define an effective electron acceleration fraction, by number, as
\begin{align}
\epsilon_{\rm acc}
=
\frac{n_{e,\rm rel}}
{n_{e,\rm rel}+n_{e,\rm rot}/\kappa_M},
\end{align}
where $n_{e,\rm rel}$ is the number density of synchrotron-emitting relativistic electrons and $n_{e,\rm rot}$ is the fiducial density of cold or weakly emitting electrons responsible for the internal Faraday rotation. The fiducial case has $\kappa_M=1$, while values $\kappa_M<1$ allow for additional Faraday-rotating material missed by the fiducial estimate. For example, $\kappa_M=0.1$ corresponds to ten times more rotating mass than in the fiducial case. If the relativistic population is injected at launch and subsequently cools, then the flare-time value should be interpreted as a lower limit on the initial accelerated fraction during ejection.

\begin{figure}
    \centering
    \includegraphics[width=1.0\linewidth]{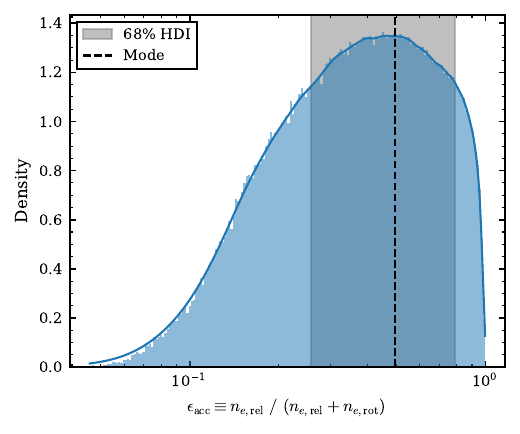}
    \caption{
    Posterior distribution of the fiducial density ratio
    $n_{e,\rm rel}/(n_{e,\rm rel}+n_{e,\rm rot})$, used as an empirical proxy for the electron acceleration efficiency by number, $\epsilon_{\rm acc}$, for $\kappa_M=1$. The grey shaded region marks the 68 per cent highest-density interval, and the dashed vertical line marks the posterior mode.
    }
    \label{fig:ne_ratio_posterior}
\end{figure}

The resulting posterior is shown in Figure~\ref{fig:ne_ratio_posterior}. The upper tail formally reaches $\epsilon_{\rm acc}\,{\sim}\,1$, but this region is dominated by samples with the smallest allowed low-energy cut-offs, $\gamma_{\rm min}\,{\sim}\,3$. Such values are already disfavoured by the rotativity argument above, and would require nearly all available electrons to enter the synchrotron-emitting population, in tension with dedicated particle-in-cell (PIC) simulations of collisionless shocks \citep[typically 0.5--2\% by number; e.g.][]{Sironi2011PIC,Gupta024PIC}. This again points to $\gamma_{\rm min}$ values above the conservative lower limit, plausibly of order tens.

Even the lower bound on the electron acceleration efficiency remains high. At the 99 per cent level, the fiducial calculation gives $\epsilon_{\rm acc}\,{\gtrsim}\,10\%$. This is above the upper end of the PIC range, but the value depends on how much Faraday-rotating material is missed by the fiducial estimate. Because this estimate includes only the single broadest Faraday-thick component, and because the L-band data may partly filter still broader Faraday-depth structure, the rotating mass could plausibly increase by a factor of a few, and perhaps by up to an order of magnitude in an extreme case. This would bring the inferred electron acceleration efficiencies closer to those expected for efficient shock acceleration in PIC simulations. By contrast, reaching the much lower electron-specific efficiencies inferred for certain large-scale X-ray binary jet interactions and some supernova-remnant shocks, $\epsilon_{\rm acc}\,{<}\,0.1\%$ \citep[e.g.][]{s26AccelEffSora2010,SNRaccelEfficiency2012,Cowie2026CirX1shocks}, would require the Faraday-rotating mass to be larger by roughly two to three orders of magnitude, corresponding to $\kappa_M\sim10^{-2}$--$10^{-3}$. Such a correction is well beyond the likely systematic uncertainty from missing Faraday-thick structure. For the more plausible range $\kappa_M\,{\gtrsim}\,0.1$, the inferred efficiencies therefore remain closer to those expected for efficient shock acceleration than to those inferred for late-time, resolved shock structures. This suggests that electron acceleration in X-ray binary jets may be especially efficient during the flaring/ejection phase, and shows that spectropolarimetry has the potential to provide a new observational probe of non-thermal particle acceleration in unresolved relativistic outflows.

\subsection{Intrinsic polarisation-angle evolution}
\label{sec:discussion_angle}

\begin{figure}
    \centering
    \includegraphics[width=1.0\linewidth]{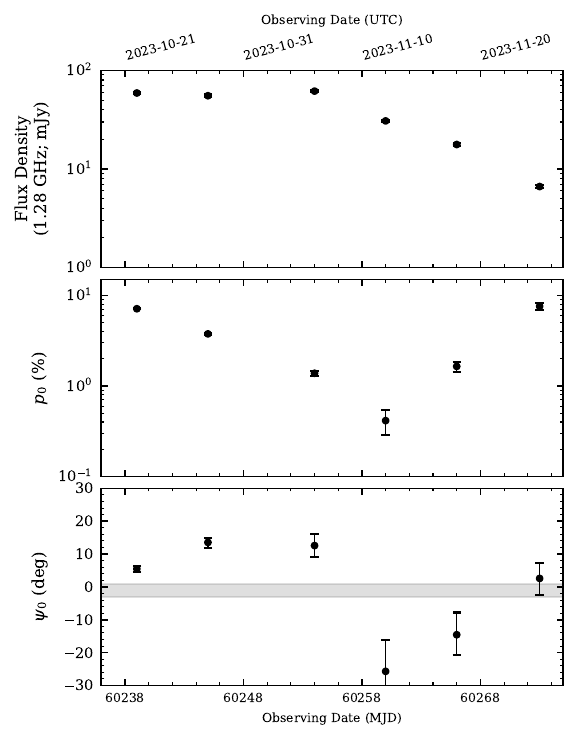}
    \caption{Post-flare polarisation evolution of the preferred Faraday-thin component. We show the Stokes $I$ flux density, intrinsic fractional polarisation, $p_0$, and intrinsic polarisation angle, $\psi_0$, after the brightest flaring activity. Although the source is simple in Faraday space during this period, both $p_0$ and $\psi_0$ evolve smoothly, indicating continued intrinsic evolution of the emitting region. The grey band marks the reference range around $\psi_0=0$, corresponding to a polarisation angle aligned with the VLBI jet position angle \citep[e.g.][]{Wood2024}.}
    \label{fig:disc_pol_zoom}
\end{figure}

While the primary focus of this work is the identification of internal Faraday effects, the intrinsic polarisation angle also constrains the jet evolution. In Section~\ref{sec:results_angle}, we showed that the largest departures from the canonical orientations occur after the brightest flaring activity. Between 2023 October 21 and 2023 November 25, the source is Faraday-simple and detected at high signal-to-noise, yet shows the most statistically significant obliquity of the outburst (Figure~\ref{fig:disc_pol_zoom}). Over the same interval, the intrinsic fractional polarisation evolves smoothly, rather than showing erratic scatter or directly tracking the Stokes $I$ light curve. The emitting region therefore appears to continue evolving after the main flaring phase, even once the Faraday structure has simplified.

One possible interpretation is that the post-flare emission traces longer-term evolution in shocks, compression regions, or ordered magnetic-field structure within the jet. Oblique polarisation angles are commonly discussed in the context of complex shocks and magnetic-field geometries in AGN jets \citep[e.g.][]{Cawthorne1966Obliqueshocks,Jorstad2007_agnpoll}, while slow time-dependent rotations of the linear polarisation angle without a corresponding jet reorientation have been described as `rotator events' \citep[particularly during gamma-ray activity in blazars;][]{Aller1981rotatorevents,Saikia1988,Kiehlmann2016rotator}. Related behaviour has also been reported in X-ray binary outbursts, particularly where changes in polarisation are not mirrored by equally clear changes in Stokes $I$ \citep{fender2003,brocksopp2007,brocksopp2013,Hannikainen2000,hughes2023}. The precise origin of such events remains uncertain, but they show that smooth angle evolution need not require a change in the large-scale jet axis.

There are, however, behaviours that are harder to explain with a single shock-interaction picture. During the flaring interval, the ST models often imply approximate orthogonality between the Faraday-thin, ISM-like component and the thick component. In the most complex epochs, the thick components themselves can also appear approximately orthogonal, and on 2023 October 06 both are oblique with respect to the jet axis, albeit at moderate to low statistical significance. At late times, once the unresolved core and downstream ejecta are spatially separated, their inferred intrinsic angles are again close to orthogonal. If these components trace ejecta launched by the same system, this could indicate different field-ordering mechanisms in different regions, such as compression versus shear \citep{Laing1980,hughes1985jetpolshocks}, or viewing-geometry effects that alter the observed angle without requiring fundamentally different intrinsic field structures.

Relativistic aberration and differential boosting provide one possible route to the latter behaviour. In a helical-field geometry, a change from relativistic to sub-relativistic bulk motion can invert the apparent polarisation-angle orientation even if the underlying magnetic-field structure remains similar \citep{Lyutikov2005_jetpol}. This offers a possible explanation for the late-time core/ejecta orthogonality. The unresolved core emission could be dominated by slower or stalled ejecta, consistent with its lack of clear angular separation from the core months after the transition to the soft state and the cessation of compact-jet activity. By contrast, the downstream resolved ejecta may still be moving fast enough for relativistic aberration to remain important, allowing a similar underlying field configuration to produce an apparently orthogonal angle. In the framework of \citet{Lyutikov2005_jetpol}, this would favour an intrinsically poloidal-dominated field geometry, with the magnetic field preferentially aligned along the jet axis.

This picture is harder to extend to the apparent orthogonality between simultaneous thick components during the flaring interval, unless those components differ substantially in bulk motion, geometry, or magnetic-field structure. Such differences are most plausible during the most erratic phase of the outburst, but these epochs are also most affected by the L-band spectropolarimetric limitations discussed above. Resolving this ambiguity will require broader-band spectropolarimetry to reduce Faraday-space filtering, together with polarimetric VLBI to test whether the components inferred from unresolved data correspond to spatially resolved ejecta, and to localise any oblique polarisation angles to specific components. For large-scale ejecta, a direct test would be to follow individual components from the early semi-ballistic phase into the later deceleration or stalling phase \citep{Carotenuto2022,Carotenuto2024}, and search for corresponding changes in intrinsic polarisation angle. Although the ejecta will be substantially fainter at these later stages, sparse but deep polarimetric observations of resolved components could distinguish bulk-relativistic viewing effects from genuinely different magnetic-field geometries. If bulk-relativistic viewing effects dominate, polarisation-angle evolution could provide a new diagnostic of jet speeds in unresolved or otherwise ambiguous sources, complementary to the use of Faraday complexity as a probe of jet plasma content.

\section{Conclusions and future directions}
\label{sec:conclusion}
We have presented time-domain MeerKAT L-band spectropolarimetry of \src\ during its 2023 outburst, focusing on the transient emergence of complex polarised structure during the brightest flaring interval. Using RM synthesis and parametric $QU$-fitting, we find that the source evolves from predominantly Faraday-simple behaviour before the flare, to requiring multiple Faraday-thick components during the flaring interval, and then back to a simple state within approximately a week of the brightest activity. The close temporal association with the radio flares, together with the stability of the foreground Faraday depth, strongly favours an origin internal to the jet ejecta rather than in the ISM or a transient external screen. Since internal rotation cancels in a pure neutral pair plasma, the data favour ejecta whose Faraday-rotating material is dominated by an electron--proton component rather than a purely symmetric electron--positron plasma. Interpreting the characteristic thickness as internal rotation through this material, and anchoring the magnetic-field and size scales with synchrotron self-absorption arguments, we infer a characteristic rotating mass of order $M_{\rm rot}\sim10^{21}{\rm\,g}$. For an assumed 12\,hr flaring interval, this corresponds to only a small fraction, $f_{\rm rot}\sim10^{-3}$, of the mass available from the accretion flow. This is much lower than the fractional masses inferred from late-time decelerating ejecta, with the comparison remaining relatively insensitive to the adopted flare duration because both estimates scale similarly with this timescale. The discrepancy therefore points towards substantial propagation-induced mass loading between the unresolved flaring phase and the later resolved ejecta.

The rotating particles are also unlikely to be dominated by the synchrotron-emitting relativistic electrons themselves. Even for a deliberately conservative low-energy cut-off of $\gamma_{\rm min}=3$, which maximises the rotativity of the non-thermal population, the synchrotron self-absorption analysis yields too few radiating electrons to reproduce the observed internal rotation. The absence of detectable circular polarisation, with $|V/I|<0.08\%$ at $3\sigma$ during the brightest epochs, also supports $\gamma_{\rm min}\,{>}\,3$, since a large mildly relativistic population should increase both intrinsic synchrotron circular polarisation and Faraday conversion. The data therefore favour a picture in which only a subset of the electrons are accelerated into the synchrotron-emitting distribution, while a colder, weakly emitting electron--proton reservoir dominates both the internal rotation and the total plasma content of the flaring ejecta.

The intrinsic polarisation-angle evolution provides a complementary view of the same jet evolution. After the brightest flaring activity, the source becomes simple again in Faraday space, but its intrinsic polarisation angle remains significantly oblique to the canonical jet-aligned orientations and evolves smoothly together with the intrinsic fractional polarisation. During the flaring interval, the fitted thick components can also appear approximately orthogonal to one another, while at late times the unresolved core and downstream resolved ejecta again show nearly orthogonal intrinsic angles. These behaviours may indicate different magnetic-field ordering mechanisms in different regions of the ejecta, for example compression versus shear, or may reflect bulk-relativistic viewing effects in helical or non-axisymmetric jet geometries \citep{Lyutikov2005_jetpol}. Following individual resolved ejecta from the early semi-ballistic phase into later deceleration or stalling would therefore test whether polarisation-angle evolution can be used as a diagnostic of bulk jet dynamics as well as magnetic-field geometry.

Beyond serving as a proof of concept for time-domain spectropolarimetry of X-ray binary jets, and potentially other bright transient radio sources, this analysis shows where the next observational gains are needed. MeerKAT L-band alone sits close to the broad-structure sensitivity limit in Faraday space, so broader-band spectropolarimetry will be needed to reduce these systematics and track the evolution of individual Faraday-thick components. Simultaneous polarimetric VLBI will also be needed to associate the unresolved Faraday components with spatially resolved ejecta. As an immediate test, we have multi-hour simultaneous MeerKAT L+S-band observations of SS~433 in hand. Although atypical, SS~433 repeatedly flares and launches discrete jet components \citep{Trushkin2003SS433radio,Jeffrey2016SS433vlbi}, making it a useful laboratory for testing the proposed time evolution of Faraday-thick structure. A systematic search through the near-decadal ThunderKAT and X-KAT archives will further test whether comparable complex episodes are present in past outbursts.

Looking ahead, next-generation radio facilities such as the DSA and the ngVLA, together with MeerKAT's transition into the SKA era, will provide the sensitivity and bandwidth needed to turn transient spectropolarimetric complexity into a routine probe of jet composition and mass loading.

\section*{Acknowledgements}
AKH thanks UKRI for support. RF thanks UKRI, the ERC and The Hintze Family Charitable Foundation for support. GRS is supported by NSERC Discovery Grant RGPIN-2021-0400. FC acknowledges support from the Royal Society through the Newton International Fellowship programme (NIF/R1/211296). TDR is an INAF IAF fellow. CMW acknowledges financial support from the Forrest Research Foundation Scholarship, the Jean-Pierre Macquart Scholarship, and the Australian Government Research Training Program Scholarship. We thank the anonymous reviewer for their constructive comments, which improved the clarity of this manuscript.

The MeerKAT telescope is operated by the South African Radio Astronomy Observatory, which is a facility of the National Research Foundation, an agency of the Department of Science and Innovation. We acknowledge the use of the Inter-University Institute for Data Intensive Astronomy (IDIA) data intensive research cloud for data processing. IDIA is a South African university partnership involving the University of Cape Town, the University of Pretoria and the University of the Western Cape.

\section*{Data Availability}
Data from MeerKAT are available through the SARAO data archive (Proposal IDs: 
SCI-20180516-PW-01 and SCI-20230907-RF-01): \url{https://archive.sarao.ac.za/}. We host machine-readable data files and analysis scripts at GitHub:~\url{https://github.com/AKHughes1994/SwJ1727_2023_Outburst}.



\bibliographystyle{mnras}
\bibliography{bibly} 

\ \par
{\itshape
\noindent\EndAffil}




\appendix

\section{Reference Calibration (1GC): 2023 October 16}
\label{sec:appendix_1GC_Oct16}
As noted in Section~\ref{sec:obs}, the 2023 October 16 epoch did not include a dedicated scan of \polcal. Although this would normally prevent a standard full-Stokes calibration, the relevant \texttt{polcal} cross-hand phase solution is effectively model-independent, provided the source is bright, linearly polarised, and circularly unpolarised. We therefore used \secondary\ to solve for the cross-hand phase, $\rho$, for this epoch. The remaining $\pm\pi$ degeneracy was resolved by exploiting the stability of $\rho$ over the surrounding epochs, selecting the solution closest to the neighbouring 3C286-derived values. After applying this correction, we corrected for parallactic-angle rotation in the usual way.

\begin{figure}
\centering
\includegraphics[width=1.0\linewidth]{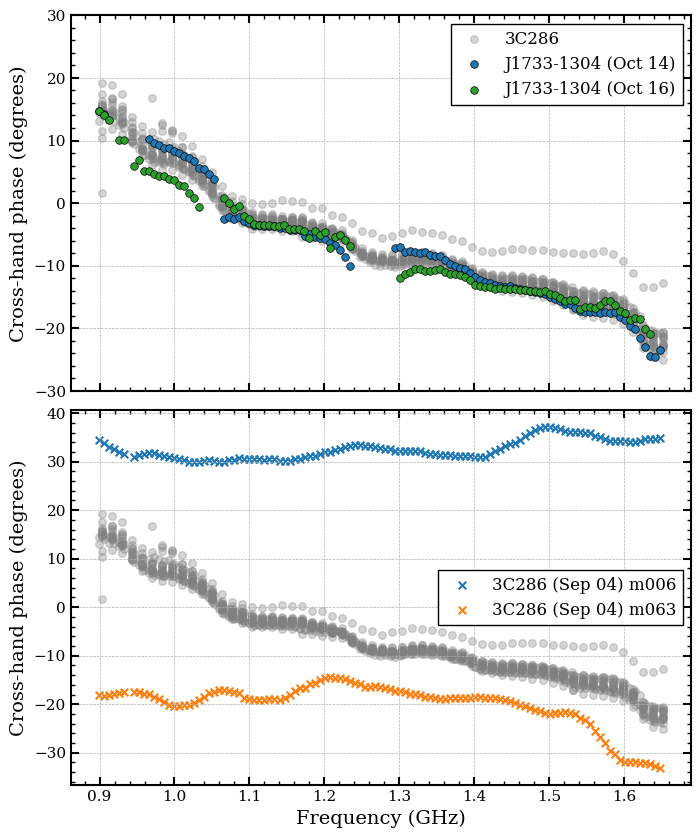}
\caption{Cross-hand phase consistency checks. Top panel shows the cross-hand phase inferred from \polcal\ and \secondary, demonstrating close agreement between the dedicated polarisation calibrator and a bright linearly polarised secondary calibrator. The 2023 October 16 solution was derived from \secondary\ because no dedicated \polcal\ scan was available. Bottom panel shows the same 2023 September 04 calibration chain solved using different reference antennas, illustrating that absolute cross-hand phase comparisons require consistent reference-antenna choices.}
\label{fig:method_crosshand_phase_poc}
\end{figure}

Figure~\ref{fig:method_crosshand_phase_poc} demonstrates that the 2023 October 16 calibration is consistent with the standard strategy. The top panel shows that the \secondary-derived solutions closely track the \polcal-derived cross-hand phases, indicating that a bright, linearly polarised secondary calibrator can provide a high-precision alternative when a dedicated \polcal\ scan is unavailable. The same comparison shows that interpolation between neighbouring \polcal\ epochs would also have been viable, since the cross-hand phase remains stable at the \relto{<}{10}\,deg level over nearly six months. The gaps in frequency occur where the feed-frame cross-hand signal becomes too low signal-to-noise to solve reliably, typically because the source rotation measure drives $U^{\prime}$ and $V^{\prime}$ close to zero. However, absolute comparisons of $\rho$ require a consistent reference antenna: as shown in the bottom panel, changing the reference antenna changes the absolute cross-hand phase because the solution inherits offsets from the preceding parallel-hand calibration. This does not imply a failed calibration, but it does mean that cross-hand phase values should only be compared, and thus interpolated across epochs, when the same reference antenna has been used. Although this is not the focus of the present work, we plan to apply these methods in a bulk spectropolarimetric reprocessing of the ThunderKAT and X-KAT data, including epochs without dedicated cross-hand phase calibrators. A more detailed assessment, including implementation directly in the visibility domain, will be presented separately.

\section{Nested sampling approach}
\label{sec:appendix_nested_sampling}
We infer the parameters of each spectropolarimetric model using nested sampling \citep{2004_skilling_Nested,2006_skilling_nested}, as implemented in \textsc{dynesty} \citep{speagle_dynesty_code,koposov_dynesty_code}. This provides posterior samples for each model while also estimating the Bayesian evidence required for model comparison. For a model $\mathcal{M}$ with parameters $\boldsymbol{\theta}$, the posterior distribution is given by Bayes' theorem,
\begin{align}
p(\boldsymbol{\theta} \mid \boldsymbol{d}, \mathcal{M})
=
\frac{\mathcal{L}(\boldsymbol{d} \mid \boldsymbol{\theta}, \mathcal{M})\pi(\boldsymbol{\theta} \mid \mathcal{M})}{\mathcal{Z}},
\end{align}
where $\boldsymbol{d}$ denotes the fractional Stokes spectra, $\mathcal{L}$ is the likelihood, $\pi$ is the prior, and $\mathcal{Z}$ is the Bayesian evidence. For each model, the predicted complex fractional polarisation is evaluated as the coherent sum of all components,
\begin{align}
p_{\rm mod}(\lambda_j^2;\boldsymbol{\theta})
=
\sum_k p_k(\lambda_j^2;\boldsymbol{\theta})
=
q_{{\rm mod},j}
+
i u_{{\rm mod},j},
\end{align}
where the sum runs over the S, T, or P components included in the model. Assuming independent Gaussian uncertainties in $q$ and $u$ in each wavelength-squared channel, the log-likelihood is
\begin{align}
\ln \mathcal{L}
=
-\frac{1}{2}
\sum_{j=1}^{N_{\rm chan}}
\left[
\frac{\left(q_j-q_{{\rm mod},j}\right)^2}{\sigma_{q,j}^2}
+
\frac{\left(u_j-u_{{\rm mod},j}\right)^2}{\sigma_{u,j}^2}
\right.
\nonumber\\
\left.
+
\ln\left(2\pi\sigma_{q,j}^2\right)
+
\ln\left(2\pi\sigma_{u,j}^2\right)
\right].
\end{align}
The evidence $\mathcal{Z}$ is then obtained by integrating the likelihood over the prior volume for the model.

As our primary goal is to identify dominant posterior modes while obtaining moderately precise evidence estimates, we employ the \textsc{DynamicNestedSampler} \citep{2019_higson_Dynamic} with the default 80/20 tuning between posterior and evidence estimation accuracy. We adopt a stopping criterion of 10\,000 effective posterior samples, where the effective sample size corresponds to the number of independent samples implied by the importance weights of the nested sampling output. For the dynamic allocation we initialise the run with 50 live points per dimension and subsequently add batches of $20 \times$ the dimensionality until convergence, terminating the initial sampling phase when the evidence uncertainty reaches $\Delta \ln \mathcal{Z} = 0.01$. For the proposal mechanism we adopt random-walk sampling \citep[\texttt{rwalk};][]{2006_skilling_nested}, as we found that slice sampling struggled near the circular prior boundaries of $\psi_0$, and for prior bounding we use the multi-ellipsoidal routine \citep[\texttt{multi};][]{multi_ellipsoid}.

For each parameter, we adopt uninformative or weakly informative priors; Table~\ref{tab:priors_summary} summarises the distributions used. We place a uniform prior on the Faraday depth $\phi_{\rm rm}$ for S- and P-components, and on the peak Faraday depth of T-components. We set the allowed range in Faraday depth using the distribution of CLEAN components from the RM-CLEAN run for each epoch. In practice, these bounds are less than \relto{\pm}{150}\radPerSqm for all epochs except 2023 October 14, which corresponds to both the brightest flare of the outburst and the broadest distribution of CLEAN components. For that epoch, we expand the prior range to \relto{\pm}{250}\radPerSqm.

\def\arraystretch{1.25}
\begin{table}
    \centering
    \begin{tabular}{cllc}
        \Xhline{3\arrayrulewidth}
        \multicolumn{1}{c}{Parameter} &
        \multicolumn{1}{c}{Prior} &
        \multicolumn{1}{c}{Range} &
        \multicolumn{1}{c}{Circular} \\
        \Xhline{3\arrayrulewidth}
        $p_0$ & Uniform (radial) & $[0,\,0.5]$ & $\times$ \\
        $\psi_0$ & $\propto \cos(4\psi_0)$ & $[-90,\,90]$\,deg & \checkmark \\
        $\phi_{\rm rm}$ & Uniform/Gaussian & epoch-dependant & $\times$ \\
        \hline
        $\sigma_\phi$ & Log-uniform & $[0.1,\,50]\,\rad\,\mathrm{m}^{-2}$ & $\times$ \\
        $N$ & Log-uniform & $[2,\,50]$ & $\times$ \\
        \hline
        $\beta$ & Uniform & $[-5,\,5]$ & $\times$ \\
        \Xhline{3\arrayrulewidth}
    \end{tabular}
    \caption{Summary of prior distributions adopted for the $QU$-fitting. The final column indicates whether the parameter is defined with a circular boundary condition.}
    \label{tab:priors_summary}
\end{table}

The Faraday-thick (T) components require two additional parameters. We therefore adopt log-uniform priors on both $\sigma_\phi$ and the shape parameter $N$. We allow $\sigma_\phi$ to vary up to $50\,\radPerSqm$. We choose a log-uniform prior to downweight larger values of $\sigma_\phi$, since this upper bound extends to roughly four times the Faraday-space scale at which MeerKAT L-band already shows substantial loss of sensitivity due to its $\lambda^2$ coverage \citep{rundick2023RMSynth}, directly analogous to the loss of sensitivity to large spatial scales set by the shortest baseline in interferometric imaging. For the shape parameter, a fixed interval $\Delta N$ produces much larger phenomenological changes in the super-Gaussian profile at small $N$ than at large $N$, so we again adopt a log-uniform prior between $N=2$ (corresponding to external Faraday dispersion) and $N=50$ (approaching the Burn slab; see Fig.~\ref{fig:method_supergaussian}). We exclude $N<2$, for which the profile approaches a Lorentzian-like form. Such profiles are not predicted by the canonical analytic depolarisation models \citep{Burn1966RMSynth,Sokoloff1998}. More complex cases, including asymmetric profiles \citep[e.g.][]{Bell2011_rmcaus,Schnitzeler2015_rmsigs}, have been explored, but we neglect them here for tractability, while intending to extend the model in future work.

\begin{figure}
    \centering
    \includegraphics[width=1.0\linewidth]{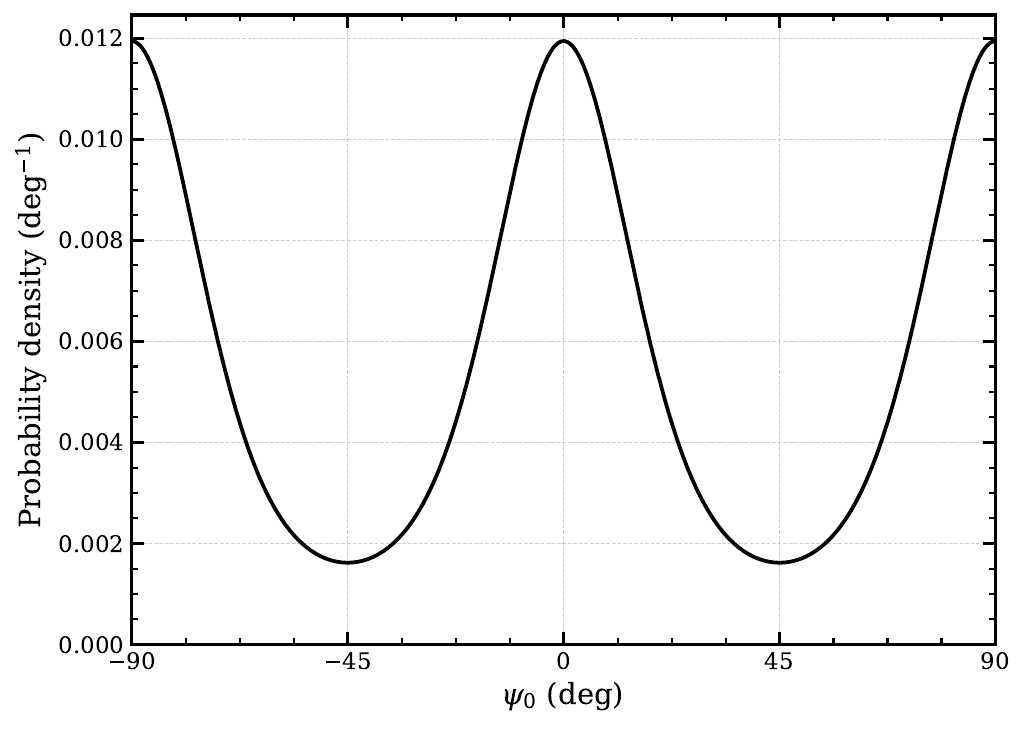}
    \caption{Weakly informative prior adopted for the intrinsic EVPA, $\psi_0$, defined as $\pi(\psi_0) \propto \cos(4\psi_0)$. The prior places greater probability density at $0$\,deg, $90\,$deg, and $180\,$deg, reflecting the observational tendency for jet polarisation vectors to align either parallel or perpendicular to the jet axis, while still allowing oblique orientations.}    
    \label{fig:appendix_angle_prior}
\end{figure}

For all components, we adopt a prior on the polarisation fraction that is uniform in radius, such that each value of $p_0$ is equally probable. In practice, this corresponds to a uniform radial prior in the $(q_0,u_0)$ plane and avoids the implicit preference for larger $p_0$ values that would arise from adopting a uniform prior directly in $(q_0,u_0)$, owing to the increasing $p_0$-area available at larger radii. For the intrinsic polarisation angle, $\psi_0$, we adopt a weakly informative prior with a circular boundary condition, periodic at $\pm90$\,deg. This choice is motivated by the known jet position angle of \src \citep[i.e. the direction of jet propagation;][]{Wood2024,Wood2025SWJ1717Ejecta,Hughes2025_lrlx}. Observational studies of both X-ray binary and AGN jets show that $\psi_0$ tends to align either parallel or perpendicular to the jet axis \citep[e.g.][]{gallo2004,stirling2004,2005Lister_mojav_pol,Lyutikov2005_jetpol,rushton2017,Hodge2018,Pushkarev_2023_mojave_pol,Kravtsov2025_cygx1_pol}, although oblique configurations are possible \citep{Cawthorne1966Obliqueshocks,Lyutikov2005_jetpol} and have been observed in a small number of sources \citep[e.g.][]{Dulwhich2009_oblique,fender2003,hughes2023}. We therefore adopt the prior $\pi(\psi_0) \propto \cos(4\psi_0)$, which places approximately six times more prior mass at $0$\,deg and $\pm90\,$deg than at $\pm45\,$deg, so that oblique configurations must be supported by the data (Fig.~\ref{fig:appendix_angle_prior}). However, this prior has negligible impact on the inferred posterior distributions.

Finally, for the spectral index $\beta$ of the P-components, we adopt a uniform prior over the range $-5 \le \beta \le 5$. This deliberately broad prior is a conservative choice, allowing the data to test whether spectral curvature alone can reproduce the observed behaviour in $q(\lambda^2)$ and $u(\lambda^2)$ before invoking additional Faraday complexity.

\section{Systematics in the spectropolarimetry of MeerKAT}
\label{sec:appendix_systematics}
In this section, we aim to show that the observed spectropolarimetric behaviour is unique to \src, both in its frequency-dependent complexity and, more importantly, in its transient emergence around the flaring epochs. We do this by comparing the behaviour of \src\ with that of the polarised calibrators included in our observations, \polcal\ and \secondary.

\begin{figure}
    \centering
    \includegraphics[width=1.0\linewidth]{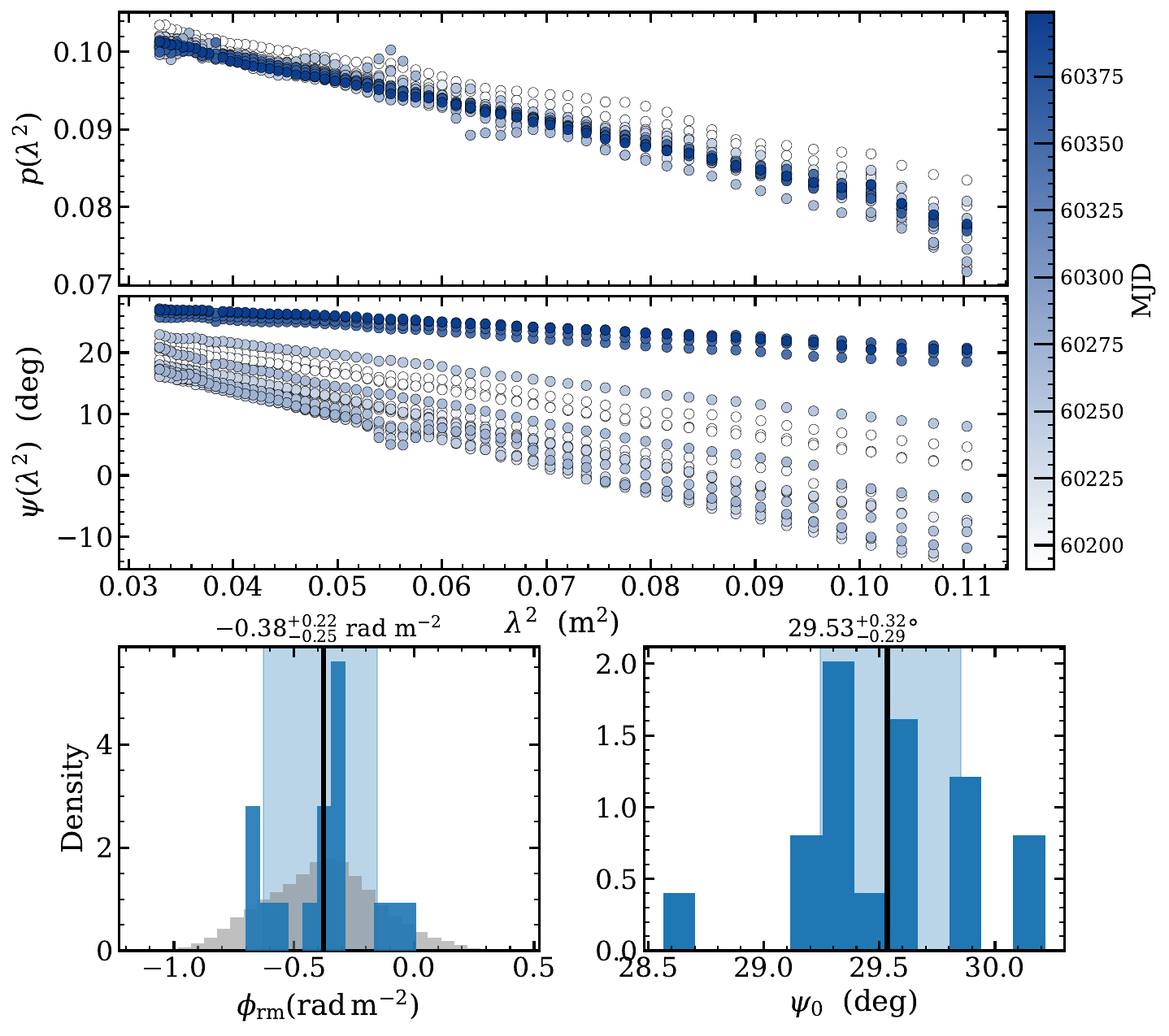}
    \caption{Summary of the spectropolarimetric properties of \polcal. (\textit{top}) Linear polarisation fraction. (\textit{middle}) Linear polarisation angle. (\textit{bottom left}) Ionospheric-corrected peak Faraday depth from our observations (blue) and from the re-sampled distributions (grey). (\textit{bottom right}) Intrinsic polarisation angle. Note the stability of these properties compared with \src.}
    \label{fig:3c286_systematics}
\end{figure}

We begin with \polcal. In Figure~\ref{fig:3c286_systematics}, we show its spectropolarimetric behaviour, including the $\lambda^2$ spectrum of the linear polarisation fraction ($p(\lambda^2)$; top panel) and the observed polarisation angle ($\psi(\lambda^2)$; middle panel), where the marker colours indicate observing time. The $p(\lambda^2)$ spectra are qualitatively similar in every epoch, and any subtle variations, for example due to secular evolution of the calibrator, are far less extreme than those observed in \src. In particular, we do not see the strong in-band depolarisation and repolarisation inversions present in the target. Similarly, while the slope and intercept of $\psi(\lambda^2)$ vary with time, this is naturally explained by changing ionospheric conditions. Crucially, unlike in \src, we do not observe angular turnovers.

In the bottom panels of Figure~\ref{fig:3c286_systematics}, we show the peak Faraday depth ($\phi_{\rm rm}$; left) and intrinsic electric vector position angle ($\psi_0$; right) inferred from RM synthesis. As described in Section~\ref{sec:results_rmsynth}, we correct for the ionospheric contribution using \textsc{spinifex}. The measured Faraday depths (blue histogram) cluster around the $-0.3$\radPerSqm\ systematic offset reported by \citet{perley_ionosphere}, motivating our adoption of this correction for \src. As an additional check, we generated 1000 realisations of the observations by perturbing the \textsc{spinifex} ionospheric correction according to its native uncertainty, assuming Gaussian errors. The resulting distribution (grey histogram) clusters even more clearly around the same offset. We therefore conservatively adopt twice the standard deviation of these realisations as the systematic uncertainty on the ionospheric Faraday-depth correction. For the intrinsic EVPA, we find tight clustering around \relto{\sim}{30}{deg}. At L-band, \polcal\ is known to show spectral dependence in its intrinsic polarisation angle beyond a pure rotation-measure term, but our recovered values remain within the expected \relto{\sim}{26-31}{deg} range found by \citet{Hugo2024}.

Taken together, these results show that the spectropolarimetric properties of \polcal\ remain stable across the observing campaign, and that we recover their expected values. Importantly, unlike for circular-feed interferometers, linear-feed instruments such as MeerKAT do not require us to impose a specific polarisation model on \polcal\ in order to recover the cross-hand phase, and thus the observed behaviour. The resulting agreement is therefore effectively model-independent, and provides strong evidence that the behaviour seen in \src\ is intrinsic rather than a product of poor calibration or broader observational systematics.

\begin{figure}
    \centering
    \includegraphics[width=1.0\linewidth]{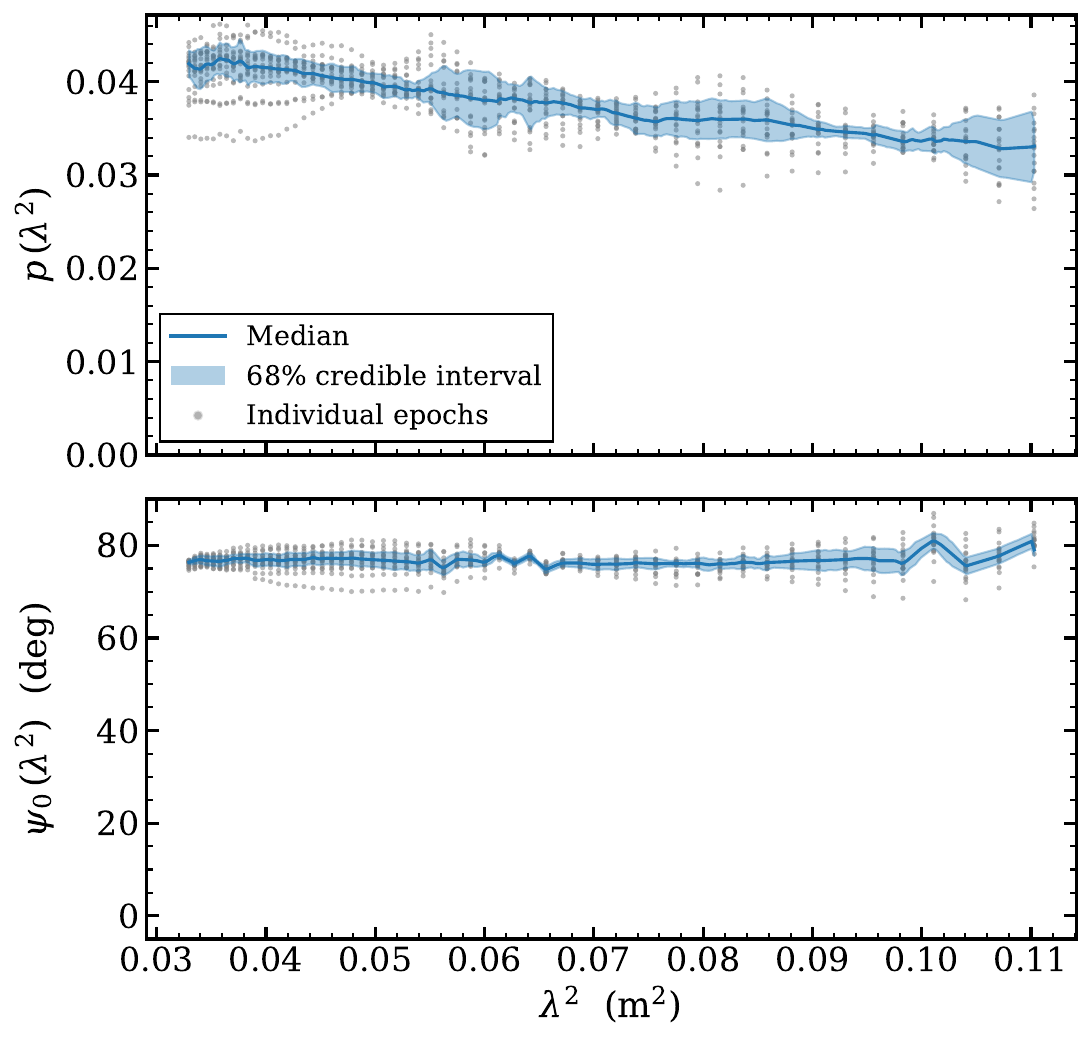}
    \caption{Spectropolarimetric diagnostic of \secondary\ across 16 scans obtained over five epochs between 2023 September 04 and 2023 October 28 (MJD 60191--60245). The top and bottom panels show the linear polarisation fraction and intrinsic polarisation angle as functions of $\lambda^2$, respectively. The solid blue line and shaded region mark the median and 68\% confidence interval, while the points show the individual scan measurements.}
    \label{fig:J1733_systematics}
\end{figure}

As a second check, we present the properties of \secondary\ across epochs bracketing the flaring interval. Unlike \polcal, \secondary\ was not used in any part of the polarisation calibration, even in a model-independent sense. It therefore provides an additional independent test of whether the unusual behaviour observed in \src\ could instead arise from residual instrumental or calibration effects. In Figure~\ref{fig:J1733_systematics}, we combine 16 scans from five observing epochs between 2023 September 04 and 2023 October 28 (MJD 60191--60245), spanning the pre-flare, flaring, and post-flare phases of the spectropolarimetric evolution of \src. The top and bottom panels present the $\lambda^2$ spectra of the linear polarisation fraction and intrinsic polarisation angle, respectively. The solid blue line and shaded region indicate the median and 68\% confidence interval, while the underlying points correspond to the individual scan measurements. For each scan, we correct the observed polarisation angle using the Faraday depth measured from RM synthesis. The corrected $\psi_0(\lambda^2)$ is broadly flat with frequency, while the polarisation fraction follows an approximately power-law-like trend. Although the individual scans are broadly consistent with one another, some scatter remains. This is naturally explained by \secondary\ being partially resolved at the higher L-band frequencies, while appearing point-like, or only marginally resolved, at the lowest frequencies. Variations in the observing time, flagging fraction, and thus, angular resolution across scans can recover slightly different mixtures of emission, introducing apparent scatter through resolution effects. One way to mitigate this would be to convolve all frequencies to a common resolution, but even in the presence of these systematics, \secondary\ does not exhibit the variability seen in \src.

\begin{figure*}
    \centering
    \includegraphics[width=0.85\linewidth]{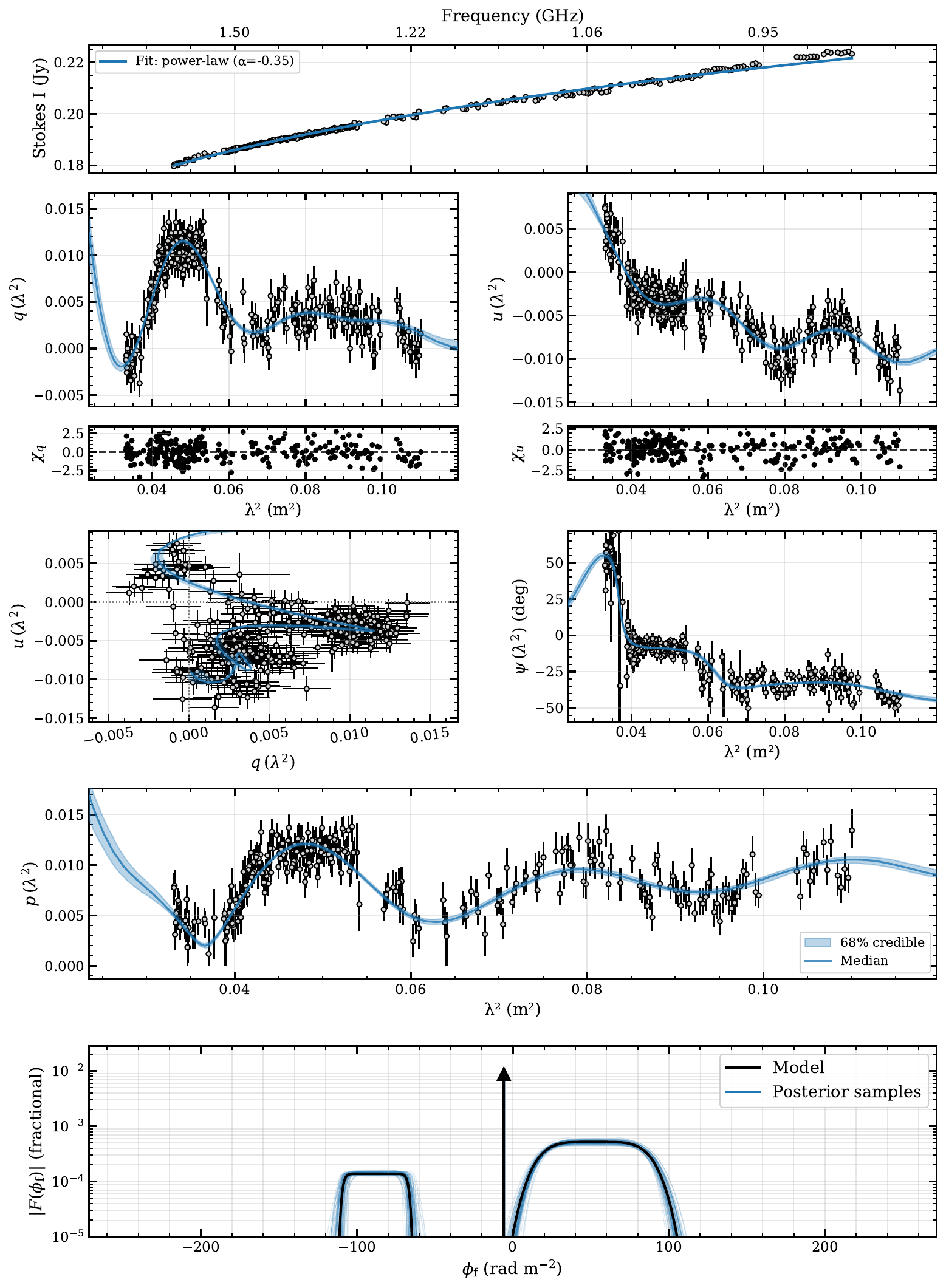}
    \caption{Comprehensive spectropolarimetric summary for \src\ on 2023 October 06 (MJD 60223), near the peak of one of the brightest flares. From top to bottom, we show the Stokes $I$ spectrum and fitted power law; the fractional Stokes parameters $q(\lambda^2)$ and $u(\lambda^2)$ with their normalised residuals; the trajectory in the $q$--$u$ plane; the corresponding polarisation angle, $\psi(\lambda^2)$; the fractional linear polarisation, $p(\lambda^2)$; and the inferred model Faraday spectrum, $|F(\phi_f)|$. Black points show the data, while the blue curves and shaded regions show the median model and 68\% credible interval from the posterior samples. The bottom panel illustrates the preferred three-component Faraday structure inferred for this epoch, consisting of a weaker Faraday-thin component consistent with the ISM and two Faraday-thick components offset to negative and positive Faraday depths, which together reproduce the strong frequency-dependent structure.}
    \label{fig:J1727_systematics}
\end{figure*}

\begin{figure*}
    \centering
    \includegraphics[width=0.85\linewidth]{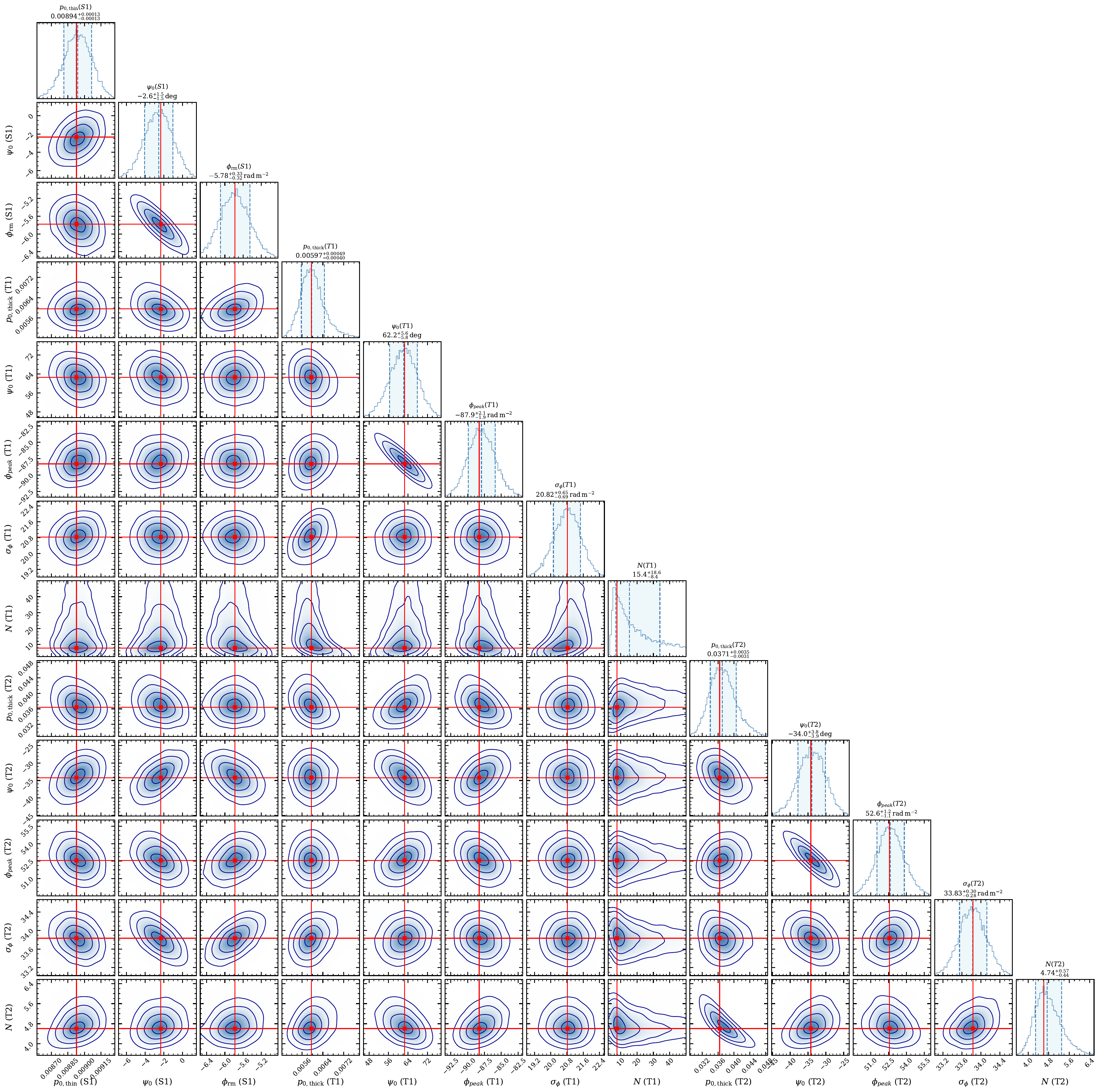}
    \caption{Posterior distributions for the favoured STT model fitted to \src\ on 2023 October 06 (MJD 60223). The diagonal panels show the marginalised one-dimensional posterior distributions for the intrinsic fractional polarisation, intrinsic polarisation angle, Faraday depth or peak Faraday depth, and, for the Faraday-thick components, the width and super-Gaussian shape parameter. The off-diagonal panels show the corresponding parameter covariances. Red lines mark the modal parameter values, while the contours enclose the credible regions of the joint posterior.}
    \label{fig:J1727_corner}
\end{figure*}

To give a complete example of the spectropolarimetric behaviour of \src, Figure~\ref{fig:J1727_systematics} shows the source near the peak of one of the brightest flares, on 2023 October 06 (MJD 60223). It includes the measured and fitted $q(\lambda^2)$ and $u(\lambda^2)$ spectra, the derived polarisation diagnostics, and the inferred model Faraday spectrum. The posterior distributions for the favoured STT model are shown in Figure~\ref{fig:J1727_corner}. Together, these figures provide a representative example of the behaviour discussed in the main text. Analogous plots for every epoch--model combination are available in the linked GitHub repository.

\begin{figure*}
    \centering
    \includegraphics[width=1.0\linewidth]{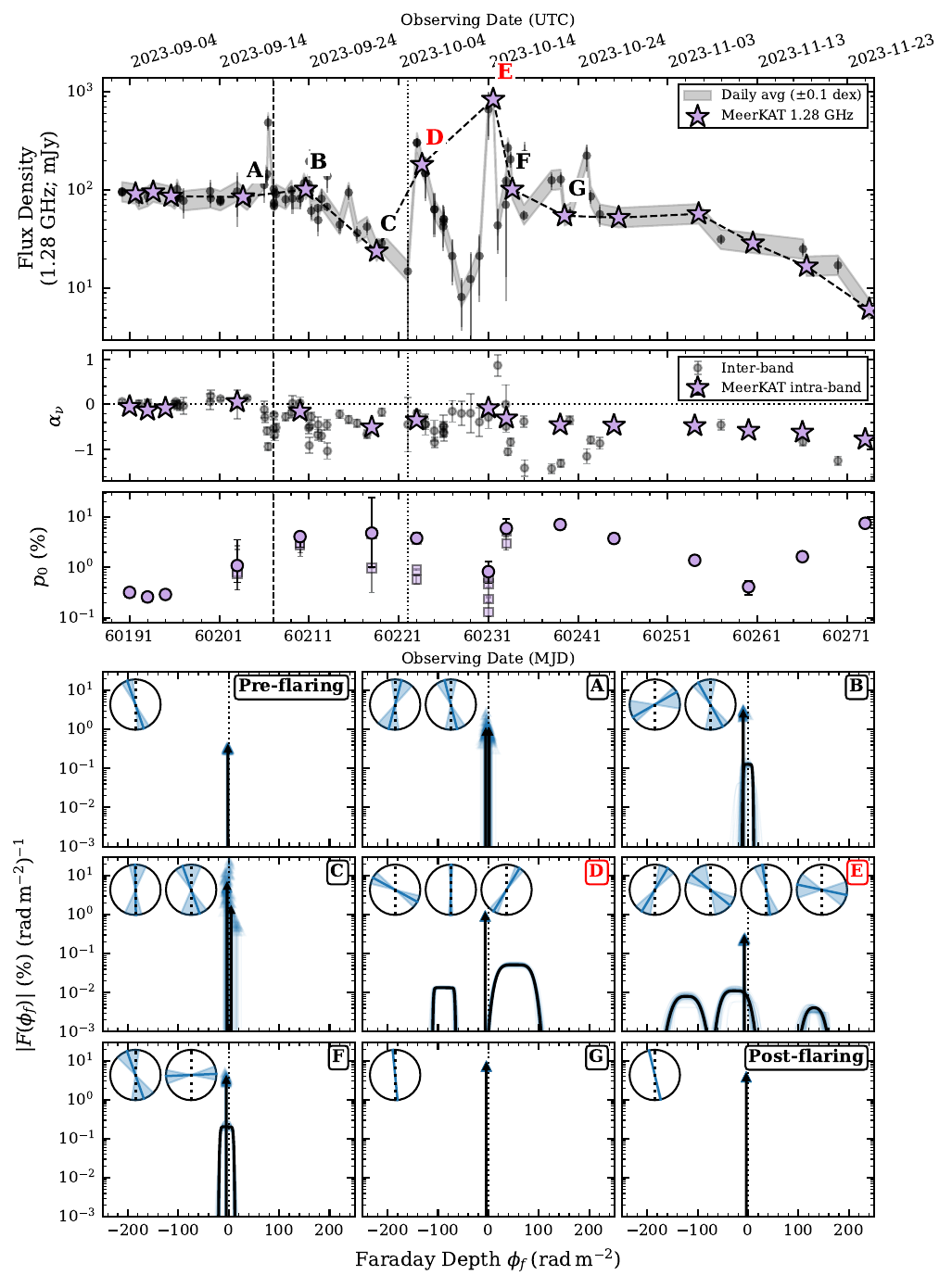}
    \caption{Same as Figure~\ref{fig:results_ILoveThisPlot}, except without the ISM-motivated S prior.}
    \label{fig:appendix_alt_approach}
\end{figure*}

Finally, Figure~\ref{fig:appendix_alt_approach} presents the results of the model-agnostic fitting approach in the same form as Figure~\ref{fig:results_ILoveThisPlot}. We keep the model classes favoured by our main method, except for 2023 October 14, where the model-agnostic approach now strongly favours STTT. The only methodological change is that we remove the ISM-motivated prior on the first S component and use uniform priors instead. The fits are qualitatively similar, although the posteriors are more prone to complex shapes and multimodality. To reduce label switching between S components and T components with $\sigma_\phi\rightarrow0$, we set the lower prior bound on $\sigma_\phi$ to 5\,\radPerSqm. The 2023 October 06 epoch, which is used for the mass estimates, recovers the same posterior as our main approach. The 2023 October 14 epoch differs more noticeably, consistent with the strong seed dependence shown in Figure~\ref{fig:results_seed_check}.

\section{Data Tables}
In this section, we include the full data tables for all observing epochs, following the format of Table~\ref{tab:qu_fitting_summary} in the main text.

\def\arraystretch{1.55}
\begin{table*}
    \centering
    \small
    \setlength{\tabcolsep}{1.0pt}
    \begin{tabular}{lccllcccccccc}
        \Xhline{5\arrayrulewidth}
        \multicolumn{1}{c}{Date} &
        \multicolumn{1}{c}{\begin{tabular}[c]{@{}c@{}}$I_0$ (mJy)\\$\alpha_\nu$\end{tabular}} &
        \multicolumn{1}{c}{Ejecta} &
        \multicolumn{1}{c}{Model} &
        \multicolumn{1}{c}{Comp.} &
        \multicolumn{1}{c}{$p_0$ (\%)} &
        \multicolumn{1}{c}{$\psi_0$ (deg)} &
        \multicolumn{1}{c}{$\phi_f$ ($\rad\,\mathrm{m}^{-2}$)} &
        \multicolumn{1}{c}{$\sigma_\phi$ ($\rad\,\mathrm{m}^{-2}$)} &
        \multicolumn{1}{c}{$N$} &
        \multicolumn{1}{c}{$\beta$} &
        \multicolumn{1}{c}{$\chi^2$ (dof)} &
        \multicolumn{1}{c}{$\ln \mathcal{Z}$} \\
        \Xhline{5\arrayrulewidth}
        \multirow[t]{4}{*}{\begin{tabular}[t]{@{}l@{}}2023-09-04 15:55\\60191.664\end{tabular}} & \multirow[t]{4}{*}{\begin{tabular}[t]{@{}c@{}}$95.98 \pm 0.07$\\$-0.047 \pm 0.005$\end{tabular}} & \multirow[t]{4}{*}{$\times$} & ${\rm \textbf{S}}$ & ${\rm S}_1$ & $0.318_{-0.016}^{+0.019}$ & $21_{-2}^{+3}$ & $-2.2_{-0.6}^{+0.6}$ & $\cdots$ & $\cdots$ & $\cdots$ & $141\,(135)$ & $688$ \\
        \cline{4-13}
         &  &  & \multirow{2}{*}{${\rm SS}$} & ${\rm S}_1$ & $0.30_{-0.18}^{+0.08}$ & $20_{-9}^{+8}$ & $-2.2_{-0.6}^{+0.7}$ & $\cdots$ & $\cdots$ & $\cdots$ & $140\,(132)$ & $682$ \\
         &  &  &  & ${\rm S}_2$ & $0.05_{-0.05}^{+0.15}$ & $14_{-26}^{+12}$ & $-1.0_{-1.8}^{+7.0}$ & $\cdots$ & $\cdots$ & $\cdots$ & $140\,(132)$ & $682$ \\
        \cline{4-13}
         &  &  & ${\rm P}$ & ${\rm P}_1$ & $0.30_{-0.03}^{+0.03}$ & $21_{-2}^{+3}$ & $-2.2_{-0.6}^{+0.6}$ & $\cdots$ & $\cdots$ & $0.3_{-0.4}^{+0.3}$ & $140\,(134)$ & $686$ \\
        \Xhline{3\arrayrulewidth}
        \multirow[t]{4}{*}{\begin{tabular}[t]{@{}l@{}}2023-09-06 15:06\\60193.630\end{tabular}} & \multirow[t]{4}{*}{\begin{tabular}[t]{@{}c@{}}$100.63 \pm 0.04$\\$-0.129 \pm 0.003$\end{tabular}} & \multirow[t]{4}{*}{$\times$} & ${\rm \textbf{S}}$ & ${\rm S}_1$ & $0.254_{-0.015}^{+0.015}$ & $-3_{-3}^{+2}$ & $-1.6_{-0.6}^{+0.6}$ & $\cdots$ & $\cdots$ & $\cdots$ & $204\,(137)$ & $689$ \\
        \cline{4-13}
         &  &  & \multirow{2}{*}{${\rm SS}$} & ${\rm S}_1$ & $0.16_{-0.07}^{+0.07}$ & $-4_{-9}^{+7}$ & $-2.2_{-0.8}^{+0.5}$ & $\cdots$ & $\cdots$ & $\cdots$ & $196\,(134)$ & $687$ \\
         &  &  &  & ${\rm S}_2$ & $0.13_{-0.08}^{+0.13}$ & $-10_{-34}^{+13}$ & $2_{-3}^{+10}$ & $\cdots$ & $\cdots$ & $\cdots$ & $196\,(134)$ & $687$ \\
        \cline{4-13}
         &  &  & ${\rm P}$ & ${\rm P}_1$ & $0.22_{-0.03}^{+0.02}$ & $-3_{-3}^{+2}$ & $-1.7_{-0.7}^{+0.5}$ & $\cdots$ & $\cdots$ & $0.8_{-0.4}^{+0.4}$ & $200\,(136)$ & $689$ \\
        \Xhline{3\arrayrulewidth}
        \multirow[t]{4}{*}{\begin{tabular}[t]{@{}l@{}}2023-09-08 14:58\\60195.624\end{tabular}} & \multirow[t]{4}{*}{\begin{tabular}[t]{@{}c@{}}$90.91 \pm 0.04$\\$-0.083 \pm 0.003$\end{tabular}} & \multirow[t]{4}{*}{$\times$} & ${\rm \textbf{S}}$ & ${\rm S}_1$ & $0.282_{-0.019}^{+0.018}$ & $-4_{-3}^{+3}$ & $-2.1_{-0.5}^{+0.7}$ & $\cdots$ & $\cdots$ & $\cdots$ & $166\,(125)$ & $614$ \\
        \cline{4-13}
         &  &  & \multirow{2}{*}{${\rm SS}$} & ${\rm S}_1$ & $0.23_{-0.02}^{+0.03}$ & $-2_{-5}^{+5}$ & $-3.0_{-0.5}^{+0.8}$ & $\cdots$ & $\cdots$ & $\cdots$ & $141\,(122)$ & $619$ \\
         &  &  &  & ${\rm S}_2$ & $0.12_{-0.03}^{+0.02}$ & $76_{-16}^{+28}$ & $33_{-9}^{+3}$ & $\cdots$ & $\cdots$ & $\cdots$ & $141\,(122)$ & $619$ \\
        \cline{4-13}
         &  &  & ${\rm P}$ & ${\rm P}_1$ & $0.19_{-0.04}^{+0.03}$ & $-5_{-3}^{+2}$ & $-2.2_{-0.6}^{+0.6}$ & $\cdots$ & $\cdots$ & $1.7_{-0.6}^{+0.5}$ & $155\,(124)$ & $619$ \\
        \Xhline{3\arrayrulewidth}
        \multirow[t]{6}{*}{\begin{tabular}[t]{@{}l@{}}2023-09-16 15:23\\60203.641\end{tabular}} & \multirow[t]{6}{*}{\begin{tabular}[t]{@{}c@{}}$88.50 \pm 0.05$\\$0.056 \pm 0.003$\end{tabular}} & \multirow[t]{6}{*}{$\times$} & ${\rm S}$ & ${\rm S}_1$ & $1.032_{-0.017}^{+0.015}$ & $0.2_{-1.0}^{+1.3}$ & $-2.3_{-0.3}^{+0.4}$ & $\cdots$ & $\cdots$ & $\cdots$ & $543\,(311)$ & $1366$ \\
        \cline{4-13}
         &  &  & \multirow{2}{*}{${\rm \textbf{SS}}$} & ${\rm S}_1$ & $0.97_{-0.12}^{+0.30}$ & $-16_{-13}^{+4}$ & $-3.4_{-0.5}^{+0.7}$ & $\cdots$ & $\cdots$ & $\cdots$ & $395\,(308)$ & $1428$ \\
         &  &  &  & ${\rm S}_2$ & $0.7_{-0.3}^{+0.4}$ & $26_{-14}^{+5}$ & $1.3_{-1.7}^{+3.0}$ & $\cdots$ & $\cdots$ & $\cdots$ & $395\,(308)$ & $1428$ \\
        \cline{4-13}
         &  &  & \multirow{2}{*}{${\rm ST}$} & ${\rm S}_1$ & $0.5_{-0.5}^{+0.2}$ & $9_{-11}^{+17}$ & $-3.0_{-0.8}^{+0.5}$ & $\cdots$ & $\cdots$ & $\cdots$ & $393\,(306)$ & $1429$ \\
         &  &  &  & ${\rm T}_1$ & $1.0_{-0.3}^{+0.3}$ & $-2_{-5}^{+4}$ & $-2_{-3}^{+5}$ & $10_{-3}^{+3}$ & $3.4_{-1.4}^{+6.0}$ & $\cdots$ & $393\,(306)$ & $1429$ \\
        \cline{4-13}
         &  &  & ${\rm P}$ & ${\rm P}_1$ & $0.75_{-0.03}^{+0.02}$ & $0.4_{-0.9}^{+1.5}$ & $-2.3_{-0.4}^{+0.4}$ & $\cdots$ & $\cdots$ & $1.18_{-0.09}^{+0.12}$ & $401\,(310)$ & $1433$ \\
        \Xhline{3\arrayrulewidth}
        \multirow[t]{8}{*}{\begin{tabular}[t]{@{}l@{}}2023-09-23 15:30\\60210.646\end{tabular}} & \multirow[t]{8}{*}{\begin{tabular}[t]{@{}c@{}}$106.32 \pm 0.05$\\$-0.153 \pm 0.003$\end{tabular}} & \multirow[t]{8}{*}{$\times$} & ${\rm S}$ & ${\rm S}_1$ & $2.229_{-0.012}^{+0.015}$ & $80.7_{-0.6}^{+0.5}$ & $-4.46_{-0.15}^{+0.12}$ & $\cdots$ & $\cdots$ & $\cdots$ & $1928\,(325)$ & $801$ \\
        \cline{4-13}
         &  &  & \multirow{3}{*}{${\rm SSS}$} & ${\rm S}_1$ & $6.8_{-1.1}^{+1.7}$ & $-79_{-3}^{+2}$ & $-4.7_{-0.6}^{+0.5}$ & $\cdots$ & $\cdots$ & $\cdots$ & $329\,(319)$ & $1577$ \\
         &  &  &  & ${\rm S}_2$ & $7.9_{-2.0}^{+0.8}$ & $-2_{-4}^{+3}$ & $-1.0_{-0.4}^{+1.1}$ & $\cdots$ & $\cdots$ & $\cdots$ & $329\,(319)$ & $1577$ \\
         &  &  &  & ${\rm S}_3$ & $1.4_{-0.2}^{+1.9}$ & $61_{-5}^{+4}$ & $3.8_{-0.6}^{+3.0}$ & $\cdots$ & $\cdots$ & $\cdots$ & $329\,(319)$ & $1577$ \\
        \cline{4-13}
         &  &  & \multirow{2}{*}{${\rm \textbf{ST}}$} & ${\rm S}_1$ & $2.72_{-0.08}^{+0.13}$ & $79_{-3}^{+3}$ & $-4.0_{-0.5}^{+0.6}$ & $\cdots$ & $\cdots$ & $\cdots$ & $327\,(320)$ & $1585$ \\
         &  &  &  & ${\rm T}_1$ & $1.82_{-0.09}^{+0.15}$ & $-9_{-3}^{+3}$ & $-5.9_{-1.3}^{+1.1}$ & $21.3_{-3.0}^{+1.2}$ & $4_{-2}^{+12}$ & $\cdots$ & $327\,(320)$ & $1585$ \\
        \cline{4-13}
         &  &  & \multirow{2}{*}{${\rm PP}$} & ${\rm P}_1$ & $4.5_{-0.6}^{+1.4}$ & $-6.2_{-1.3}^{+1.3}$ & $-2.9_{-0.5}^{+0.3}$ & $\cdots$ & $\cdots$ & $-3.8_{-0.3}^{+0.4}$ & $325\,(321)$ & $1591$ \\
         &  &  &  & ${\rm P}_2$ & $7.3_{-0.5}^{+1.5}$ & $79.2_{-0.6}^{+0.7}$ & $-2.9_{-0.5}^{+0.3}$ & $\cdots$ & $\cdots$ & $-2.56_{-0.16}^{+0.19}$ & $325\,(321)$ & $1591$ \\
        \Xhline{3\arrayrulewidth}
        \multirow[t]{6}{*}{\begin{tabular}[t]{@{}l@{}}2023-10-01 12:50\\60218.535\end{tabular}} & \multirow[t]{6}{*}{\begin{tabular}[t]{@{}c@{}}$25.77 \pm 0.03$\\$-0.504 \pm 0.006$\end{tabular}} & \multirow[t]{6}{*}{$\times$} & ${\rm S}$ & ${\rm S}_1$ & $4.95_{-0.05}^{+0.07}$ & $3.4_{-1.0}^{+1.0}$ & $-3.8_{-0.3}^{+0.3}$ & $\cdots$ & $\cdots$ & $\cdots$ & $451\,(327)$ & $1050$ \\
        \cline{4-13}
         &  &  & \multirow{2}{*}{${\rm \textbf{SS}}$} & ${\rm S}_1$ & $4.84_{-0.14}^{+0.15}$ & $2_{-3}^{+3}$ & $-3.9_{-0.5}^{+0.5}$ & $\cdots$ & $\cdots$ & $\cdots$ & $412\,(324)$ & $1060$ \\
         &  &  &  & ${\rm S}_2$ & $0.47_{-0.16}^{+0.30}$ & $-31_{-27}^{+41}$ & $18_{-12}^{+6}$ & $\cdots$ & $\cdots$ & $\cdots$ & $412\,(324)$ & $1060$ \\
        \cline{4-13}
         &  &  & \multirow{2}{*}{${\rm ST}$} & ${\rm S}_1$ & $4.8_{-0.4}^{+0.2}$ & $2_{-4}^{+3}$ & $-3.9_{-0.4}^{+0.6}$ & $\cdots$ & $\cdots$ & $\cdots$ & $411\,(322)$ & $1061$ \\
         &  &  &  & ${\rm T}_1$ & $0.5_{-0.2}^{+0.8}$ & $4_{-41}^{+9}$ & $0_{-4}^{+19}$ & $0.5_{-0.4}^{+4.0}$ & $5_{-3}^{+16}$ & $\cdots$ & $411\,(322)$ & $1061$ \\
        \cline{4-13}
         &  &  & ${\rm P}$ & ${\rm P}_1$ & $4.49_{-0.11}^{+0.08}$ & $3.2_{-1.0}^{+1.0}$ & $-3.9_{-0.3}^{+0.3}$ & $\cdots$ & $\cdots$ & $0.45_{-0.08}^{+0.07}$ & $415\,(326)$ & $1064$ \\
        \Xhline{5\arrayrulewidth}
    \end{tabular}
    \caption{Summary of the $QU$-fitting results for all epochs. Parameter values are posterior modes with 68 per cent highest-density intervals. The Stokes~$I$ column gives the flux density at 1.28 GHz and the spectral index $\alpha_\nu$. The $\phi_f$ column gives $\phi_{\rm rm}$ for Faraday-thin and power-law components, and $\phi_{\rm peak}$ for Faraday-thick components. The bolded models are the favoured models.}
    \label{tab:qu_fitting_appendix}
\end{table*}

\def\arraystretch{1.55}
\begingroup
\renewcommand{\thetable}{\ref{tab:qu_fitting_appendix}}
\begin{table*}
    \centering
    \small
    \setlength{\tabcolsep}{1.0pt}
    \begin{tabular}{lccllcccccccc}
        \Xhline{5\arrayrulewidth}
        \multicolumn{1}{c}{Date} &
        \multicolumn{1}{c}{\begin{tabular}[c]{@{}c@{}}$I_0$ (mJy)\\$\alpha_\nu$\end{tabular}} &
        \multicolumn{1}{c}{Ejecta} &
        \multicolumn{1}{c}{Model} &
        \multicolumn{1}{c}{Comp.} &
        \multicolumn{1}{c}{$p_0$ (\%)} &
        \multicolumn{1}{c}{$\psi_0$ (deg)} &
        \multicolumn{1}{c}{$\phi_f$ ($\rad\,\mathrm{m}^{-2}$)} &
        \multicolumn{1}{c}{$\sigma_\phi$ ($\rad\,\mathrm{m}^{-2}$)} &
        \multicolumn{1}{c}{$N$} &
        \multicolumn{1}{c}{$\beta$} &
        \multicolumn{1}{c}{$\chi^2$ (dof)} &
        \multicolumn{1}{c}{$\ln \mathcal{Z}$} \\
        \Xhline{5\arrayrulewidth}
        \multirow[t]{17}{*}{\begin{tabular}[t]{@{}l@{}}2023-10-06 14:21\\60223.599\end{tabular}} & \multirow[t]{17}{*}{\begin{tabular}[t]{@{}c@{}}$196.40 \pm 0.04$\\$-0.3466 \pm 0.0014$\end{tabular}} & \multirow[t]{17}{*}{$\times$} & ${\rm S}$ & ${\rm S}_1$ & $0.788_{-0.007}^{+0.009}$ & $6.5_{-0.3}^{+0.3}$ & $-6.539_{-0.004}^{+0.013}$ & $\cdots$ & $\cdots$ & $\cdots$ & $3215\,(549)$ & $1486$ \\
        \cline{4-13}
         &  &  & \multirow{6}{*}{${\rm SSSSSS}$} & ${\rm S}_1$ & $2.96_{-0.14}^{+0.20}$ & $-26.3_{-1.7}^{+1.7}$ & $-6.0_{-0.3}^{+0.7}$ & $\cdots$ & $\cdots$ & $\cdots$ & $841\,(534)$ & $2602$ \\
         &  &  &  & ${\rm S}_2$ & $0.13_{-0.13}^{+0.80}$ & $-20_{-6}^{+13}$ & $10.3_{-3.0}^{+1.0}$ & $\cdots$ & $\cdots$ & $\cdots$ & $841\,(534)$ & $2602$ \\
         &  &  &  & ${\rm S}_3$ & $23_{-3}^{+3}$ & $-21_{-3}^{+3}$ & $11.6_{-0.6}^{+1.2}$ & $\cdots$ & $\cdots$ & $\cdots$ & $841\,(534)$ & $2602$ \\
         &  &  &  & ${\rm S}_4$ & $29_{-3}^{+4}$ & $47_{-3}^{+4}$ & $16.8_{-1.1}^{+0.9}$ & $\cdots$ & $\cdots$ & $\cdots$ & $841\,(534)$ & $2602$ \\
         &  &  &  & ${\rm S}_5$ & $8.5_{-0.9}^{+1.2}$ & $88_{-5}^{+3}$ & $28.5_{-1.1}^{+1.2}$ & $\cdots$ & $\cdots$ & $\cdots$ & $841\,(534)$ & $2602$ \\
         &  &  &  & ${\rm S}_6$ & $0.89_{-0.10}^{+0.09}$ & $42_{-8}^{+5}$ & $60.4_{-1.9}^{+1.2}$ & $\cdots$ & $\cdots$ & $\cdots$ & $841\,(534)$ & $2602$ \\
        \cline{4-13}
         &  &  & \multirow{3}{*}{${\rm \textbf{STT}}$} & ${\rm S}_1$ & $0.893_{-0.012}^{+0.013}$ & $-2.3_{-1.8}^{+1.3}$ & $-5.8_{-0.3}^{+0.3}$ & $\cdots$ & $\cdots$ & $\cdots$ & $689\,(539)$ & $2709$ \\
         &  &  &  & ${\rm T}_1$ & $0.60_{-0.05}^{+0.04}$ & $63_{-6}^{+5}$ & $-88.3_{-1.7}^{+2.0}$ & $20.8_{-0.7}^{+0.7}$ & $8_{-4}^{+16}$ & $\cdots$ & $689\,(539)$ & $2709$ \\
         &  &  &  & ${\rm T}_2$ & $3.6_{-0.3}^{+0.4}$ & $-34_{-4}^{+4}$ & $52.6_{-1.2}^{+1.1}$ & $33.8_{-0.3}^{+0.3}$ & $4.6_{-0.4}^{+0.6}$ & $\cdots$ & $689\,(539)$ & $2709$ \\
        \cline{4-13}
         &  &  & \multirow{4}{*}{${\rm STTT}$} & ${\rm S}_1$ & $0.857_{-0.020}^{+0.015}$ & $-9.7_{-3.0}^{+1.5}$ & $-4.2_{-0.3}^{+0.3}$ & $\cdots$ & $\cdots$ & $\cdots$ & $658\,(534)$ & $2709$ \\
         &  &  &  & ${\rm T}_1$ & $0.67_{-0.08}^{+0.08}$ & $60_{-8}^{+9}$ & $-87_{-3}^{+3}$ & $20.5_{-0.7}^{+0.7}$ & $6_{-3}^{+9}$ & $\cdots$ & $658\,(534)$ & $2709$ \\
         &  &  &  & ${\rm T}_2$ & $2.0_{-0.2}^{+0.5}$ & $-37_{-15}^{+9}$ & $54_{-2}^{+2}$ & $18.2_{-1.4}^{+1.3}$ & $3.2_{-1.2}^{+8.0}$ & $\cdots$ & $658\,(534)$ & $2709$ \\
         &  &  &  & ${\rm T}_3$ & $3.9_{-1.1}^{+2.0}$ & $-25_{-35}^{+8}$ & $50.1_{-1.3}^{+2.0}$ & $33.7_{-1.0}^{+1.2}$ & $4.8_{-0.5}^{+0.6}$ & $\cdots$ & $658\,(534)$ & $2709$ \\
        \cline{4-13}
         &  &  & \multirow{3}{*}{${\rm PPP}$} & ${\rm P}_1$ & $0.74_{-0.07}^{+0.09}$ & $-54.0_{-1.8}^{+1.9}$ & $-2.07_{-0.30}^{+0.07}$ & $\cdots$ & $\cdots$ & $-0.6_{-0.5}^{+0.6}$ & $1292\,(542)$ & $2413$ \\
         &  &  &  & ${\rm P}_2$ & $4.2_{-0.6}^{+0.9}$ & $1.8_{-2.0}^{+1.7}$ & $-2.07_{-0.30}^{+0.07}$ & $\cdots$ & $\cdots$ & $4.62_{-0.13}^{+0.05}$ & $1292\,(542)$ & $2413$ \\
         &  &  &  & ${\rm P}_3$ & $3.4_{-0.7}^{+0.9}$ & $-89.1_{-1.7}^{+1.8}$ & $-2.07_{-0.30}^{+0.07}$ & $\cdots$ & $\cdots$ & $4.992_{-0.030}^{+0.008}$ & $1292\,(542)$ & $2413$ \\
        \Xhline{3\arrayrulewidth}
        \multirow[t]{17}{*}{\begin{tabular}[t]{@{}l@{}}2023-10-14 12:47\\60231.533\end{tabular}} & \multirow[t]{17}{*}{\begin{tabular}[t]{@{}c@{}}$875.03 \pm 0.09$\\$-0.0809 \pm 0.0007$\end{tabular}} & \multirow[t]{17}{*}{$\times$} & ${\rm S}$ & ${\rm S}_1$ & $0.3044_{-0.0020}^{+0.0015}$ & $12.36_{-0.17}^{+0.20}$ & $-6.675_{-0.010}^{+0.040}$ & $\cdots$ & $\cdots$ & $\cdots$ & $4525\,(537)$ & $1583$ \\
        \cline{4-13}
         &  &  & \multirow{4}{*}{${\rm SSSS}$} & ${\rm S}_1$ & $0.65_{-0.12}^{+0.13}$ & $-18_{-2}^{+3}$ & $-5.2_{-0.5}^{+0.6}$ & $\cdots$ & $\cdots$ & $\cdots$ & $1865\,(528)$ & $2859$ \\
         &  &  &  & ${\rm S}_2$ & $0.50_{-0.13}^{+0.14}$ & $49_{-4}^{+2}$ & $-1.9_{-1.0}^{+1.3}$ & $\cdots$ & $\cdots$ & $\cdots$ & $1865\,(528)$ & $2859$ \\
         &  &  &  & ${\rm S}_3$ & $0.045_{-0.003}^{+0.003}$ & $-77_{-13}^{+8}$ & $119_{-2}^{+3}$ & $\cdots$ & $\cdots$ & $\cdots$ & $1865\,(528)$ & $2859$ \\
         &  &  &  & ${\rm S}_4$ & $0.044_{-0.002}^{+0.003}$ & $21_{-9}^{+9}$ & $156_{-2}^{+2}$ & $\cdots$ & $\cdots$ & $\cdots$ & $1865\,(528)$ & $2859$ \\
        \cline{4-13}
         &  &  & \multirow{4}{*}{${\rm \textbf{SSTT}}$} & ${\rm S}_1$ & $0.270_{-0.008}^{+0.012}$ & $2_{-3}^{+3}$ & $-4.4_{-0.6}^{+0.5}$ & $\cdots$ & $\cdots$ & $\cdots$ & $922\,(524)$ & $3326$ \\
         &  &  &  & ${\rm S}_2$ & $0.043_{-0.005}^{+0.014}$ & $-35_{-23}^{+21}$ & $24_{-5}^{+6}$ & $\cdots$ & $\cdots$ & $\cdots$ & $922\,(524)$ & $3326$ \\
         &  &  &  & ${\rm T}_1$ & $0.50_{-0.03}^{+0.05}$ & $5_{-4}^{+5}$ & $-111.2_{-1.8}^{+1.5}$ & $48.6_{-0.7}^{+0.8}$ & $10_{-2}^{+4}$ & $\cdots$ & $922\,(524)$ & $3326$ \\
         &  &  &  & ${\rm T}_2$ & $0.21_{-0.03}^{+0.03}$ & $-88_{-15}^{+8}$ & $125_{-3}^{+6}$ & $17.6_{-0.9}^{+0.9}$ & $2.3_{-0.3}^{+0.3}$ & $\cdots$ & $922\,(524)$ & $3326$ \\
        \cline{4-13}
         &  &  & \multirow{4}{*}{${\rm STTT}$} & ${\rm S}_1$ & $0.267_{-0.007}^{+0.012}$ & $2_{-3}^{+2}$ & $-4.6_{-0.5}^{+0.6}$ & $\cdots$ & $\cdots$ & $\cdots$ & $922\,(522)$ & $3332$ \\
         &  &  &  & ${\rm T}_1$ & $0.50_{-0.05}^{+0.04}$ & $5_{-5}^{+5}$ & $-111.3_{-1.8}^{+1.5}$ & $48.7_{-0.8}^{+0.8}$ & $10_{-3}^{+4}$ & $\cdots$ & $922\,(522)$ & $3332$ \\
         &  &  &  & ${\rm T}_2$ & $0.047_{-0.009}^{+0.013}$ & $-36_{-24}^{+21}$ & $24_{-5}^{+6}$ & $0.4_{-0.3}^{+3.0}$ & $4_{-2}^{+15}$ & $\cdots$ & $922\,(522)$ & $3332$ \\
         &  &  &  & ${\rm T}_3$ & $0.20_{-0.03}^{+0.04}$ & $90_{-10}^{+13}$ & $125_{-5}^{+4}$ & $17.7_{-0.9}^{+0.8}$ & $2.3_{-0.3}^{+0.3}$ & $\cdots$ & $922\,(522)$ & $3332$ \\
        \cline{4-13}
         &  &  & \multirow{4}{*}{${\rm PPPP}$} & ${\rm P}_1$ & $5.1_{-0.7}^{+0.5}$ & $31.8_{-0.9}^{+0.9}$ & $-6.66_{-0.02}^{+0.09}$ & $\cdots$ & $\cdots$ & $1.03_{-0.12}^{+0.12}$ & $1899\,(527)$ & $2782$ \\
         &  &  &  & ${\rm P}_2$ & $10.3_{-1.0}^{+1.0}$ & $-57.0_{-0.9}^{+1.0}$ & $-6.66_{-0.02}^{+0.09}$ & $\cdots$ & $\cdots$ & $2.68_{-0.07}^{+0.06}$ & $1899\,(527)$ & $2782$ \\
         &  &  &  & ${\rm P}_3$ & $13.3_{-0.3}^{+0.3}$ & $33.4_{-0.9}^{+0.9}$ & $-6.66_{-0.02}^{+0.09}$ & $\cdots$ & $\cdots$ & $4.47_{-0.05}^{+0.04}$ & $1899\,(527)$ & $2782$ \\
         &  &  &  & ${\rm P}_4$ & $7.6_{-0.4}^{+0.4}$ & $-56.4_{-0.8}^{+1.0}$ & $-6.66_{-0.02}^{+0.09}$ & $\cdots$ & $\cdots$ & $4.997_{-0.011}^{+0.003}$ & $1899\,(527)$ & $2782$ \\
        \Xhline{5\arrayrulewidth}
    \end{tabular}
    \caption[]{Continued.}
\end{table*}
\endgroup

\def\arraystretch{1.55}
\begingroup
\renewcommand{\thetable}{\ref{tab:qu_fitting_appendix}}
\begin{table*}
    \centering
    \small
    \setlength{\tabcolsep}{1.0pt}
    \begin{tabular}{lccllcccccccc}
        \Xhline{5\arrayrulewidth}
        \multicolumn{1}{c}{Date} &
        \multicolumn{1}{c}{\begin{tabular}[c]{@{}c@{}}$I_0$ (mJy)\\$\alpha_\nu$\end{tabular}} &
        \multicolumn{1}{c}{Ejecta} &
        \multicolumn{1}{c}{Model} &
        \multicolumn{1}{c}{Comp.} &
        \multicolumn{1}{c}{$p_0$ (\%)} &
        \multicolumn{1}{c}{$\psi_0$ (deg)} &
        \multicolumn{1}{c}{$\phi_f$ ($\rad\,\mathrm{m}^{-2}$)} &
        \multicolumn{1}{c}{$\sigma_\phi$ ($\rad\,\mathrm{m}^{-2}$)} &
        \multicolumn{1}{c}{$N$} &
        \multicolumn{1}{c}{$\beta$} &
        \multicolumn{1}{c}{$\chi^2$ (dof)} &
        \multicolumn{1}{c}{$\ln \mathcal{Z}$} \\
        \Xhline{5\arrayrulewidth}
        \multirow[t]{7}{*}{\begin{tabular}[t]{@{}l@{}}2023-10-16 15:57\\60233.665\end{tabular}} & \multirow[t]{7}{*}{\begin{tabular}[t]{@{}c@{}}$109.89 \pm 0.03$\\$-0.3047 \pm 0.0017$\end{tabular}} & \multirow[t]{7}{*}{$\times$} & ${\rm S}$ & ${\rm S}_1$ & $1.883_{-0.016}^{+0.014}$ & $64.8_{-0.2}^{+0.2}$ & $-6.8197_{-0.0007}^{+0.0030}$ & $\cdots$ & $\cdots$ & $\cdots$ & $5859\,(537)$ & $-216$ \\
        \cline{4-13}
         &  &  & \multirow{2}{*}{${\rm SS}$} & ${\rm S}_1$ & $10_{-2}^{+5}$ & $48.6_{-1.5}^{+2.0}$ & $-5.1_{-0.5}^{+0.4}$ & $\cdots$ & $\cdots$ & $\cdots$ & $545\,(534)$ & $2434$ \\
         &  &  &  & ${\rm S}_2$ & $9_{-2}^{+5}$ & $-51_{-3}^{+2}$ & $-2.6_{-0.6}^{+0.7}$ & $\cdots$ & $\cdots$ & $\cdots$ & $545\,(534)$ & $2434$ \\
        \cline{4-13}
         &  &  & \multirow{2}{*}{${\rm \textbf{ST}}$} & ${\rm S}_1$ & $3.4_{-0.6}^{+0.4}$ & $17_{-4}^{+5}$ & $-4.6_{-0.5}^{+0.7}$ & $\cdots$ & $\cdots$ & $\cdots$ & $532\,(532)$ & $2442$ \\
         &  &  &  & ${\rm T}_1$ & $5.6_{-0.6}^{+0.5}$ & $-89.6_{-1.9}^{+2.0}$ & $-4.1_{-0.3}^{+0.4}$ & $13.4_{-2.0}^{+0.8}$ & $5_{-3}^{+8}$ & $\cdots$ & $532\,(532)$ & $2442$ \\
        \cline{4-13}
         &  &  & \multirow{2}{*}{${\rm PP}$} & ${\rm P}_1$ & $6.0_{-2.0}^{+0.9}$ & $-2.7_{-1.5}^{+3.0}$ & $-4.2_{-0.3}^{+0.4}$ & $\cdots$ & $\cdots$ & $-2.1_{-0.4}^{+0.3}$ & $539\,(533)$ & $2438$ \\
         &  &  &  & ${\rm P}_2$ & $5.2_{-2.0}^{+0.8}$ & $78.4_{-1.6}^{+1.2}$ & $-4.2_{-0.3}^{+0.4}$ & $\cdots$ & $\cdots$ & $-0.31_{-0.19}^{+0.30}$ & $539\,(533)$ & $2438$ \\
        \Xhline{3\arrayrulewidth}
        \multirow[t]{4}{*}{\begin{tabular}[t]{@{}l@{}}2023-10-22 11:32\\60239.481\end{tabular}} & \multirow[t]{4}{*}{\begin{tabular}[t]{@{}c@{}}$59.05 \pm 0.03$\\$-0.453 \pm 0.004$\end{tabular}} & \multirow[t]{4}{*}{$\times$} & ${\rm \textbf{S}}$ & ${\rm S}_1$ & $7.11_{-0.03}^{+0.03}$ & $5.5_{-0.4}^{+0.3}$ & $-4.15_{-0.09}^{+0.09}$ & $\cdots$ & $\cdots$ & $\cdots$ & $294\,(277)$ & $1172$ \\
        \cline{4-13}
         &  &  & \multirow{2}{*}{${\rm SS}$} & ${\rm S}_1$ & $7.3_{-1.7}^{+1.9}$ & $6_{-3}^{+3}$ & $-4.13_{-0.15}^{+0.40}$ & $\cdots$ & $\cdots$ & $\cdots$ & $291\,(274)$ & $1168$ \\
         &  &  &  & ${\rm S}_2$ & $0.4_{-0.4}^{+1.5}$ & $-79_{-38}^{+7}$ & $-3.7_{-0.7}^{+2.0}$ & $\cdots$ & $\cdots$ & $\cdots$ & $291\,(274)$ & $1168$ \\
        \cline{4-13}
         &  &  & ${\rm P}$ & ${\rm P}_1$ & $7.15_{-0.04}^{+0.05}$ & $5.5_{-0.3}^{+0.3}$ & $-4.16_{-0.09}^{+0.09}$ & $\cdots$ & $\cdots$ & $-0.024_{-0.017}^{+0.030}$ & $293\,(276)$ & $1168$ \\
        \Xhline{3\arrayrulewidth}
        \multirow[t]{4}{*}{\begin{tabular}[t]{@{}l@{}}2023-10-28 12:22\\60245.516\end{tabular}} & \multirow[t]{4}{*}{\begin{tabular}[t]{@{}c@{}}$55.41 \pm 0.03$\\$-0.464 \pm 0.004$\end{tabular}} & \multirow[t]{4}{*}{$\times$} & ${\rm \textbf{S}}$ & ${\rm S}_1$ & $3.75_{-0.02}^{+0.03}$ & $13.5_{-0.6}^{+0.5}$ & $-3.66_{-0.17}^{+0.14}$ & $\cdots$ & $\cdots$ & $\cdots$ & $361\,(323)$ & $1368$ \\
        \cline{4-13}
         &  &  & \multirow{2}{*}{${\rm SS}$} & ${\rm S}_1$ & $3.5_{-1.1}^{+0.7}$ & $18_{-4}^{+4}$ & $-4.90_{-0.17}^{+1.10}$ & $\cdots$ & $\cdots$ & $\cdots$ & $346\,(320)$ & $1368$ \\
         &  &  &  & ${\rm S}_2$ & $0.5_{-0.4}^{+0.3}$ & $-24_{-19}^{+13}$ & $-3.7_{-0.9}^{+5.0}$ & $\cdots$ & $\cdots$ & $\cdots$ & $346\,(320)$ & $1368$ \\
        \cline{4-13}
         &  &  & ${\rm P}$ & ${\rm P}_1$ & $3.88_{-0.04}^{+0.05}$ & $13.5_{-0.6}^{+0.5}$ & $-3.68_{-0.16}^{+0.15}$ & $\cdots$ & $\cdots$ & $-0.16_{-0.03}^{+0.04}$ & $346\,(322)$ & $1370$ \\
        \Xhline{3\arrayrulewidth}
        \multirow[t]{4}{*}{\begin{tabular}[t]{@{}l@{}}2023-11-06 10:22\\60254.432\end{tabular}} & \multirow[t]{4}{*}{\begin{tabular}[t]{@{}c@{}}$61.70 \pm 0.04$\\$-0.473 \pm 0.004$\end{tabular}} & \multirow[t]{4}{*}{$\times$} & ${\rm \textbf{S}}$ & ${\rm S}_1$ & $1.38_{-0.03}^{+0.03}$ & $12.6_{-1.7}^{+1.2}$ & $-0.7_{-0.3}^{+0.4}$ & $\cdots$ & $\cdots$ & $\cdots$ & $464\,(311)$ & $1196$ \\
        \cline{4-13}
         &  &  & \multirow{2}{*}{${\rm SS}$} & ${\rm S}_1$ & $1.1_{-0.3}^{+0.3}$ & $9_{-26}^{+4}$ & $-1.9_{-0.7}^{+0.6}$ & $\cdots$ & $\cdots$ & $\cdots$ & $439\,(308)$ & $1204$ \\
         &  &  &  & ${\rm S}_2$ & $1.0_{-0.6}^{+0.5}$ & $28_{-15}^{+12}$ & $1.4_{-1.5}^{+3.0}$ & $\cdots$ & $\cdots$ & $\cdots$ & $439\,(308)$ & $1204$ \\
        \cline{4-13}
         &  &  & ${\rm P}$ & ${\rm P}_1$ & $1.16_{-0.04}^{+0.06}$ & $12.5_{-1.5}^{+1.4}$ & $-0.7_{-0.4}^{+0.4}$ & $\cdots$ & $\cdots$ & $0.70_{-0.16}^{+0.12}$ & $439\,(310)$ & $1206$ \\
        \Xhline{3\arrayrulewidth}
        \multirow[t]{4}{*}{\begin{tabular}[t]{@{}l@{}}2023-11-12 11:26\\60260.476\end{tabular}} & \multirow[t]{4}{*}{\begin{tabular}[t]{@{}c@{}}$30.78 \pm 0.02$\\$-0.576 \pm 0.005$\end{tabular}} & \multirow[t]{4}{*}{$\times$} & ${\rm \textbf{S}}$ & ${\rm S}_1$ & $0.41_{-0.04}^{+0.05}$ & $-26_{-5}^{+3}$ & $-4.6_{-0.7}^{+0.6}$ & $\cdots$ & $\cdots$ & $\cdots$ & $358\,(331)$ & $1199$ \\
        \cline{4-13}
         &  &  & \multirow{2}{*}{${\rm SS}$} & ${\rm S}_1$ & $0.5_{-0.2}^{+0.5}$ & $-57_{-10}^{+29}$ & $-4.9_{-0.5}^{+0.8}$ & $\cdots$ & $\cdots$ & $\cdots$ & $354\,(328)$ & $1194$ \\
         &  &  &  & ${\rm S}_2$ & $0.3_{-0.3}^{+0.5}$ & $1_{-31}^{+23}$ & $-3.4_{-2.0}^{+5.0}$ & $\cdots$ & $\cdots$ & $\cdots$ & $354\,(328)$ & $1194$ \\
        \cline{4-13}
         &  &  & ${\rm P}$ & ${\rm P}_1$ & $0.31_{-0.07}^{+0.07}$ & $-27_{-4}^{+3}$ & $-4.7_{-0.5}^{+0.7}$ & $\cdots$ & $\cdots$ & $1.1_{-0.6}^{+0.8}$ & $356\,(330)$ & $1199$ \\
        \Xhline{3\arrayrulewidth}
        \multirow[t]{4}{*}{\begin{tabular}[t]{@{}l@{}}2023-11-18 11:11\\60266.467\end{tabular}} & \multirow[t]{4}{*}{\begin{tabular}[t]{@{}c@{}}$17.727 \pm 0.017$\\$-0.616 \pm 0.006$\end{tabular}} & \multirow[t]{4}{*}{$\times$} & ${\rm \textbf{S}}$ & ${\rm S}_1$ & $1.64_{-0.09}^{+0.07}$ & $-15_{-2}^{+3}$ & $-4.0_{-0.5}^{+0.6}$ & $\cdots$ & $\cdots$ & $\cdots$ & $351\,(345)$ & $1066$ \\
        \cline{4-13}
         &  &  & \multirow{2}{*}{${\rm SS}$} & ${\rm S}_1$ & $1.7_{-1.0}^{+1.9}$ & $-16_{-9}^{+7}$ & $-4.1_{-0.4}^{+0.7}$ & $\cdots$ & $\cdots$ & $\cdots$ & $349\,(342)$ & $1061$ \\
         &  &  &  & ${\rm S}_2$ & $0.5_{-0.5}^{+3.0}$ & $6_{-56}^{+26}$ & $-3.5_{-1.0}^{+2.0}$ & $\cdots$ & $\cdots$ & $\cdots$ & $349\,(342)$ & $1061$ \\
        \cline{4-13}
         &  &  & ${\rm P}$ & ${\rm P}_1$ & $1.50_{-0.13}^{+0.13}$ & $-14_{-2}^{+2}$ & $-4.0_{-0.5}^{+0.7}$ & $\cdots$ & $\cdots$ & $0.4_{-0.3}^{+0.3}$ & $349\,(344)$ & $1064$ \\
        \Xhline{3\arrayrulewidth}
        \multirow[t]{4}{*}{\begin{tabular}[t]{@{}l@{}}2023-11-25 10:55\\60273.455\end{tabular}} & \multirow[t]{4}{*}{\begin{tabular}[t]{@{}c@{}}$6.642 \pm 0.019$\\$-0.772 \pm 0.017$\end{tabular}} & \multirow[t]{4}{*}{$\times$} & ${\rm \textbf{S}}$ & ${\rm S}_1$ & $7.6_{-0.3}^{+0.2}$ & $2.6_{-2.0}^{+1.6}$ & $-4.0_{-0.6}^{+0.4}$ & $\cdots$ & $\cdots$ & $\cdots$ & $275\,(337)$ & $726$ \\
        \cline{4-13}
         &  &  & \multirow{2}{*}{${\rm SS}$} & ${\rm S}_1$ & $7.5_{-4.0}^{+1.6}$ & $3_{-4}^{+4}$ & $-4.4_{-0.3}^{+0.8}$ & $\cdots$ & $\cdots$ & $\cdots$ & $274\,(334)$ & $722$ \\
         &  &  &  & ${\rm S}_2$ & $0.5_{-0.5}^{+5.0}$ & $1_{-83}^{+15}$ & $-3.4_{-1.5}^{+6.0}$ & $\cdots$ & $\cdots$ & $\cdots$ & $274\,(334)$ & $722$ \\
        \cline{4-13}
         &  &  & ${\rm P}$ & ${\rm P}_1$ & $7.6_{-0.4}^{+0.4}$ & $2.6_{-2.0}^{+1.7}$ & $-4.1_{-0.5}^{+0.5}$ & $\cdots$ & $\cdots$ & $-0.01_{-0.20}^{+0.15}$ & $275\,(336)$ & $724$ \\
        \Xhline{5\arrayrulewidth}
    \end{tabular}
    \caption[]{Continued.}
\end{table*}
\endgroup

\def\arraystretch{1.55}
\begingroup
\renewcommand{\thetable}{\ref{tab:qu_fitting_appendix}}
\begin{table*}
    \centering
    \small
    \setlength{\tabcolsep}{1.0pt}
    \begin{tabular}{lccllcccccccc}
        \Xhline{5\arrayrulewidth}
        \multicolumn{1}{c}{Date} &
        \multicolumn{1}{c}{\begin{tabular}[c]{@{}c@{}}$I_0$ (mJy)\\$\alpha_\nu$\end{tabular}} &
        \multicolumn{1}{c}{Ejecta} &
        \multicolumn{1}{c}{Model} &
        \multicolumn{1}{c}{Comp.} &
        \multicolumn{1}{c}{$p_0$ (\%)} &
        \multicolumn{1}{c}{$\psi_0$ (deg)} &
        \multicolumn{1}{c}{$\phi_f$ ($\rad\,\mathrm{m}^{-2}$)} &
        \multicolumn{1}{c}{$\sigma_\phi$ ($\rad\,\mathrm{m}^{-2}$)} &
        \multicolumn{1}{c}{$N$} &
        \multicolumn{1}{c}{$\beta$} &
        \multicolumn{1}{c}{$\chi^2$ (dof)} &
        \multicolumn{1}{c}{$\ln \mathcal{Z}$} \\
        \Xhline{5\arrayrulewidth}
        \multirow[t]{4}{*}{\begin{tabular}[t]{@{}l@{}}2024-02-10 03:38\\60350.152\end{tabular}} & \multirow[t]{4}{*}{\begin{tabular}[t]{@{}c@{}}$3.50 \pm 0.02$\\$-0.69 \pm 0.04$\end{tabular}} & \multirow[t]{4}{*}{$\times$} & ${\rm \textbf{S}}$ & ${\rm S}_1$ & $7.7_{-0.5}^{+0.4}$ & $89_{-2}^{+3}$ & $-1.8_{-0.6}^{+0.6}$ & $\cdots$ & $\cdots$ & $\cdots$ & $89\,(93)$ & $189$ \\
        \cline{4-13}
         &  &  & \multirow{2}{*}{${\rm SS}$} & ${\rm S}_1$ & $8_{-3}^{+3}$ & $-90_{-5}^{+6}$ & $-1.6_{-0.7}^{+0.5}$ & $\cdots$ & $\cdots$ & $\cdots$ & $87\,(90)$ & $185$ \\
         &  &  &  & ${\rm S}_2$ & $1.1_{-1.1}^{+3.0}$ & $0_{-26}^{+16}$ & $-0.7_{-1.6}^{+7.0}$ & $\cdots$ & $\cdots$ & $\cdots$ & $87\,(90)$ & $185$ \\
        \cline{4-13}
         &  &  & ${\rm P}$ & ${\rm P}_1$ & $7.9_{-0.8}^{+0.7}$ & $89_{-2}^{+3}$ & $-1.7_{-0.7}^{+0.5}$ & $\cdots$ & $\cdots$ & $-0.1_{-0.4}^{+0.3}$ & $89\,(92)$ & $186$ \\
        \Xhline{3\arrayrulewidth}
        \multirow[t]{2}{*}{\begin{tabular}[t]{@{}l@{}}2024-02-10 03:38\\60350.152\end{tabular}} & \multirow[t]{2}{*}{\begin{tabular}[t]{@{}c@{}}$1.328 \pm 0.016$\\$-0.62 \pm 0.07$\end{tabular}} & \multirow[t]{2}{*}{$\checkmark$} & \multirow[t]{2}{*}{${\rm \textbf{S}}$} & ${\rm S}_1$ & $4.9_{-1.2}^{+1.2}$ & $-14_{-6}^{+7}$ & $-1.5_{-0.6}^{+0.7}$ & $\cdots$ & $\cdots$ & $\cdots$ & $74\,(93)$ & $106$ \\
        & & & & \multicolumn{9}{c}{} \\
        \Xhline{3\arrayrulewidth}
        \multirow[t]{4}{*}{\begin{tabular}[t]{@{}l@{}}2024-02-19 04:21\\60359.182\end{tabular}} & \multirow[t]{4}{*}{\begin{tabular}[t]{@{}c@{}}$2.912 \pm 0.019$\\$-0.61 \pm 0.04$\end{tabular}} & \multirow[t]{4}{*}{$\times$} & ${\rm \textbf{S}}$ & ${\rm S}_1$ & $11.4_{-0.6}^{+0.5}$ & $87_{-2}^{+2}$ & $-1.9_{-0.5}^{+0.6}$ & $\cdots$ & $\cdots$ & $\cdots$ & $80\,(91)$ & $175$ \\
        \cline{4-13}
         &  &  & \multirow{2}{*}{${\rm SS}$} & ${\rm S}_1$ & $12_{-2}^{+4}$ & $88_{-4}^{+6}$ & $-1.8_{-0.6}^{+0.5}$ & $\cdots$ & $\cdots$ & $\cdots$ & $79\,(88)$ & $172$ \\
         &  &  &  & ${\rm S}_2$ & $1.0_{-1.0}^{+3.0}$ & $-2_{-14}^{+90}$ & $-0.5_{-1.7}^{+8.0}$ & $\cdots$ & $\cdots$ & $\cdots$ & $79\,(88)$ & $172$ \\
        \cline{4-13}
         &  &  & ${\rm P}$ & ${\rm P}_1$ & $12.2_{-0.9}^{+0.9}$ & $87_{-3}^{+2}$ & $-1.9_{-0.5}^{+0.6}$ & $\cdots$ & $\cdots$ & $-0.4_{-0.2}^{+0.3}$ & $79\,(90)$ & $173$ \\
        \Xhline{3\arrayrulewidth}
        \multirow[t]{2}{*}{\begin{tabular}[t]{@{}l@{}}2024-02-19 04:21\\60359.182\end{tabular}} & \multirow[t]{2}{*}{\begin{tabular}[t]{@{}c@{}}$0.932 \pm 0.018$\\$-0.94 \pm 0.11$\end{tabular}} & \multirow[t]{2}{*}{$\checkmark$} & \multirow[t]{2}{*}{${\rm \textbf{S}}$} & ${\rm S}_1$ & $8.5_{-1.8}^{+1.4}$ & $13_{-6}^{+5}$ & $-1.7_{-0.6}^{+0.7}$ & $\cdots$ & $\cdots$ & $\cdots$ & $154\,(105)$ & $44$ \\
        & & & & \multicolumn{9}{c}{} \\
        \Xhline{3\arrayrulewidth}
        \multirow[t]{4}{*}{\begin{tabular}[t]{@{}l@{}}2024-02-25 03:22\\60365.141\end{tabular}} & \multirow[t]{4}{*}{\begin{tabular}[t]{@{}c@{}}$2.274 \pm 0.015$\\$-0.66 \pm 0.04$\end{tabular}} & \multirow[t]{4}{*}{$\times$} & ${\rm \textbf{S}}$ & ${\rm S}_1$ & $8.1_{-0.7}^{+0.6}$ & $-85_{-3}^{+3}$ & $-1.4_{-0.6}^{+0.6}$ & $\cdots$ & $\cdots$ & $\cdots$ & $112\,(95)$ & $143$ \\
        \cline{4-13}
         &  &  & \multirow{2}{*}{${\rm SS}$} & ${\rm S}_1$ & $8_{-7}^{+3}$ & $-85_{-7}^{+5}$ & $-1.4_{-0.6}^{+0.6}$ & $\cdots$ & $\cdots$ & $\cdots$ & $112\,(92)$ & $139$ \\
         &  &  &  & ${\rm S}_2$ & $0.9_{-0.9}^{+6.0}$ & $-89_{-19}^{+15}$ & $-0.4_{-1.7}^{+5.0}$ & $\cdots$ & $\cdots$ & $\cdots$ & $112\,(92)$ & $139$ \\
        \cline{4-13}
         &  &  & ${\rm P}$ & ${\rm P}_1$ & $8.2_{-1.0}^{+0.9}$ & $-85_{-3}^{+3}$ & $-1.5_{-0.6}^{+0.7}$ & $\cdots$ & $\cdots$ & $-0.1_{-0.5}^{+0.4}$ & $112\,(94)$ & $141$ \\
        \Xhline{3\arrayrulewidth}
        \multirow[t]{2}{*}{\begin{tabular}[t]{@{}l@{}}2024-02-25 03:22\\60365.141\end{tabular}} & \multirow[t]{2}{*}{\begin{tabular}[t]{@{}c@{}}$1.208 \pm 0.018$\\$-0.71 \pm 0.09$\end{tabular}} & \multirow[t]{2}{*}{$\checkmark$} & \multirow[t]{2}{*}{${\rm \textbf{S}}$} & ${\rm S}_1$ & $9.1_{-1.2}^{+1.3}$ & $9_{-4}^{+4}$ & $-1.5_{-0.5}^{+0.7}$ & $\cdots$ & $\cdots$ & $\cdots$ & $97\,(95)$ & $90$ \\
        & & & & \multicolumn{9}{c}{} \\
        \Xhline{3\arrayrulewidth}
        \multirow[t]{4}{*}{\begin{tabular}[t]{@{}l@{}}2024-03-31 00:44\\60400.031\end{tabular}} & \multirow[t]{4}{*}{\begin{tabular}[t]{@{}c@{}}$1.44 \pm 0.02$\\$-0.57 \pm 0.09$\end{tabular}} & \multirow[t]{4}{*}{$\times$} & ${\rm \textbf{S}}$ & ${\rm S}_1$ & $1.8_{-1.3}^{+0.8}$ & $-6_{-16}^{+14}$ & $-1.3_{-0.6}^{+0.7}$ & $\cdots$ & $\cdots$ & $\cdots$ & $104\,(89)$ & $96$ \\
        \cline{4-13}
         &  &  & \multirow{2}{*}{${\rm SS}$} & ${\rm S}_1$ & $1.3_{-1.3}^{+1.8}$ & $-2_{-14}^{+14}$ & $-1.3_{-0.6}^{+0.6}$ & $\cdots$ & $\cdots$ & $\cdots$ & $103\,(86)$ & $92$ \\
         &  &  &  & ${\rm S}_2$ & $1.2_{-1.2}^{+1.9}$ & $-17_{-82}^{+17}$ & $1_{-2}^{+12}$ & $\cdots$ & $\cdots$ & $\cdots$ & $103\,(86)$ & $92$ \\
        \cline{4-13}
         &  &  & ${\rm P}$ & ${\rm P}_1$ & $0.5_{-0.5}^{+1.2}$ & $-5_{-18}^{+15}$ & $-1.3_{-0.7}^{+0.6}$ & $\cdots$ & $\cdots$ & $0.6_{-1.0}^{+4.0}$ & $104\,(88)$ & $96$ \\
        \Xhline{3\arrayrulewidth}
        \multirow[t]{2}{*}{\begin{tabular}[t]{@{}l@{}}2024-03-31 00:44\\60400.031\end{tabular}} & \multirow[t]{2}{*}{\begin{tabular}[t]{@{}c@{}}$0.861 \pm 0.014$\\$-1.05 \pm 0.10$\end{tabular}} & \multirow[t]{2}{*}{$\checkmark$} & \multirow[t]{2}{*}{${\rm \textbf{S}}$} & ${\rm S}_1$ & $9.2_{-2.0}^{+1.6}$ & $3_{-5}^{+7}$ & $-1.3_{-0.5}^{+0.7}$ & $\cdots$ & $\cdots$ & $\cdots$ & $70\,(81)$ & $58$ \\
        & & & & \multicolumn{9}{c}{} \\
        \Xhline{5\arrayrulewidth}
    \end{tabular}
    \caption[]{Continued.}
\end{table*}
\endgroup

\bsp	
\label{lastpage}
\end{document}